%% file: __article_CubeInterestingness.tex
\documentclass{article}
\usepackage[utf8]{inputenc}
\usepackage[english]{babel}
\usepackage[dvipsnames]{xcolor,colortbl}
\usepackage{url}
\usepackage{graphicx}
\usepackage{amsfonts}
\usepackage{amsmath,amssymb}
\usepackage{amsthm}
\usepackage[linesnumbered, ruled]{algorithm2e}

\usepackage{typed-checklist} 

\theoremstyle{definition}

\theoremstyle{remark}
\newtheorem*{remark}{Remark}

\DeclareGraphicsExtensions{.png,.jpg,.pdf}

\newcommand{\silence}[1]{}

\newcommand{\forcepagebreak}{\clearpage \newpage}

\newif\ifDRAFT
\DRAFTfalse       

\ifDRAFT
    \newcommand{\isDraft}[1]{#1}
\else
    \newcommand{\isDraft}[1]{}
\fi

\ifDRAFT
    \def\inlineRem#1{{\noindent\bf\fcolorbox{red}{white}{\color{blue}{$\|$~{#1}~$\|$}}}}
\else
    \newcommand{\inlineRem}[1]{}
\fi

\ifDRAFT
    \def\rem#1{\colorbox{yellow}{\scriptsize\color{brown} @@} \marginpar{\scriptsize\color{brown} #1}}
\else
    \newcommand{\rem}[1]{}
\fi

\ifDRAFT
    \def\sideNew#1{\marginpar{\indent\color{ForestGreen}{\footnotesize{\textbf{#1}}}}}
\else
    \newcommand{\sideNew}[1]{}
\fi

\def\sticky#1{\colorbox{yellow}{\scriptsize\color{blue} @@} \marginpar{\scriptsize\color{red} #1}}

\title{Cube Query Interestingness: Novelty, Relevance, Peculiarity and Surprise}
\author{
    Dimos Gkitsakis$^{[0009-0006-9559-819X]}$, Spyridon Kaloudis \\
	University of Ioannina, Ioannina 45110, Greece \\
    \textsf{ \{dgkits@cs.uoi.gr, kaloudis.sp@gmail.com\} }\\
    \and
    Eirini Mouselli\footnote{Work done with Univ. Ioannina.}\\
    Natech S.A.\\
    Ioannina, Greece\\
    \textsf{e.mouselli@natechsa.com}\\
    \and
    Veronika Peralta$^{[0000-0002-9236-9088]}$, Patrick Marcel$^{[0000-0003-3171-1174]}$\\
	University of Tours, Blois, France \\
    \textsf{firstname.lastname@univ-tours.fr}\\
    \and
    Panos Vassiliadis $^{[0000-0003-0085-6776]}$  \\
	University of Ioannina, Ioannina 45110, Greece \\
	\textsf{pvassil@cs.uoi.gr}
}

\begin{document}

\maketitle

\begin{abstract}
In this paper, we discuss methods to assess the interestingness of a query in an environment of data cubes. We assume a hierarchical multidimensional database, storing data cubes and level hierarchies. We start with a comprehensive review of related work in the fields of human behavior studies and computer science. We define the interestingness of a query as a vector of scores along different dimensions, like novelty, relevance, surprise and peculiarity and complement this definition with a taxonomy of the information that can be used to assess each of these dimensions of interestingness. We provide both syntactic (result-independent) and extensional (result-dependent) checks, measures and algorithms for assessing the different dimensions of interestingness in a quantitative fashion. We also report our findings from a user study that we conducted, analyzing the significance of each dimension, its evolution over time and the behavior of the study's participants.   
\end{abstract}

\input{01_CubeInterestIntro}

\forcepagebreak
\input{02_related}

\forcepagebreak
\input{03_model}
\forcepagebreak

\input{04_taxonomy}

\forcepagebreak
\input{05_novelty}

\forcepagebreak
\input{06_relevance}

\forcepagebreak
\input{07_peculiarity}

\forcepagebreak
\input{08_surprise}
\forcepagebreak
\input{09_kaloudis-exps}
\forcepagebreak
\input{10_userStudy}

\forcepagebreak
\input{11_conclusions}

\bibliographystyle{alpha}
\bibliography{localBib}

\ifDRAFT
    \forcepagebreak
    \input{99_CommentsKept}
    \forcepagebreak
    \input{00_ToDo}
\fi    
\end{document}

%% file: 01_CubeInterestIntro.tex
\section{Introduction}\label{sec:intro}
\textit{How interesting is a (data cube) query?} What are the fundamental characteristics that make a (data cube) query interesting for a user?

Assessing query interestingness is important for at least two common scenarios: (a) \emph{a-priori} interestingness prediction, and, (b) \emph{a-posteriori} interestingness evaluation. 
\begin{itemize}
    \item A-priori prediction of query interestingness occurs in the case where a recommender system is in the process of automatically generating candidate queries, in order to provide the user with an overview of the information space, as well as with suggestions on how to explore it, or how to follow up on previous query in an on-going query session.
    \item A-posteriori evaluation of query interestingness is relevant in the case where a large number of queries have already been issued (possibly by other users too), they are cached and readily available, and we need to pick the ones that seem the most significant either in order to recommend them to a user, or, because they highlight best the user actions and goals in the query session.
\end{itemize}

The above are by no means an exhaustive enumeration of cases where the evaluation of query interestingness is important. The common thread in both cases, however, is that both for reasons of efficiency and computational overhead, and for reasons of cognitive load of the person who is involved in the process, it is imperative that a small subset of queries, out of a large number of candidates, are picked for further processing. 

In our deliberations, we focus on data organized in cubes due to 
(a) their extreme relevance to the problem, as analysts explore data in query sessions via Business Intelligence tools, 
(b) their simplicity -- as the simplest possible database setting in terms of how data are presented to the end-users,
(c) their most focused setup, also due to the simplicity of the underlying schema, but also because the queries follow a pattern of filtering and grouping with very specific joins between the dimension and fact tables, 
and, (d)  the richness of information content, due to the presence of hierarchically structured dimensions that allow manipulating, examining and understanding the data from multiple layers of abstraction.
In other words, cubes are relevant to the problem, simple, allow focused query sessions to take place and demonstrate information richness. This last property is also what differentiates cube queries from regular, relational ones: the presence of a hierarchical multidimensional space allows comparisons at multiple levels of granularity that would otherwise be very hard to express or detect in a plain relational environment.
\footnote{The observant reader might have already forecasted that after having successfully addressed the problem in such a setting, generalizing it to arbitrary database schemata, queries and user intention is the next step; the simile is like solving the problem in vitro in a lab, before addressing it in an industrial factory.} 

Therefore, in our work, we assume an OLAP environment, consisting of cubes, dimensions, levels, and aggregate cube queries posed in the context of user sessions. We will also assume the ability to register, extract, or simply approximate user goals, beliefs and profiles.

What is then the assessment of interestingness for cube queries? To address the question, we will first frame the assessment aspect: we regard assessment as the process where an assessor (person or software) examines specific properties of an object that is evaluated (in our case: cube queries), within a certain context (in our case, as we will demonstrate, the multidimensional space, the query history, the goals, beliefs and interests of the user), for its degree of support/fulfillment of a property (in our case: interestingness aspects) via a method that objectively quantifies the above degree of support via a numerical score or label that is interpretable via a reference scale of assessment.

Intuitively speaking, we need to establish the different properties/dimensions of interestingness and introduce algorithms to numerically assess the objects of study (cube queries) for their ``performance" with respect to these properties, in the context of a specific user (with his own characteristics) and a specific session.

Which are these properties, then? Based on the study of the related literature, both in the area of psychology, and in the area of computer science, we have concluded that \textit{interestingness is not a single entity, but rather, a vector of scores along several dimensions}~\cite{DBLP:conf/adbis/MarcelPV19}:

\begin{itemize}
\item Relevance: the extent to which a piece of information (here: 
the results of the query) is related to the overall information goals and 
preferences of the user.
\item Surprise: the extent to which the result of the query 
contradicts and revises the user's prior beliefs.
\item Novelty: the extent to which the information presented to the users 
is new, and previously unknown to them.
\item Peculiarity: the extent to which the  query is different, and not in accordance with the previous queries of the session or history.
\end{itemize}


\textit{In this paper, we provide a systematic taxonomy of the dimensions of interestingness, and their relationship with the case of data cubes in hierarchically structured multidimensional spaces, and, we propose specific measures and algorithms for assessing the different dimensions of cube query interestingness in a quantitative fashion.}\\

\textbf{Contributions and Roadmap}. The contributions of this paper, and the way they are laid out in this document are as follows:
\begin{itemize}
    \item In Section~\ref{sec:related}, we perform a comprehensive review of related work. We do not just survey the related work in the field of computer science, but lay the foundations of our work from the literature in the field of psychology and studies of human behavior. Thus, foundational concepts like interest, curiosity, novelty and surprise act as the starting point for our deliberations.
    \item In Section~\ref{sec:background}, we provide a formal framework of the data space within which we operate, along with a reference example, to be used throughout the paper.
    \item We define the interestingness of a query as a vector of scores along difference dimensions, like novelty, relevance, surprise and peculiarity. To assess these scores, we need metrics and algorithms. Before presenting such tools, however, in Section~\ref{section:_GoBack}, we provide a taxonomy of what information can be exploited, or, equivalently, is needed, for each of the dimensions of interestingness.
    \item In the context of the aforementioned taxonomy, for all the interestingness dimensions, we provide both syntax-based (result-independent) and extension-based (result-dependent) measures and algorithms, structured as follows: novelty is discussed in Section~\ref{sec:novelty}, relevance in Section~\ref{sec:relevance}, peculiarity in Section~\ref{sec:peculiarity}, and, surprise in Section~\ref{sec:Surprise}. 
    \item We assess the proposed framework in terms of effectiveness and efficiency. Concerning the efficiency of the proposed metrics, we present an experimental evaluation in Section~\ref{sec:exps}. Concerning the effectiveness of the framework of interestingness dimensions, in Section~\ref{sec:userStudy}, we present the results of a user study that we conducted, analyzing the significance of each dimension, its evolution over time and the behavior of the study's participants.  We demonstrate that although no particular dimension dominates the overall interest for a query, surprise and relevance seem to be more significant. Novelty seems to gain some significance later in the user deliberations, whereas surprise progressively loses significance as the time progresses. 
    \item Finally, we conclude our deliberations in the final section, with points for future work.
\end{itemize}

%% file: 02_related.tex
\sideNew{Sec. \ref{sec:related} NEW!}
\section{Related work} \label{sec:related}
In this Section, we start by surveying the different aspects of interestingness in the field of psychology and the study of human behavior. Then, we move on to survey how computer science has attempted to address the issue. 

\subsection{Interestingness from the viewpoint of the study of human behavior}

How can we define interestingness? In this subsection, we frame an answer to this 
question from the viewpoint of the study of human behavior. 

\paragraph{Interest} To the best of our knowledge, there is no formal definition of 
interestingness. We define interestingness as the property of an object, event or piece of information to be of interest to an individual. Of course this delegates the definition to the task of defining interest.

\textit{Online Definitions}. Online definitions of interest return "the feeling of 
wanting to know or learn about something or someone."\footnote{\texttt{https://en.oxforddictionaries.com/definition/interest}}, "the feeling 
of wanting to give your attention to something or of wanting to be 
involved with and to discover more about something"\footnote{\texttt{https://dictionary.cambridge.org/dictionary/english/interest}}, "a: 
feeling that accompanies or causes special attention to an object or 
class of objects; concern; b: something that arouses such attention; 
c: a quality in a thing arousing interest"\footnote{\texttt{https://www.merriam-webster.com/dictionary/interest}} - in other words, 
it appears that \textit{interest is mostly characterized by the urge of 
	learning more about a subject}. Our exploration of Wikipedia\footnote{ 
	\texttt{https://en.wikipedia.org/wiki/Interest\_(emotion)}} revealed a 
consistent definition of interest as "\,Interest is a feeling or emotion 
that causes attention to focus on an object, event, or process. In 
contemporary psychology of interest, the\emph{ term is used as a 
	general concept that may encompass other more specific psychological 
	terms, such as curiosity and to a much lesser degree surprise}". 
Practically, this means that \emph{the \textbf{interestingness of a piece of information} is the \textbf{degree} to which this piece of information ignites the emotion of \textbf{curiosity} (which in terms, means the desire to \textbf{acquire more knowledge on the issue}), or (less importantly) \textbf{surprise} (i.e., the detection -and adaptation to- a \textbf{discrepancy} between \textbf{newly acquired information} and \textbf{preexisting cognitive schemas}}).

\textit{Interest from the viewpoint of psychology}. In more technical terms, psychology characterizes interest along similar dimensions. In \cite{2008Silvia}, interest is characterized as an emotion whose function is to motivate learning and exploration. The author explains that it is hard to structure characteristics of interest due to between-people variability (different people are interested in different things) and within-person variability as interest changes over time. As emotions come from appraisals, i.e., the way people evaluate events, the author argues that interest comes from two appraisals: (a)  the evaluation of an event's \textbf{novelty} and \textbf{complexity} ("Intuition and decades of research (Berlyne, 1960) show that new, complex, and unexpected events can cause interest") and (b) the evaluation of the \textbf{comprehensibility} of an event.

In \cite{2014RouSu}, the authors provide a definition of interest from the viewpoint of psychology: "We define interests as trait-like preferences for activities, contexts in which activities occur, or outcomes associated with preferred activities that motivate goal-oriented behaviors and orient individuals toward certain environments." The definition highlights two aspects of interests: (a) they are trait-like and (b) they are contextualized, because of an object or activity of interest. According to the authors, interest not only determines choices that people make, but also the success they achieve. In \cite{Su201911}, a new theory combines two aspects. On the one hand, interest, referred to as situational interest, is defined as "momentary feelings of curiosity, fascination, and enjoyment triggered by an environment or a task" along with "cognitive evaluations of the value or importance of the environment or task". On the other hand, interests are also traits, referred to as dispositional  or individual interests. The authors show how the two aspects can be combined and emphasize the contextualization of interest, i.e., the need for an object of interest in relation to an environment.

On the basis of the aforementioned aspects of interest, we further explore the related concepts of novelty, curiosity and surprise, in order to 
determine more concretely what interest is all about. In the context of these deliberations, peculiarity also emerges as an important factor too.

\paragraph{Curiosity} \emph{Are you hungry for new information?} According to \cite{Litm05}, "curiosity may be defined as the desire to know, to see, 
or to experience that motivates exploratory behaviour directed towards 
the acquisition of new information". \cite{Litm05} gives a vivid 
presentation of how antagonizing theories on curiosity can converge to a 
unifying model. Specifically, the curiosity-drive theory treats 
curiosity as a need to acquire information in order to close a knowledge 
gap between information that is known and information that is unknown to 
them. This is inline with one of the most highly cited works in the area \cite{Loewenstein1994}. The optimal stimulation theory suggests that the exploration for 
information that takes place concerns pleasurable states of arousal. The 
combination of these two models into a single, "interest/deprivation" 
model (where curiosity stems from 'deprivation of information' or from 
'interest' (towards pleasing emotions)) is also discussed. \cite{Litm05} 
also makes a connection of this unifying model to the different neural 
circuits of wanting and liking which are correlated but distinct and 
discusses the issue of indifference/boredom/lack of curiosity.

\textit{So how is curiosity related to a person's interest after all? 
	Depending on whether we are hungry to eliminate our ignorance, or simply 
	enjoying learning something new, the answer can differ. In the first 
	case, when the 'wanting' of information is intense, a concrete answer to 
	an underlying question, the 'solving of a puzzle' and, in summary, the 
	\textbf{closing of the knowledge gap}, are the issues that have to be addressed. In the second case, 
	\textbf{novelty via new and unusual stimuli} ('tell me something I don't know') 
	seems to be the answer (and esp., the cure for boredom when both the 
	'wanting' and 'liking' motives are low). }

We refer the interested reader to \cite{Kidd15} for a recent survey on the developments in the area of understanding curiosity; \cite{Loewenstein1994} albeit older gives a nice categorization of the efforts encountered up to its time (including a historic overview starting from Aristotle and St. Augustin, to Bentham, Kant, Freud and Pavlov) and also offers the information gap theory which seems to withstand criticism up to now.

\paragraph{Novelty} \cite{FoMG10} discusses novelty from the viewpoint of psychology with 
respect to when people characterize events as novel, and how the mental 
processing of these events takes place. Interestingly, people are 
predominately correlating positive feelings to the opposite of novelty, 
familiarity. Yet, this does not necessarily mean that novelty is 
correlated with negative feelings; in fact, it appears that both our 
attraction/aversion to a novel event, as well as the characterization of 
the event itself as novel or not, depend on several other factors 
(predisposition being a major one). 

Novelty occurs when an event (in our case: demonstrated information) 
does not fit existing mental categories. People are not necessarily 
negatively predisposed to such a situation, due to their inherent 'motive to 
know' (as already mentioned for curiosity, closing knowledge gaps can 
produce pleasant feelings). Then, people try to understand it and in 
order to do so, they apply a typical mental reaction: they try to relate 
it to events or information with which they are already familiar. Practically this 
means that people try first to abstract the incoming input and 
categorize it in larger, pre-existing mental categories (practically 
searching for similarities with these larger categories). If this 
attempt fails, the focus is shifted to details and dissimilarities from 
more detailed mental categories, on the grounds of detailed aspects. 
Notably, the above process is not followed in the case of threat, where 
people immediately focus to the details, as typically happens when 
self-protective motives predominate. 

\emph{Overall, novelty is strongly 
	correlated to curiosity and occurs when a person fails to include the 
	demonstrated information / event / object into a \textbf{pre-existing mental 
		category}. The processing of novel information starts from trying 
	to align it with high-level, abstract phenomena that promote the 
	\textbf{detection of commonalities}, and later, esp., if the commonalities are 
	not there, with a drilling into the \textbf{details that cause dissimilarities}. }

\paragraph{Surprise} Surprise is the third aspect of interest that we discuss. 
\cite{ReMN12} defines surprise as "A peculiar state of mind, usually of 
brief duration, caused by unexpected events of all kinds \ldots (via) 
\ldots an evolved mechanism whose function is (a) to detect 
discrepancies between cognitive schemas and newly acquired information, 
and (b) if they are detected, to instigate processes that enable the 
short- and long-term adaptation to them." Practically speaking, the main 
idea is that our beliefs about objects, events and their sequences are 
structured in so-called 'schemas' and whenever a significant discrepancy 
(above a certain threshold) is detected between the underlying belief 
schema and a new input (new information, in our case), the surprise 
mechanism elicits a surprise reaction that involves (a) analysis and 
evaluation of the event, (b) the possible reaction to it, and (c) the 
revision and adaptation of the schema, to remove the discrepancy. 
Therefore, \emph{surprise occurs when our previous beliefs are disconfirmed or 
	contradicted}. In fact, there are two types of surprise depending on what 
kind of belief is challenged: (a) misexpected events occur when a belief 
is directly challenged (e.g., I originally believed that sales in Athens 
are approximately 100K and they turn out to be less than 50K, which I 
deem as an important discrepancy), and, (b) unexpected events occur when 
an implied belief is challenged, due to the challenging of background or 
contextual beliefs (e.g., I expected to see a drop in the sales of wine, 
because the price had gone up, but instead consumption turned out to be 
steady). 

Surprise is different from novelty: whereas surprise involves new information  
that challenges the things we already know, novelty involves new information concerning things that we 
did not previously know.

An important lesson coming from the study of the mechanisms of surprise 
is that when attempting to enrich our data exploration systems with a 
forecasting of what can be surprising for the user, it is important to 
(a) try to structure the \textbf{beliefs of the users} (practically: the values they 
expect to see) for the explored data in a structured schema (which can 
include rules, inferences, probabilities, \ldots , based on factual 
data, the history of what they have seen before, explicitly stated 
assumptions that the users make, etc.), and, (b) to incorporate mechanisms 
of adapting this schema to new information, as it progressively 
demonstrates itself.

\paragraph{Peculiarity} Both \cite{Loewenstein1994} and \cite{Kidd15} discuss the efforts of D.E. Berlyne \cite{berlyne54} to establish a taxonomy on curiosity. The taxonomy classifies curiosity as perceptual (typically encountered in animals) vs epistemic (mostly encountered in humans, aimed at acquiring knowledge) on the one hand, as well as specific (targeted at a particular piece of information) vs diversive (not associated with specific rewards or punishments). Diversive curiosity - which is not part of the above-mentioned dimensions of interestingness -- has been heavily criticized as concerns its essence as curiosity or not (see \cite{Loewenstein1994} for a discussion), yet it reveals a new possibility, the one of seeking information beyond a specific task, the one that \cite{Litm05} tries to unify into a single theory with the closing of an information gap. \cite{Kidd15} makes an interesting observation on information tradeoff tasks: \emph{"The optimal strategy requires adjudication between exploration (sampling to improve knowledge and, therefore, future choices)
and exploitation (choosing known best options). Sampling typically gives a lower immediate payoff but can provide information that improves choices in the future, leading to greater overall performance."}
In other words, the idea of sampling the information space for a broader understanding of what lies in it, might provide delayed rewarding, but overall greater performance. This view is further enhanced by the authors discussing how a longer time horizon strengthens the propensity of subjects to explore, as opposed to behaviors in the knowledge that the context will dramatically change soon, in which case subjects opt for more immediate rewards.
Thus, investing into understanding the information space in its entirety seems to be an inherent aspect of curiosity and thus interestingness, and to cover this aspect, \emph{we introduce \textbf{peculiarity} as a dimension of interest in our deliberations: whereas relevance is targeting towards pursuing a specific, exploitative goal, peculiarity and novelty aim to strengthen the understanding of the broader information space: novelty in terms of information not previously known, and peculiarity in terms of information significantly different than what is already known}.

\paragraph{What is not interestingness} Another way to address the issue is to frame the problem via an answer to the negative existential question -- here, this question is "what is NOT (cube) query interestingness?" From our point of view, any metric, or quality dimension, or, in general, any property of a query, falls out of scope with respect to our understanding of query interestingness if it does not help the user close an (intentional) information gap, or deeply understand the (broader) information space. To achieve this goal, we assume that the utility of such a property will be depending on the "current state of the user session" as this is expressed by the combination of user goals, beliefs, interests and query history.

\subsection{Earlier proposals of interestingness measures} 

Various interestingness measures were proposed in the different  areas of data exploration.
In this subsection we discuss interestingness 
measures proposed for 
(i) pattern mining,\rem{pattern mining may need updates, especially for surprise}  
(ii) recommendation,
and, (iii) interactive exploration of multidimensional datasets.


\subsubsection{Interestingness criteria for pattern mining}
In \cite{DBLP:journals/csur/GengH06}, the authors point out that interestingness
is a broad concept and identify from the literature 
9  criteria to determine whether or not a pattern is interesting:
conciseness, generality/coverage, reliability, peculiarity, diversity, novelty, surprisingness, utility and actionability/applicability. Specifically:
 \begin{itemize}
 \item   Conciseness. A pattern is concise if it contains relatively few attribute-value pairs, while a set of patterns is concise if it contains relatively few patterns.
 \item Generality/Coverage. A pattern is general if it covers a relatively large subset of a dataset.
 \item  Reliability. A pattern is reliable if the relationship des\-cribed by the pattern occurs in a high percentage of applicable cases.
 \item Peculiarity. A pattern is peculiar if it is far away from other discovered patterns according to some distance measure.
 \item  Diversity. A pattern is diverse if its elements differ significantly from each other, while a set of patterns is diverse if the patterns in the set differ significantly from each other. Diversity is a common factor for measuring the interestingness of summaries.
 \item  Novelty. A pattern is novel to a person if he or she did not know it before and is not able to infer it from other known patterns.
 \item  Surprisingness. A pattern is surprising (or unexpected) if it contradicts a person's existing knowledge or expectations. The difference between surprisingness and novelty is that a novel pattern is new and not contradicted by any pattern already known to the user, while a surprising pattern contradicts the user's previous knowledge or expectations. 
 \item Utility. A pattern is of utility if its use by a person contributes to reaching a goal.
 \item Actionability/Applicability. A pattern is actionable (or applicable) in some domain if it enables decision making about future actions in this domain.
 \end{itemize}
In \cite{DBLP:journals/csur/GengH06}, the authors categorize these criteria in 3 groups:
i) objective measures, based only on the raw data (generality, reliability, peculiarity, diversity, conciseness), like for instance the classical support, ii) subjective measures, considering both the data and the user (surprise and novelty), like for instance the informational content \cite{DBLP:conf/ida/Bie13}, and iii) semantic measures, based on the semantics and explanations of the patterns (utility and actionability), like for instance measures based on user preferences \cite{DBLP:series/sci/YaoCY06}. 

According to De Bie \cite{DBLP:conf/ida/Bie13}, subjective
interestingness is particularly well adapted for 
\textit{exploratory} data mining, whose goal is to pick patterns that will result in the best updates of the user's belief state, while presenting a minimal strain on the user's resources. The data mining process consists of extracting patterns and presenting first those that are subjectively surprising, and then refining the belief.
De Bie \cite{DBLP:conf/ida/Bie13} introduced a formal framework 
for defining measures of surprise for exploratory data mining,
using an information-theoretic approach.
The framework consists of quantifying 
the interactive exchange of information between data and user, 
accounting for the \textit{user's prior belief state}.
Of course, in this context, one challenge is how to define and update the belief of the user.
Approximating the belief that the user would attach to the result being expected
is modeled as a background distribution, namely,
a probability measure over the exploration results.
This background distribution, which initially can e.g.,
be uniform over all the exploration  results, 
is updated after each 
result is presented to the user. 

\subsubsection{Interestingness criteria for recommendations}

There is a long discussion about interestingness in the 
area of evaluating recommender systems \cite{DBLP:journals/tois/HerlockerKTR04,DBLP:journals/jmlr/GunawardanaS09,DBLP:journals/tiis/KaminskasB17}.
We mention \cite{DBLP:journals/tiis/KaminskasB17} as an excellent 
recent survey on the topic.
The survey presents 4 criteria (diversity, serendipity, novelty, and coverage), in addition to the traditional accuracy, for evaluating the quality of a recommendation.
 \begin{itemize}
 \item Diversity. The average/aggregated pairwise distance between items in the recommendation list, according to some distance measure.
 \item Serendipity. It refers to the process of "finding valuable or pleasant things that are not looked for". It  consists of two components: surprise and relevance.
 A common practice is to compare the generated recommendations with recommendations produced by a primitive baseline system, as the goal of a serendipitous recommender is to suggest items that are difficult to predict.
 \item Novelty. A novel recommended item is one that is previously unknown to the user.
 \item Coverage. It reflects the degree to which the generated recommendations cover the catalog of available items.
 \end{itemize}

\cite{DBLP:journals/tiis/KaminskasB17} defines novelty for recommender systems as "A novel recommended item is one that is previously unknown to the user" and then moves one to discuss the difference of novelty with (a) serendipity (a serendipitous item must be both novel and surprising) and (b) unexpectedness (an unexpected item does not have to be novel to the user, but only relevant and different from the user's expectations of what would be recommended to them).\\

Query recommendation techniques (see e.g., \cite{DBLP:journals/tkde/EirinakiAPS14,DBLP:journals/dss/AligonGGMR15}) 
are usually evaluated with interestingness measures coming from the literature on recommender systems exposed above. 
We mention the more  OLAP-specific \emph{foresight} measure \cite{DBLP:journals/dss/AligonGGMR15}, that quantifies how 
distant is the recommendation from the current point of 
exploration. 


\subsubsection{Interestingness criteria for interactive exploration of multidimensional datasets}

Started with the seminal papers by Sunita Sarawagi et al. \cite{DBLP:conf/edbt/SarawagiAM98},
various interestingness criteria have been proposed to qualify
an interesting property or pattern for a subset of the data in a dataset,
often called insight, highlights, findings, discoveries, etc., 
typically characterized by an interestingness score \cite{DBLP:conf/vldb/Sarawagi00,DBLP:journals/is/GkesoulisVM15,DBLP:journals/tvcg/WangSZCXMZ20,DBLP:conf/sigmod/ElMS20, DBLP:conf/sigmod/MiloS20}. 
Two works addressed the classification of these criteria \cite{DBLP:journals/csur/GengH06,DBLP:conf/adbis/MarcelPV19}.

In \cite{DBLP:journals/csur/GengH06}, the authors also review interestingness measures for what they call summaries, i.e., aggregated cross-tabs  corresponding to the result of an OLAP query, where numeric values (i.e., measures) are aggregated by several criteria (i.e., dimensions).
Out of the 9 criteria defined for pattern interestingness, 4 are adapted to summaries: 
\begin{itemize}
 \item  Diversity. Whether a summary is diverse is determined by two factors: the proportional distribution of classes in the population, and the number of classes.
 \item   Conciseness and Generality. Concise summaries are easily understood and remembered, and thus more interesting than complex ones. Then, a summary is more concise if it is more general (i.e., aggregated). 
 \item Peculiarity. A cell in a summary is peculiar if it is differs from the other cells in the summary.
 \item  Surprisingness/Unexpectedness. A summary is surprising if it deviates from user's expectations. For example, variance can be calculated by replacing observed probabilities by expected probabilities.  
 \end{itemize}

According to the classification of \cite{DBLP:journals/csur/GengH06}, the first three criteria are objective and the last one is subjective.

In our previous work  \cite{DBLP:conf/adbis/MarcelPV19}, 
we have previously identified four main dimensions that differ in what is contrasted  to generate interestingness:
 (i)  \emph{peculiarity} (P): the similarity of a cube query to a user's history is assessed (either at the level of the query expression or at the level of the query results); 
(ii) \emph{novelty} (N): a cube query is contrasted to a user's exploration history;
(iii)  \emph{relevance} (R): a cube query is contrasted to a user's exploration goal; and
(iv) \emph{surprise} (S): the result of a cube query is contrasted to a user's belief. 
We adopt  this classification to review the various interestingness measures proposed.

\paragraph{Peculiarity}
\rem{this part is quite large and may be summarized}

It appears that peculiarity has attracted most of the attention in the literature.
The main measures defined in this dimension concern either (i)
the significance, (ii) the coverage, or (iii) the coherency 
of the insights.

The \textit{significance}
of an insight 
\cite{DBLP:conf/sigmod/TangHYDZ17,DBLP:conf/chi/ZgraggenZZK18,DBLP:conf/sigmod/DingHXZZ19,DBLP:journals/isf/FranciaMPR22,DBLP:journals/vldb/AbuzaidKSGXSASM21,DBLP:conf/edbt/ChansonLMRT22}
allows to quantify its importance among its \textit{peer data}.
This importance is often related to the data distribution.
\cite{DBLP:journals/isf/FranciaMPR22} performs a preliminary ad-hoc attempt to measure 
significance via the difference in z-scores
of the data obtained in two consecutive exploration steps.
Recently, a trend is to turn insights into hypothesis testing \cite{DBLP:conf/chi/ZgraggenZZK18,DBLP:conf/sigmod/DingHXZZ19,DBLP:conf/edbt/ChansonLMRT22},
which has many advantages:
(i) using the p-value for the insight significance,
(ii) defining false discoveries (type-1 errors, e.g.,
visualizations supporting a non-significant insight)
and false omissions (type-2 errors, e.g., visualizations not supporting a significant insight),
(iii) defining credibility (e.g., percentage of  visualizations supporting an insight).
However, since the risk of type-1 error increases as more than one hypothesis are considered at once,
a correction is needed in the statistical test to ensure 
that non-spurious insights are reported \cite{DBLP:conf/chi/ZgraggenZZK18}.

Discovery-driven analysis \cite{DBLP:conf/vldb/Sarawagi99,DBLP:conf/vldb/Sarawagi00,DBLP:conf/edbt/SarawagiAM98,DBLP:conf/vldb/SatheS01} for measuring cell interestingness in the context of cube exploration is mostly based on peculiarity-related measures for individual cells.
Discovery-driven analysis guides the exploration of a datacube by providing users with interestingness values for measuring the peculiarity of the cells in a data cube, according to statistical models, e.g., based on the maximum entropy principle,
and leveraging the intrinsic structure of multidimensional information.
From an initial user query, the system automatically calculates 3 kinds of interestingness values for each cell in the query result: (i) $SelfExp$ measures the difference between the observed and anticipated values (the latter are calculated statistically by computing the mean of subsets of attributes), (ii) $InExp$ is obtained as the maximum of $SelfExp$ over all cells that are under this cell (those that result from a drill down), and (iii) $PathExp$ is calculated as the maximum of $SelfExp$ over all cells reachable by drilling down along a given path.
The DIFF, INFORM and RELAX advanced OLAP operators proposed
in \cite{DBLP:conf/vldb/Sarawagi99,DBLP:conf/vldb/Sarawagi00,DBLP:conf/vldb/SatheS01} use such interestingness values to recommend relevant cells for explaining drops or increases, or for recommending areas of a cube that should surprise the user, based on their history with the cube.

Klemettinen et al. \cite{KLEMETTINEN1999} use skewness, as a peculiarity measure of asymmetry in data distribution, for discovering interesting paths and guiding the navigation in a data cube. Given a cuboid, the possible drill-downs are explored, measuring skewness and generating skew-based navigation rules for the more significant paths. Skewness is computed observing the underlying facts (the raw data that is aggregated), looking for outliers or substantial differences with other facts. Based on skewness, Kumar et al. \cite{KUMAR2008} propose interestingness measures based on the unexpectedness of skewness in navigation rules and navigation paths.

Fabris and Freitas \cite{DBLP:conf/sbbd/FabrisF01} defined interestingness measures for attribute-value pairs in a data cube: the $I_1$ measure reflects the difference between the observed probability of an attribute-value pair and the average probability in the summary and the $I_2$ measure reflects the degree of correlation among two attributes. Both measures can be seen as value-based conciseness.

Two also recent works \cite{DBLP:conf/sigmod/ZhaoSZBUK17,DBLP:conf/sigmod/SalimiGS18}
are concerned with detecting the validity of insights gained by users when examining query answers. As with other works measuring peculiarity by leveraging the nature of OLAP cubes, this is again achieved by statistical tests comparing data at different levels of detail.

Measuring  the \textit{coverage} of the insight  consists of quantifying how
the subject of an insight represents the \textit{entire dataset} \cite{DBLP:conf/sigmod/TangHYDZ17,DBLP:conf/sigmod/DingHXZZ19,DBLP:conf/sigmod/00040HZ21}.
In most cases, anti-monotonic conditions are checked to
prune insights, like, for instance: if the subject of insight A is a superset of the subject of insight B, 
then the impact of A should be no less than the impact of B.

Characterizing the \textit{coherency} of an insight
compares the insight with others in the \textit{exploration session}, to
check whether a given exploratory operation is coherent at a certain point. 
For instance, in \cite{DBLP:conf/sigmod/ElMS20}
heuristic classification rules are used to express
general properties of the operations sequence
(e.g., a group-by on a continuous, numerical attribute is incoherent)
or on the input dataset's semantics
(e.g., if the user focuses on flight delays, aggregating on the ``departure-delay time'' columns is preferred).
Other works use distances between exploration actions to measure 
how coherent a sequence of actions is; for instance, in  \cite{DBLP:conf/edbt/ChansonLMRT22} 
a weighted Hamming distance of relational query parts is used.

\paragraph{Novelty}

Interestingness measures of the novelty dimension
are used to characterize data in terms of either
being new observations or operations in terms of  \textit{favoring  
going further} in the exploration.
In its simplest expression, novelty can simply be measured as
a Boolean indicating whether some data have already been seen \cite{DBLP:journals/isf/FranciaMPR22}.
However, more advanced definitions exist.
For instance, in \cite{DBLP:conf/sigmod/MiloS20},
a diversity measure is computed as the 
minimal Euclidean distance between the current observation and all the previous displays obtained.
In \cite{DBLP:conf/cikm/PersonnazABFS21},  curiosity
is inversely proportional to the number of times a result is encountered.

\paragraph{Relevance}

Interestingness measures of the relevance dimension
are used to characterize data in terms of 
the user being \textit{familiar} with them.
This dimension seems to be the one that attracted less attention. 
In \cite{DBLP:conf/cikm/PersonnazABFS21},
a familiarity measure is defined as the
concentration ratio of target objects in a set.
It is implemented as a variant of the
Jaccard index between objects encountered
during the exploration and a given
 target set of familiar objects.
 This measure is expected to increase as the exploration of the dataset goes on,
to avoid over-exploiting a set of familiar objects.

\paragraph{Surprise}

Chanson et al.  \cite{DBLP:conf/dolap/ChansonCDLM19},  propose  a way to measure subjective interestingness for exploratory OLAP, inspired by De Bie's work \cite{DBLP:conf/ida/Bie13}. The user belief is inferred based on the user’s past interactions over a data cube, the cube schema and the other users’ past activities. This belief is expressed by a probability distribution over all the query parts potentially accessible to the user. Surprise is then measured as in De Bie's work. Francia et al. \cite{DBLP:journals/isf/FranciaMPR22} propose to measure surprise as the proportion of values that have not been seen frequently, presented in models (e.g., clustering) extracted from the data under observation. In a quite different setting, Sintos et al. \cite{DBLP:journals/pvldb/SintosAY19} use the term surprise to refer to the extent of the incorrectness of a value in a data set -- practically measuring the amount of false information of two values before and after a data cleaning procedure. 




\paragraph{Combining interestingness measures}

Many works combine various interestingness measures, often measures
from different dimensions. 
As to how they are combined, there is no consensual approach. For instance a ratio is used in \cite{DBLP:conf/ida/Bie13},
a weighted sum is used in \cite{DBLP:conf/sigmod/ElMS20,DBLP:journals/isf/FranciaMPR22,DBLP:conf/cikm/PersonnazABFS21},
and a product is used in \cite{DBLP:conf/sigmod/TangHYDZ17,DBLP:conf/edbt/ChansonLMRT22}.
Djedaini et al. \cite{DBLP:conf/adbis/DjedainiLMP17, DjedainiDLMPV19} use supervised classification techniques for learning two interest measures for OLAP queries:  \textit{focus}, that indicates to what extent a query is well detailed and related to other queries in an exploration, indicating that the user investigates in details precise facts and learns from this investigation  \cite{DBLP:conf/adbis/DjedainiLMP17}, and
 \textit{contribution}, that highlights to what extent a query is important for an exploration, contributing to its interest and quality \cite{DjedainiDLMPV19}.

\subsection{Comparison to related work}
There are several axes of comparison to related work for this paper.

\paragraph{What is it so important that makes cube queries special?} As already mentioned in the introduction, the presence of multidimensional spaces with dimensions that are hierarchically structured provides a very specific environment, where cubes at different levels of detail can be related, although potentially defined with different schemata or selection conditions. This facilitates the assessment of all the different dimensions of interestingness at a much deeper level, as we can relate cube queries that would otherwise be unrelated.

\paragraph{Given the fact that there is so much previous literature in the field of data and knowledge management on interestingness, why is there a need for a new paper?} 
A second point that differentiates our work from the rest of the literature has to do that we follow a basic-principles approach, starting from the fundamentals of interest and its dimensions in psychology, to establish the ground upon which our modeling takes place. Moreover, in Section~\ref{section:_GoBack} we also provide a structured taxonomy of how the analysts' goals, beliefs and interests as well as the computational environment relates to the evaluation of the different aspects of interestingness. To the best of our knowledge, this is the first time that such a structuring (also involving the multi-level hierarchical dimensions of the data space) takes place.

\paragraph{Comparison to our own previous work}
Compared to our previous work on \textit{cell} interestingness~\cite{DBLP:conf/adbis/MarcelPV19}, apart from the basic dimensions of interestingness, the two papers have very little to share. In \cite{DBLP:conf/adbis/MarcelPV19} we deal with the problem of evaluating interestingness of individual cells rather than queries, which means we are restricted to the coordinates of the cells, rather than taking into consideration the semantics of the queries. However, a query is much more than a composition of its result cells, esp., if the interestingness of the query is to be assessed before deciding if we will execute it. To answer the reasonable question on why a recommender system might a-priori generate several candidate queries, we believe it is sufficient to mention that different queries rank differently according to different interestingness dimensions: therefore, several candidate queries may qualify based on different criteria. A trade-off of performance and interestingness might also affect the recommendation of queries.

Moreover, in \cite{DBLP:conf/dolap/GkitsakisKMPMV23}, we have presented a preliminary version of the present work as a first effort ever to explicitly handle the issue of assessing the interestingness of cubes and cube queries. The present paper extends \cite{DBLP:conf/dolap/GkitsakisKMPMV23} with
(i) an extensive review of related work (the current section), (ii) a taxonomy of the problem's parameters, presented in Section~\ref{section:_GoBack}, that allows us to clarify the problem and organize the algorithms assessing cube query interestingness in a principled way, (iii) several algorithms and metrics not mentioned in \cite{DBLP:conf/dolap/GkitsakisKMPMV23} for lack of space, and, (iv) a user study, to evaluate the effectiveness of the proposed algorithms and assess the significance and evolution over time of the assessed metrics.

%% file: 03_model.tex
\section{Formal Background \& Reference Example}\label{sec:background}

In our deliberations, we assume the formal model of \cite{PV21} (practically, extending \cite{DBLP:journals/is/VassiliadisMR19}) for the definition of the multidimensional space, cubes and cube queries. We follow a simplified apodosis of the formalities here to allow for a concise description. 

\subsection{Formal Background}

\textbf{Multidimensional space}. Data are defined in the context of a multidimensional space. The multidimensional space includes a finite set of dimensions. Dimensions  provide the context for factual measurements and will be structured in terms of dimension levels, which are 
abstraction levels that aid in observing the data at different levels of granularity. For example, the dimension $Time$ is structured on the basis of the dimension levels $Day$, $Month$, 
$Year$, $All$.

A \textit{dimension level} $L$ includes a name and a finite set of values, $dom$($L$), as its domain. Following the 
traditional OLAP terminology, the values that belong to the domains of the levels are called \textit{dimension members}, or simply \textit{members} (e.g., the values Paris, Rome, Athens are members of the domain of level $City$, and, subsequently, of dimension $Geography$).

A \textit{dimension} is a non-strict partial order of a finite set of \textit{levels}, obligatorily including (a) a most detailed level at the lowest possible level of coarseness, and (b) an upper bound, which is called $ALL$, with a single value 'All'. We denote the partial order of dimensions with $\preceq$, i.e., $D.L_{low}$ $\preceq$ $ D.L_{high}$ signifies that $D.L_{low}$ is at a lower level of coarseness than $D.L_{high}$ in the context of dimension $D$ -- e.g., $Geo.City$ $\preceq$ $Geo.Country$. 

We can map the members at a lower level of coarseness to values at a higher level of coarseness via an \textit{ancestor function} $anc_{L^l}^{L^h}()$. Given a member of a level $L_{l}$ as a parameter, say $v_{l}$, the function $anc_{L^l}^{L^h}()$ returns the corresponding ancestor value, for $v_{l}$, say $v_{h}$, at the level $L_{h}$, i.e., $v_{h}$ = $anc_{L^l}^{L^h}(v_{l})$. The inverse of an ancestor function is not a function, but a mapping of a high level value to a set of \textit{descendant values} at a lower level of coarseness (e.g., $Continent$ Europe is mapped to the set of all European cities at the $City$ level), and is denoted via the notation $desc_{L^h}^{L^l}()$. For 
example $Europe$ = $anc_{City}^{Continent}(Athens)$. See \cite{PV21} for more constraints and explanations.


\textbf{Cubes}. Facts are structured in cubes. A \textit{cube} $C$ is defined with respect to several dimensions, fixed at specific levels and also includes a number of \textit{measures} to hold the measurable aspects of its 
facts. Thus the schema of a cube is a set of attributes, including a set of dimension levels (over different dimensions) and a set of measures that include factual measurements for the data stored in the cube. Thus, the schema of a cube $schema(C)$, is a tuple, say $[D_{1}.L_{1}, ..., D_{n}.L_{n}, M_{1}, ..., M_{m}]$, with the combination of the dimension levels acting as primary key and context for the measurements and a set of measures as placeholders for the (aggregate) measurements. If all the dimension levels of a cube schema are the lowest possible levels of their dimension, the cube is a 
\textit{detailed cube}, typically denoted via the notation $C^{0}$ with a schema $[D_{1}.L^{0}_{1}, ..., D_{n}.L^{0}_{n}, M^{0}_{1}, ..., M^{0}_{m}]$. The results of a query $q$ is a set of cells that we denote as $q.cells$.

Each record of a cube $C$ under a schema
$[D_1.L_1, \dots, D_n.L_n, M_1, \dots, M_m]$, also known as a \textit{cell}, is a tuple $c$ = $[l_{1}, \ldots , l_{n}, m_{1}, \ldots , m_{m}]$,  such that $l_{i} \in dom(D_{i}.L_{i})$ and $m_{j}$ $\in$ $dom(M_{j})$. The vector $[l_{1}, \ldots , l_{n}]$ signifies the \textit{coordinates} of a cell. Equivalently, a $cell$ can be thought as a point in the multidimensional space of the cube's dimensions annotated with, or hosting, a set of measures. 

A cube $c$ includes a finite set of cells as its extension, which we denote as $q.cells$.

\textbf{Queries}. A cube query is a cube too, specified by (a) the detailed cube over which it is imposed, (b) a selection condition that isolates the facts that qualify for further processing, (c) the grouping levels, which determine the 
coarseness of the result, and (d) an aggregation over some or all measures of the cube that accompanies the grouping levels in the final result.

\begin{center}
    $q$ = $<C^{0}$, $\phi$, $[L_{1}, ..., L_{n},
M_{1}, ..., M_{m}]$, $[agg_{1}(M^{0}_{1} )$, ...,$agg_{m}(M^{0}_{m})]$ $>$
\end{center}

We assume (again, intentionally simplifying the model of \cite{DBLP:journals/is/VassiliadisMR19}):

\begin{itemize}

\item Selection conditions which are conjunctions of atomic filters of the form $L$ = $value$, or in general $L~\in~\{v_1,\dots,v_k\}$. Although our theoretical framework covers the latter, as the most general case, typically, the encountered expressions in practice are of the former, special-case, format. In any case, what is important is the property that selection conditions of this form can eventually be translated to their \textit{equivalent} selection conditions at the detailed level, via the conjunction of the detailed equivalents of the atoms of $\phi$. Specifically, assuming an atom $L~\in~\{v_1,\dots,v_k\}$, then $L^{0}$ $\in$ $\{desc_L^{L^0}(v_{1}) \cup ... \cup desc_L^{L^0}(v_{k})\}$, eventually producing an expression $L^{0}$ $\in$ $\{v'_{1} ,...,v'_{k'}\}$ is its detailed equivalent, called \textit{detailed proxy}. The reason for deriving $\phi^{0}$ is that $\phi^{0}$, as the conjunction of the respective atomic filters at the most detailed level, is directly applicable over $C^{0}$ and produces exactly the same subset of the multidimensional space as $\phi$, albeit at a most detailed level of granularity. For example, assume $Year~\in~\{2018,2019\}$, its detailed proxy is $Day~\in~\{2018/01/01,\ldots,2019/12/31\}$. We assume a single atomic filter per dimension. For a dimension $D$ that is not being explicitly filtered by any atom, one can equivalently assume a filter of the form $D.ALL~=~all$.

\item We define a grouping level for each dimension (remember that every dimension $D$ includes a single-valued level $D.ALL$, practically signifying the exclusion of the dimension from the grouping -- i.e., we group for 
all the members of the dimension).

\item Aggregation functions $agg_{i}$ belong to the set of frequently used aggregate functions like $\{sum,max, min, count, ...\}$ with the respective well-known semantics.

\end{itemize}

The semantics of the query are:

(i) apply $\phi^0$, the detailed equivalent of the selection condition over $C^{0}$ and produce a subset 
of the detailed cube, say $q^{0}$, known as the
detailed area of the query,

(ii) map each dimension member to its ancestor value at the level specified by the grouping levels and group the tuples with the same coordinates in the same same-coordinate group, 

(iii) for each same-coordinate group, apply the aggregate functions to the measures of its cells, thus producing a single value per aggregate measure.

A cube query is also a cube under the schema $[L_{1}, 
..., L_{n},M_{1}, ..., M_{m}]$, with the set of cells of the query result (denoted as $q.cells$) as its extension.\\


%

\textbf{Signatures and detailed areas}. We will use the term \textit{signature} to refer to sets of coordinates that specify an area of interest in the multidimensional space. Specifically:
\begin{itemize}
    \item The signature of a cell c, denoted as $c^+$, is its coordinates, that uniquely identify the area of the multidimensional space that pertains to it.
    \item The signature of an atomic filter $\mathbin{\alpha: L \in \{v_1,\ldots,v_k\}}$ is the value set $\{v_1,\ldots,v_k\}$ and it is denoted as $\alpha^+$.
    \item The signature of a selection condition of the form $\mathbin{\phi: \alpha_1 \wedge \dots \wedge \alpha_n}$ (assuming a single atom per dimension) is the expression $\phi^+: \alpha_1^+ \times \dots \times \alpha_n^+$. In other words, we compute the Cartesian product of the values of the involved atom signatures.
    \item The signature of a query $q$, $q^+$ is the \textit{set} of coordinates computed as follows: (a) compute the signature, i.e., the set of coordinates pertaining to $\phi^0$, the detailed equivalent of its selection condition; (b) within each of these coordinates, replace the (detailed) value of each dimension by its ancestor value at the level of the schema of the query. This guarantees that the resulting coordinates will be the coordinates of the query result.
\end{itemize}

The \textit{detailed signatures} of the above categories are produced by replacing the respective values of their regular signatures with the expression $desc_L^{L^0}(\cdot)$, computing the respective set of descendant values and taking their union.
The \textit{detailed signature of a query} is (simply) the set of coordinates that pertain to the signature of $\phi^0$. 

The \textit{detailed proxies} of expressions are the respective expressions transformed at the most detailed level for each of the involved dimensions. The \textit{detailed proxy of a query}

\begin{center}
    $q$ = $<C^{0}$, $\phi$, $[L_{1}, ..., L_{n},
M_{1}, ..., M_{m}]$, $[agg_{1}(M^{0}_{1} )$, ...,$agg_{m}(M^{0}_{m})]$ $>$
\end{center}

is the query (i.e., an expression again)
\begin{center}
    $q^0$ = $<C^{0}$, $\phi^0$, $[L_{1}^0, ..., L_{n}^0,
M_{1}^0, ..., M_{m}^0]$, $[agg_{1}(M^{0}_{1} )$, ...,$agg_{m}(M^{0}_{m})]$ $>$
\end{center}

\textit{Detailed areas} are sets of cells, pertaining to an aggregate cell, or set of cells, like, e.g., the result of a query. 

The\textit{ detailed area} of a cell $c: <v_1, \ldots, v_n>$ is the set of descendant cells that can be obtained by replacing each of its coordinates, say $v_i$, by $desc_L^{L^0}(v_i)$ and taking the Cartesian product of each such value set. 

The \textit{detailed area} of the query $q$ is the set of cells of the result $q^0$, $q^0.cells$.\\

\textbf{History}. A \textit{session} $S$ is a list of cube queries $S$ = $\{q_{1}, ..., q_{n}\}$ that have been recorded. We assume the 
knowledge of the syntactic definition of the queries, and possibly, but 
not obligatorily, their result cells.

A \textit{session history of a user} is a list of sessions, resulting 
in a list of queries, following the order of their sessions.

The \textit{cell history}, or simply, 
history, of a session history is the set of cells that belong to the queries of the session history. The history of detailed equivalents is the set of detailed equivalents of the cells of the query history.

\subsection{Reference Example}\label{sec:refEx}\sideNew{Sec. \ref{sec:refEx}: NEW!}

In this example, we work with the loan cube from the PKDD 1999 Discovery Challenge\footnote{The example comes from the Discovery Challenge of PKDD 1999 \url{https://sorry.vse.cz/~berka/challenge/pkdd1999/berka.htm} and now can be found at \url{https://github.com/sabirakhtar/PKDD-99-Discovery-Challenge}
}. 
The cube has anonymized data from Czech banks that concern loans that have been granted to customer. The dimensions of the data cube concern (a) the customer \textit{Accounts}, with a hierarchy of levels: $Account$ $\preceq$ $District$ $\preceq$ $Region$ $\preceq$ $ALL$, (b) the \textit{Status} of a loan, with levels $Status$ $\preceq$ $ALL$, and, (c) \textit{Date} with a hierarchy $Day$ $\preceq$ $Month$ $\preceq$ $Year$ $\preceq$ $ALL$. For simplicity, we use a single measure $Amt$, referring to the amount of the loan that was granted.

Assume now that at a certain time point, four queries have been issued already, and a new one, to which we refer as $q$, is also submitted to the system. The desideratum is to compute the interestingness dimensions of the query. Coming back to our opening remarks in the Introduction, this can occur due to several possible reasons. In a clear a-priori case, the new query $q$ is generated by the system, and is candidate to be recommended to the user for execution. Before executing it however, and thus without any knowledge of what is included in the result, the recommender system needs to predict what it will contribute to the user's understanding on the data space. To this end, we need \textit{syntactic} metrics and algorithms, that take only the query expression into consideration, to predict interestingness. In an a-posteriori case, the result of $q$ has already been computed. Thus, we can use \textit{extensional} algorithms and metrics that exploit this result and compare it to the cached results of the previous queries in order to compute its interestingness.

\begin{figure*}[hbt]
\centering
\includegraphics[width=\textwidth]{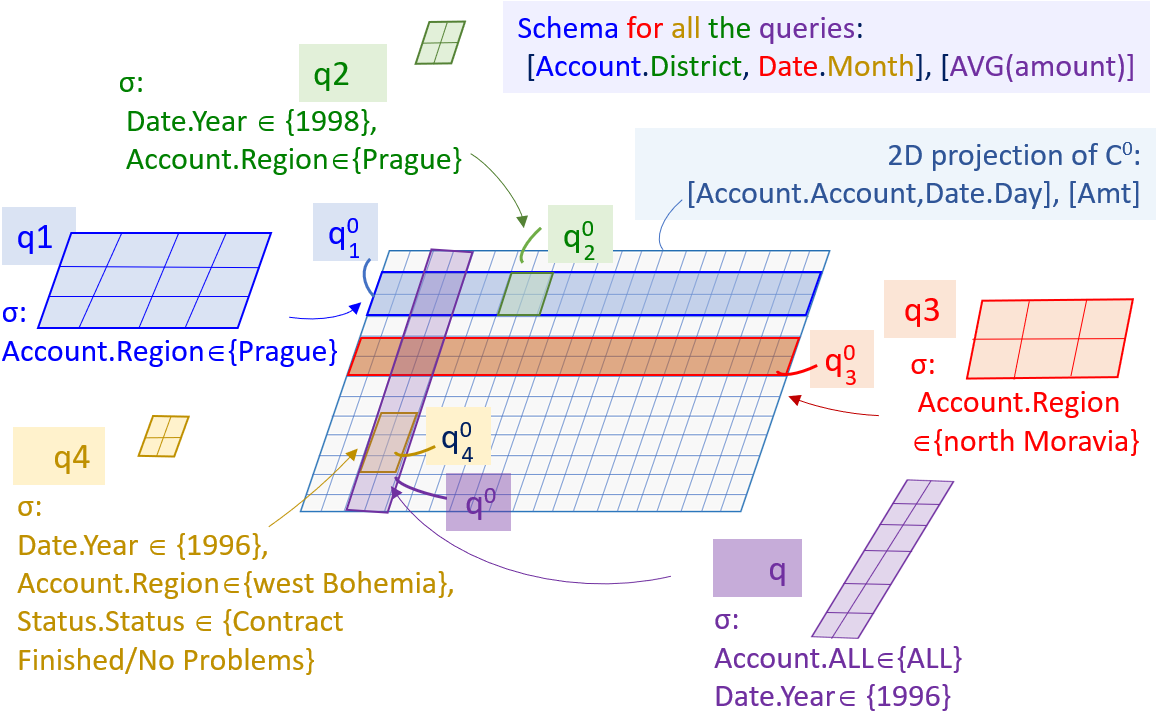}
\caption{The setup of our reference example.}\label{fig:RefEx}
\end{figure*}

\sloppy In Figure~\ref{fig:RefEx}, we visually present the general setup of the problem. We assume a basic cube defined at the most primitive levels of detail: $C^0$: $[Account.Account, Status.Status, Date.Day, Amt]$. Then, for ease of diagrammatic depiction, we have all 5 queries of the figure defined at the schema $[Account.District, Date.Month, AvgAmt]$, practically expressed via the following formula: $q_i$ = $<C^{0}$, $\phi_i$, $ [Account.District$, $Status.ALL$, $Date.Month$, $AvgAmt]$, $[avg(Amt)]>$, with each $\phi_i$ having a different expression, as depicted in the figure.

\sloppy The center of Figure~\ref{fig:RefEx} depicts a 2D projection of the space of the basic cube $C^0$ along the 2 dimensions of the query schemata, $Account$ and $Date$ (we omit $Status$ to simplify the figure). Each of the queries has (a) a detailed proxy
$q_i^0$ = $<C^{0}$, $\phi_i^0$, $ [Account.Account, Status.Status, Date.Day, AvgAmt]$, $[avg(Amt)]>$
and (b) a detailed area of cells, depicted as a band in the 2D projection of the multidimensional space. Some detailed areas are completely contained inside others: for example $q_2^0$ is completely contained within $q_1^0$. The detailed area of $q$, $q^0$, has all sorts of relationships with the detailed areas of the other queries: (a) $q_4^0$ is completely contained within $q^0$, (b) $q^0$ has a partial overlap with $q_3^0$ and $q_1^0$, (c) there is no relationship between $q^0$ and $q_2^0$, whatsoever.
There are no identical queries, either.

%% file: 04_taxonomy.tex
\section{Taxonomies for the assessment of cube query interestingness}\label{section:_GoBack}\sideNew{Sec. \ref{section:_GoBack} NEW!}

\begin{figure*}[bt]
\centering
\includegraphics[width=\linewidth]{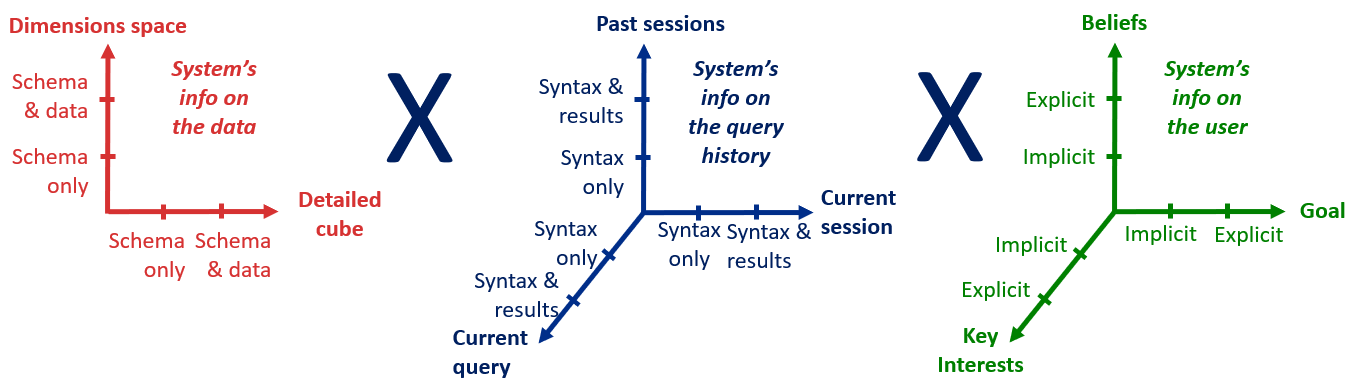}
\caption{The 8 dimensions that determine the different 
levels of information that a system must have in order to assess the 
interestingness of a cube.}\label{fig:dimensions}
\end{figure*}

In this paper, we propose dimensions of interestingness and algorithms to assess them. Specifically, we define the interestingness of a cube query $q$, $I(q)$ as a vector of scores along the fundamental interestingness dimensions, (i) novelty, (ii) relevance, (iii) surprise, and (iv) peculiarity:

\[I(q) = \langle novelty(q), relevance(q), surprise(q), peculiarity(q) \rangle
\]

Before proceeding to the individual interestingness dimensions and the respective algorithms, we provide two taxonomies that allow to mentally structure the problem and organize the algorithms accordingly. We will introduce \textit{a taxonomy of types of returned scores and internal workings} for the proposed algorithms. Before that, however, we start this session with presenting a \textit{taxonomy for the types of input} the algorithms will need. 

\subsection{Taxonomy of the input information needed to assess interestingness}\label{sec:inputTax}

We identify 8 fundamental dimensions of needed (equiv.: 
potentially available) information that will be needed in order for our 
\textit{cube query interestingness scoring system} to address the goal 
of automatically computing a score of interestingness for a given cube 
query. These 8 dimensions are further organized in 3 major families.

\textit{For all families and dimensions, an implicit value of our 
knowledge is the} \textbf{no-knowledge }\textit{value}. To avoid 
repetition, we will not refer to this level of knowledge again, although 
it is quite possible that several dimensions will be of no-knowledge 
value in a scoring system in practical situations.\\

\textbf{\color{red}{The Data Family}}. The Data Family of dimensions is concerned 
with what kind of information about the underlying data is available to 
our scoring system. Specifically, we identify two dimensions of 
interest, concerning the hierarchies of the multidimensional space and 
the factual cubes that are available.

\begin{itemize}
\item \textit{Dimension space}. The dimension space characterizes our 
level of knowledge/information on the dimensions of the multidimensional 
space within which the queries are going to be configured and posed. We 
identify two potential levels of knowledge (a) knowledge only of the 
schema (i.e, dimension \& level names, hierarchical relationships) of 
the multidimensional space, or, (b) knowledge of both the schema and the 
values (aka dimension members in the OLAP literature) of the involved 
levels. 
\item \textit{(Detailed) Cubes}. The detailed cube space 
characterizes our level of knowledge/information on the factual data of 
the multidimensional space -- i.e., the detailed cubes over which the 
queries are going to be posed. We identify two potential levels of 
knowledge (a) knowledge only of the schema (i.e, dimensions \& levels) 
of the cubes' schemata, or, (b) knowledge of both the schema and the 
values i.e., cube cells of the involved cubes.
\end{itemize}

\textbf{\color{blue}{The Query History Family}}. The Query History Family of 
dimensions is concerned with what kind of information about the queries 
being and having been issued by the user is available to our scoring 
system. Specifically, we identify three dimensions of interest, 
concerning (a) the knowledge of the current query being posed, (b) the 
knowledge of the user's current session, and, (c) the knowledge about 
the overall history of queries of the user.

\begin{itemize}
\item \textit{Current Query}. The current query being posed to the 
query answering system (e.g., an OLAP server) can be known by our 
scoring system at two levels of information: (a) syntax only, where only 
the query specification is available (e.g., before the 
query having been answered, or in order to save space or speed up computations without 
using the result cells), or, (b) both query specification and results 
are known to the scoring system.
\item \textit{Current Session. }Similarly to the current query, the 
current session comprises a list of queries that are known to the 
scoring system. Like the case of the current query we may either know 
only the syntax of the queries, or both the syntax and the cells of the 
session queries' results.
\item \textit{Past Sessions. }Similarly to the current session 
dimension, the past sessions dimension generalizes it to include 
previous sessions of the user (or other users, similar to the one being 
assessed) too. Again, we may either know only the syntax of the queries, 
or both the syntax and the cells of the queries' results.
\end{itemize}

\textbf{\color{Green}{The User Profile Family}}. The User Profile Family of 
dimensions is concerned with what kind of information about the user is 
available to our scoring system. Specifically, we identify three 
dimensions of interest, concerning (a) the knowledge of the user Key 
Interests, (b) the knowledge of the user's Beliefs about the data 
values, and, (c) the knowledge about the current user Goals that are 
available to our scoring system.

\begin{itemize}
\item \textit{Key User Interests}. The user's recurring interests -- 
as close to a user profile as we can get-- comprise the context for this 
dimension. The Key Interests can be considered a static aspect of the 
user profile and we will assume they take the form of a set of Key 
Performance Indicators (KPIs)\cite{DBLP:journals/dke/MateTM17}, which practically comprise a query and a 
labeling schema for the results of a query on the basis of expected 
values for them. So practically, every cell of a KPI query result is 
mapped to a finite set of values (e.g., bad/med/good, or a Likert scale 
of stars) on the basis of rules that compare it to an expected value and 
assign a performance score on the basis of the discrepancy of the actual 
vs the expected value of the cell. We discriminate two levels of 
knowledge the system can have on the KPI's of the user: (a) implicit, 
i.e., this kind of information is not explicitly specified by the user, 
but approximated and estimated by other information available to the 
system like the history of past queries, that somehow mark a range of 
preferences on what interests the user on a regular basis, or, (b) 
explicit, directly stated by the user (e.g., in this case an explicit 
specification of KPI's).
\item \textit{User Beliefs}. The beliefs about the data that the user 
has, are captured by this dimension. In other words, assuming a query is 
posed, the beliefs of the user is the set of expected values for the 
query cells that the user expects to see. These can be (a) implicitly 
estimated, e.g., derived from the history of past queries by 
extrapolating values on the basis of similar values the user has seen in 
the past, or identified by some relevant KPI's carrying expected values 
for certain aggregate cells, or, (b) explicitly known, e.g., extracted 
from the history of past queries, in case a certain cell has been 
presented to the user in the (recent) past.
\item \textit{User Goals}. The user goals are the current information 
goals that the user has towards fulfilling an information need, or 
exploring the data space and discovering new information. The Goals are 
dynamic characteristics of the user, temporally local and transient 
(i.e., they concern an information need of the current time) and they 
are related closely to the intentions of the Intentional model \cite{DBLP:conf/dolap/VassiliadisM18} like describe, explain, analyze, etc. We discriminate two levels 
of knowledge the system can have on the goals of the user: (a) implicit, 
i.e., this kind of information is not explicitly specified by the user, 
but approximated and estimated by other information available to the 
system, or, (b) explicit, directly stated by the user (e.g., in case he 
is firing intentional queries). 
\end{itemize}

\textbf{How are the dimensions of interestingness related to the dimensions of the problem}. Concerning the relationship of the aspects of interestingness with the dimensions of available knowledge, we can make a few, first coarse observations:

\begin{itemize}
\item Peculiarity is related to the history of past queries and their 
results. A query can be peculiar if (a) it does not fit nicely in the 
set of previous queries in terms of its syntax (and thus, of the area of 
the multidimensional space that it covers), or (b) if its results show 
values quite different than the values one had seen in previous, similar 
queries.
\item Surprise is related to the beliefs the user already has. There 
beliefs can be anywhere in the range of (a) concrete values of past 
query cells, all the way to (b) some probability distribution on the 
expected measure values (or labels) a given cell can have.
\item Novelty is affected by the presence of the history of past queries.
\item Relevance is an aspect related to the static profile (KPI's, 
preferences, interests) as well as to the dynamic profile (current goal) 
of the user. The static profile refers to what the user is typically 
interested in, and is an approximation of the user needs on a recurring 
basis, whereas the current goal is a more to the point description of 
the specific info need of the user at this moment in time.
\end{itemize}

In the rest of our deliberations, unless explicitly stated otherwise, 
the system works under the following assumptions:

\begin{itemize}
\item the dimensions' schema and data are both known;
\item the cubes' schema is known (but not necessarily the data);
\item the syntax of the current query is known (but not necessarily its 
results);
\item the two dimensions of the past are unioned into a single 
dimension, history; no assumptions can be made for its knowledge by 
default;
\item similarly, no assumptions are by default made for the three 
dimensions of the system's information on the user. 
\end{itemize}

\subsection{Internal taxonomy of algorithms}\label{sec:internalTax}
Apart from the aforementioned taxonomy used to characterize the type of information needed to be able to assess interestingness (pretty much amounting to the type of \textit{input} information the assessment algorithms need), we can also discriminate algorithms with respect to the type of problem they solve and the returned value they compute (practically, the \textit{output} of the algorithm), as well as, the way the algorithms perform the checks and potential constraints the algorithms might have (practically, characterizing the \textit{how} of the algorithm).

We use the following terminology:
\begin{enumerate}
    \item \textit{Decision} vs \textit{Enumeration} Problem: the decision problems answer a Boolean check (e.g., whether a cube is novel or not), whereas the enumeration problems report which subsets of cells are part of a solution (e.g., which part of a new cube is already covered, or novel). In all our subsequent deliberations, unless explicitly specified otherwise, we work on the enumeration problem.
    \item \textit{Full} vs \textit{partial} Assessment: full assessment means that the checks made return a true/false answer on whether a new query is interesting or not; partial assessment means that the checks return an interestingness score (in fact: a score for a particular interestingness dimension) as a real number (typically in the interval [0 \dots 1]). Naturally, a partial assessment that returns 1, also implies full interestingness.
    \item \textit{Syntactic} vs \textit{Extensional} Assessment: syntactic assessment is based only on query definitions, whereas extensional also assumes the presence of the cells of the query result(s).
    \item \textit{Same-Level vs \textit{Detailed}} Assessment: same-level (equiv., immediate) assessment assumes that two cubes are at the same level of aggregation; detailed (equiv., indirect or derivable) assessment means that the comparison of two cubes will be done at levels lower than their definition -- typically, we will use the most detailed level as the common ground upon which the constituting detailed cells for two cubes can be compared.

\end{enumerate}

%% file: 05_novelty.tex
\section{Novelty}\label{sec:novelty}
Novelty assesses the amount of previously unknown information delivered 
to the user via a query. Due to this inherent characteristic, we need to 
either explicitly know, or at least estimate the prior knowledge of the data that the user has.

Naturally, a system is not in a position to actually have knowledge of 
the user's memory or knowledge. Knowledge can come to the user via 
external channels, not related to the query answering and thus, the 
system necessarily has 
``knowledge" of just a subset of the user's 
actual knowledge. At the same time, one should also take account of the 
effects of time that erases, hides or distorts the remembrance of facts 
encountered in the past. Although in our following deliberations we will 
not directly address the above problems, we will occasionally offer 
insights on how to handle some of them. However, when we use the term 
``knowledge" we simplify and approximate the situation, by assuming 
that the system knows what the user has seen, or what the user 
has explicitly stated that she believes.

Explicit knowledge is primarily attained by knowing the history of user 
queries (and assuming that the user remembers it). A second way to 
approximate what the user remembers is to exploit the registered beliefs 
of the user that have a low level of confidence, by making 
the rational assumption that since she has expressed practically 
uncertain beliefs about some cells, she does not know their values.

Novelty is mostly goal-independent, i.e., it is not affected by neither 
the current (goal) or the typical (key interests) informational needs of 
the user.

Overall, in terms of our taxonomic dimensions, novelty is mostly related (a) to \textit{history}, and, (b) to \textit{registered values for beliefs with confidence below a certain threshold}. We will examine the different alternatives in the respective subsections.

\begin{figure*}[tbp]
    \centering
    \includegraphics[width=\linewidth]{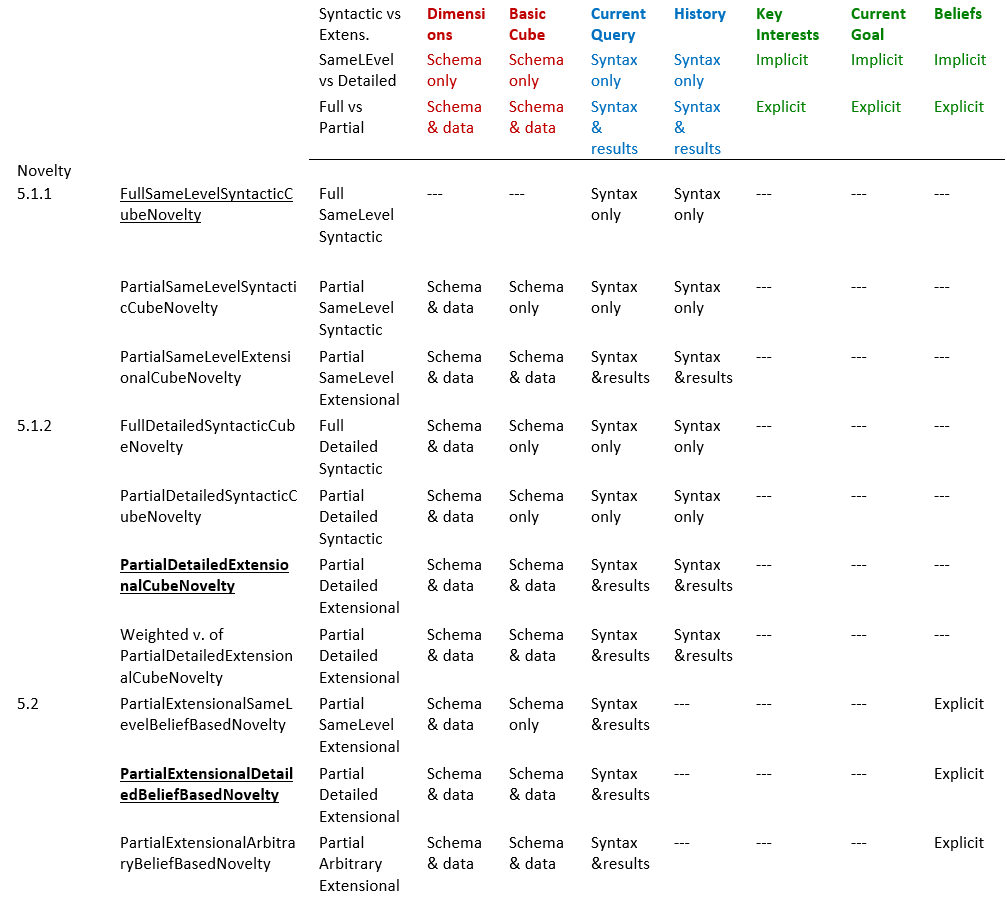}
    \caption{List of Novelty algorithms, characterized with respect to the reference taxonomy (underline: implemented, bold: experimented)}
    \label{fig:ListoAlgoNov}
\end{figure*}

\silence{
The fundamental taxonomy for novelty is as follows:

\begin{table}[htbp]
\centering
\begin{tabular}{|m{2cm}|m{2cm}|m{2cm}|m{2cm}|m{2cm}|}
\hline
 & Boolean Immediate \textit{(true/false)} & Partial Immediate \textit{
(pct)} & Boolean Derivable \textit{(true/false)} & Partial Derivable
 \textit{(pct)} \\
\hline
\textbf{Syntactic} \textit{(only q.~def's are used)} & \textbf{Does 
q belong to Q? \textit{(only via a q. syntax check)}} & What pct of 
q.cells belong to Q.cells?\textit{(only via a q. syntax check)} & Can 
q.cells be computed from Q.cells?\textit{(only via a q. syntax check)
} & Which pct of q.cells can be computed from Q.cells? \textit{(only 
via a q. syntax check)} \\
\hline
\textbf{Extensional} \textit{(can use q. or Q results too)} & Do all 
the cells of q belong to Q.cells? & What pct of q.cells belong to 
Q.cells? & Can q.cells be computed from Q.cells? & Which pct of q.cells 
can be computed from Q.cells? \\
\hline
\end{tabular}
\end{table}

} 

 \subsection{Novelty assessment in the presence of a query history}\label{sec:NoveltyInHistory}
\sideNew{Sec. \ref{sec:NoveltyInHistory}: New after FSLSN}

First, we will assess the novelty of a cube query $q$ assuming a query 
history $Q$ = $\{q_{1}, \ldots, q_{n}\}$ exists. 

\subsubsection{Same-Level Assessment of Novelty}
Assume that we only check $q$ against members of $Q$ whose schema is at the same level with $Q$. We also require the same detailed measures and aggregate functions to be used, otherwise the comparison is referring to essentially different measures, and also different numbers, and, therefore, novelty is guaranteed.\\

\textbf{Full Same-Level Syntactic Assessment of Novelty}. In this case, the question to be answered is: Given $q$ and $Q$ = $\{q_{1}, \ldots, q_{n}\}$, is there any $q_i~\in~Q$ such that $q~=~q_i$? 

In this case, the solution is a trivial \emph{syntactic} check: we iterate through the syntactic definitions of the queries of $Q$ and check whether there is any query that is identical to $q$. Then, \textit{Full Same-Level Syntactic Novelty} (FSLSN) is defined as a Boolean flag:

\[FSLSN =  \begin{cases}
        0 & \text{if a } q_i = q \text{ exists } \\
        1 & \text{otherwise}
     \end{cases}
\]

The check is full, syntactic and same-level. \\ 

\textbf{Partial Same-Level Syntactic Assessment of Novelty}.\sideNew{New} In this case, the question to be answered is: Given $q$ and $Q$ = $\{q_{1}, \ldots, q_{n}\}$, can we identify which part of the results of $q$ is already covered by the queries of $Q$ without actually computing them?\\

The answer to the question is given by Algorithm \textsf{ComputePartialSameLevelCubeCoverage} in \cite{PV21} that takes $q$ and $Q$ as inputs and divides the coordinates of the result of $q$ in two sets: a set of cell coordinates that are covered by existing queries, $q^{cov}$ and its complement, $q^{nov}$, a set of cell coordinates that are novel. Then, 
\textit{Partial Same-Level Syntactic Novelty} (PSLSN)
is the fraction of novel cells of the total population of cells of $q$ (which is also the union of $q^{nov}$ and $q^{cov}$). At the syntactic level, we only need the coordinates (signatures) of the cells, without having to compute their measures.

\[PSLSN = 
\frac{ |q^{nov^{+}}|} {|q^{nov^{+}}|~\bigcup~|q^{cov^{+}}|}
\]


The check is partial, syntactic and same-level. \\ 

\textbf{Partial Extensional Same-Level Assessment of Novelty}.\sideNew{New} In this case, the question to be answered is the same: Given $q$ and $Q$ = $\{q_{1}, \ldots, q_{n}\}$, can we identify which part of the results of $q$ are already covered by the queries of $Q$? However, in this case, we assume that the results of the queries are available and the check takes this into consideration.\\

The premise to the question is given by Theorem \textsf{Same-Level-Intersection} in \cite{PV21} that takes two queries $q^1$ and $q^2$ with the same schema, and decides whether their selection conditions make them  eligible for a check on their intersection. Then, Algorithm \textsf{EnumerateSameCellsviaResultComparison} in \cite{PV21} returns the cells that are covered and the cells that are not.\\

A simple adaptation of the Algorithm \textsf{ComputePartialSameLevelCubeCoverage} in \cite{PV21} that works with signatures, to work with cells produces the novelty of the new query $q$. The formula for \textit{Partial Same-Level Extensional Cube Novelty} is the same with the one of \textit{Partial Same-Level Syntactic Cube Novelty} and the difference is only in efficiency (which of the two variants is faster is open to experimental evaluation).

The check is partial, extensional and same-level. \\ 

\subsubsection{Detailed Assessment of Novelty}\label{sec:DetAssNov}
\sideNew{Sec. \ref{sec:DetAssNov}: New  until PDEN}

Assume now that instead of checking cubes defined at the same level, we compare cubes with respect to their constituting cells at the most detailed level.\\

\textbf{Full Syntactic Detailed Assessment of Novelty}. In this case, the question to be answered is: Given $q$ and $Q$ = $\{q_{1}, \ldots, q_{n}\}$, is there any $q_i~\in~Q$ such that $q_i^{0^+}$, the \emph{detailed signature of $q_i$} (i.e., the coordinates of the most detailed cells over which $q_i$ is computed), is a superset of $q^{0^+}$, \emph{the detailed signature of $q$}?\\

The premise to the question is provided by the Theorem on \textsf{Foundational Containment} in \cite{PV21} stating when a certain query $q_i$ can foundationally contain a new query $q$. A simple iteration over the contents of the query set $Q$ can reveal whether such a query exists or not. \textit{Full Syntactic Detailed Novelty} (FSDN) determines whether a query $q$ is novel with respect to a previous query $q_i$ $\in$ $Q$.

\[FSDN =  
    \begin{cases}
        0 & \text{if } \exists~q_i~\in~Q \text{ that foundationally contains } q \\
        1 & \text{otherwise}
     \end{cases}
\]

The check is full, syntactic and detailed. 

\hrulefill
\begin{remark}
Alternatives for better efficiency. What if, instead of computing the detailed area of each query $q_i$ in $Q$ separately and on-demand, we compute (ideally: proactively, and storing it) the detailed expression $q_i^0$? This will slow down the query execution by a tiny bit, but will improve the performance of the algorithm that checks for novelty.
\end{remark}
\hrulefill

\textbf{Partial Detailed Syntactic Assessment of Novelty}. In this case, the question to be answered is: Given $q$ and $Q$ = $\{q_{1}, \ldots, q_{n}\}$, can we identify which part of the results of the detailed area of $q^0$ are already covered by the detailed areas of the queries of $Q$, by comparing solely the signatures of the queries?\\


\begin{algorithm}[ht]
\DontPrintSemicolon 
\KwIn{A query $q$; the query history $Q$, i.e., a set of queries $q_i$, all with the same aggregate functions over the same detailed measures}
\KwOut{The subset of the coordinates of $q^{0}$, say $q^{cov^{0^+}}$ that are also part of the union of the coordinates of the queries in $Q$, i.e., the union of $q_i^0$, and its complement $q^{nov^{0^+}}$}
\Begin{
produce $q^{0^+}$ and $q_i^{0^+}$ for all $q_i$\;
populate the hashmap(cell signature) $Q^0$ $\gets$ $\bigcup_{i} q_i^{0^+}$\;
$q^{cov^{0^+}} \gets \emptyset $\;
$q^{nov^{0^+}} \gets q^{0^+}$ \;

\ForAll {$c^{0^+} \in q^{0^+}$}{
    \If {$c^{0^+}~\in~Q^{0^+}$} {
        remove $c^{0^+}$ from $q^{nov^{0^+}}$ and add it to $q^{cov^{0^+}}$\;
    }
}

\Return{$q^{cov^{0^+}}$, $q^{nov^{0^+}}$ }
}
\caption{\sf{Signature-based syntactic enumeration of covered detailed cells}}
\label{algo:EnumDetailedQueryContainmentSignatureBased}
\end{algorithm}

Algorithm~\ref{algo:EnumDetailedQueryContainmentSignatureBased} computes the union of the signatures of the detailed proxies of the queries in the query list and intersects it with the detailed signature of the query under question. The resulting \textit{Partial Detailed Syntactic Novelty} (PDSN) is the fraction of the detailed not covered (i.e., novel) cells over the entire detailed area of $q$. 

\[PDSN =  
\frac{ |q^{nov^{0^+}}|} {|q^{nov^{0^+}}|~\bigcup~|q^{cov^{0^+}}|}
\]

The check is partial, syntactic and detailed. \\ 

\textbf{Partial Detailed Extensional Assessment of Novelty}. In this case, the question to be answered is practically the same, albeit with a different means to compute the answer, specifically, cells instead of signatures: Given $q$ and $Q$ = $\{q_{1}, \ldots, q_{n}\}$, can we identify which part of the results of the detailed area of $q^0$ are already covered by the detailed areas of the queries of $Q$?





\begin{algorithm}[ht]
\DontPrintSemicolon 
\KwIn{A query $q$; the query history $Q$ expressed as a set of queries $q_i$, all with the same aggregate functions over the same detailed measures}
\KwOut{The subset of the cells of $q^{0}$, say $q^{cov}$ that are also part of the union of the results of the queries in $Q$, i.e., the union of $q_i^0$, and its complement $q^{nov}$}
\Begin{
produce $q^{0}.cells$ \;
produce $q_i^{0}.cells$ for all $q_i$\;
populate the hashmap(cell signature) $Q^0$ $\gets$ $\bigcup_{i} q_i^{0}.cells$ \;
$q^{cov^0} \gets \emptyset $\;
$q^{nov^0} \gets q^{0}.cells $ \;

\ForAll {$c^0 \in q^0.cells$}{
    \If {$c^0~\in~Q^0$} {
        remove $c^0$ from $q^{nov^0}$ and add it to $q^{cov^0}$\;
    }
}

\Return{$q^{cov^0}$, $q^{nov^0}$ }
}
\caption{\sf{Cell-based extensional enumeration of covered detailed cells}}
\label{algo:EnumDetailedQueryContainment}
\end{algorithm}

Algorithm~\ref{algo:EnumDetailedQueryContainment} computes the union of the detailed areas of the queries in the query list and intersects it with the detailed area of the query under question. We remark that only the queries in the history concerning the same measures and aggregation functions than $q$ are passed to the algorithm. The resulting \textit{Partial Detailed Extensional Novelty} (PDEN) is the fraction of the detailed not covered (i.e., novel) cells over the entire detailed area of $q$.

\[PDEN =  
\frac{ |q^{nov^0}|} {|q^{nov^0}|~\bigcup~|q^{cov^0}|}
\]

The check is  (a) partial (practically a normalized score), (b) extensional (via cells), and, (c) detailed, i.e., with respect to the detailed levels of the involved cubes.

The complexity of Algorithm~\ref{algo:EnumDetailedQueryContainment} is mainly determined by the cost of answering the detailed queries in Lines 2 and 3 that produce the cells for $q^0$ and $q_i^0$ for both the input query and the query history. The complexity of these actions is: (a) linear with respect to the size of the query history, and, (b) linear with respect to the cube size, assuming that the cube query is linear with respect to the cube size. The rest of the algorithm, requires a linear in-memory pass of the result to populate $Q^0$ and a linear lookup for each cell of $q^0.cells$ to cross-check if it belongs to $Q^0$. Again, this cost is linear, yet, we consider it insignificant comparing it to the  time needed for query answering. Therefore, the overall cost of the algorithm is linear with respect to the size of the query history and to the cube size.

\hrulefill
\begin{remark}
Observe that, since the check is done at the most detailed level, the only thing we care about is that the measures and aggregate functions are the same. Selection conditions can be arbitrary. The same applies for the grouper levels: to the extent that we assess novelty with respect to the detailed cells, the grouping levels of the compared queries can be arbitrary.
\end{remark}
\begin{remark}
It is easy to introduce a \textbf{weighted variation} of the above algorithm. 
Observe that the Algorithm~\ref{algo:EnumDetailedQueryContainment} computes the union of the detailed areas of the queries with set semantics. We can produce a weighted variant if we introduce the following variations to the algorithm:
\begin{itemize}
    \item Each cell is accompanied by a counter of its occurrences; so, every time we perform the union of $Q^0$ with the next $q_i^0.result$, for every detailed cell that is already part of $Q^0$, we increase its counter by one. Let us denote the number of occurrences of each cell $c$ with $c.weight$.
    \item Given a set of cells, $C$, we can compute its total weight, $C.weight$, as the sum the weights of its constituent cells.
    \item Then, $WeightedDetailedNovelty$ (WDN) is the total weight of $q^{nov^0}$ over the sum of the total weights of $q^{nov^0}$ and $q^{cov^0}$.
\end{itemize}
\[WDN =  
\frac{ q^{nov^0}.weight} {q^{nov^0}.weight + q^{cov^0}.weight}
\]
This way, cells that are more frequently encountered count more (thus, increasing the denominator and reducing the total novelty of the new query, if it includes such cells in its result). In case $q^{cov^0}$ is empty, novelty takes the value of 1.
\end{remark}
\begin{remark}
Variations of the above formula on the total weight can also be devised, to normalize the weights of the cells. Also, the same theme can be applied to (a) signatures and (b) same-level checks, too.
\end{remark}
\hrulefill

\subsection{Novelty assessment in the presence of belief statements}\label{sec:noveltyBelief}

Assume we do not have explicit knowledge of the user history, or key interests, but we do have an estimation of probabilities for the likely values of some cells in the multidimensional space.

Assume that for certain cells, it has been possible to either deduce or explicitly have the user register probabilities per expected value for the value \textit{m = c.M}, of a cell c and a certain measure $M$. So, some cells in the multidimensional space are annotated with a set of \textit{cell expected-value statements}, which are statements of the form 
\begin{multline*} 
p(M \in [l_i \dots u_i] | c) = p_i, p_i~\in~[0 .. 1], [l_i \dots u_i] \\
\text{ is a range of values of } dom(M)
\end{multline*}
or of the form
\begin{multline*} 
p(M \in \mathbf{s_i} | c) = p_i, p_i~\in~[0 .. 1], \mathbf{s_i} \\ 
\text{ is a discrete finite set of values of } dom(M)
\end{multline*}

For uniformity of notation we will use the syntactic form \[p(M \in m_i | c) = p_i\] to denote either a range or a finite set of values for the value-set of the expressed belief. The distinction makes no difference for the evaluation of novelty. 

We refer to the set of statements of the above form for a cell $c$, the \textit{probable active domain} of $c$, or $dom^{pa}(c)$. A \textit{well-formed probable active domain} of a cell has the property that all it's statements' probabilities sum up to 1. However, requiring well-formed probable active domains is too restrictive, in the sense that maybe some probabilities are unknown, or hard to evaluate; thus, we do not require it as a necessary property for the sequel.

We call a cell c to be $\Pi-known$ if, within the statements of the probable active domain of $c$, there exists a probability $p_i$ which is equal or higher to a threshold $\Pi$. Otherwise, if all the probabilities of $c$ are below $\Pi$ the cell is called $\Pi-unknown$.

\emph{The intuition behind this treatment lies on the observation that if a user has a set of beliefs about the behavior of a cell, with a high amount of certainty (i.e., the probability is above a certain threshold), then we cannot consider the cell to be ``unknown" to the user.} The result of a query might be surprising, if it is far from the expected value, but the existence of this area of the multidimensional space is not novel to the user.

To give a practical example, assume the following user beliefs
\begin{eqnarray*}
p(sales \in [100..200) ~|~ city=Athens, year=2020) = 30\%\\
p(sales \in  [80..100) ~|~ city=Athens, year=2020) = 70\%
\end{eqnarray*}
\noindent assuming all other dimensions set to ALL. For a particular cell therefore, concerning the sales in Athens for 2020, we have a probability distribution for the range of its values. Let's also assume that we have agreed that if a user has a belief  higher or equal to 50\% for a cell's measure, then the user ``knows" the cell; this means setting a value of $\Pi$ = 50\%. Given the above belief set, and the existence of a belief with probability 70\% (i.e., higher than $\Pi$), we can say that this particular cell is indeed 50\%-known, and thus consider it not novel.

Let \textit{B} a set of beliefs expressed as cell expected-value 
statements\textit{ }for a set of cells $C^{B}$. Assume now a query 
$q$, and its resulting cells $C = q.cells$ = $\{c_1, \ldots, c_n\}$. Assume 
also a threshold $\Pi$. Then, the $\Pi-direct~novelty$ of $q$ is the 
percentage of cells of $q.cells$ that are $\Pi-unknown$. 
We can distinguish three cases for computing belief-based novelty, depending on the level that the cells of $C^{B}$ have been defined.

\subsubsection{Same-Level Belief-Based Novelty}\label{sec:SLBBN}
\sideNew{Sec. \ref{sec:SLBBN}: New}
In this case, the set of beliefs $B$ is expressed over a set of cells $C^B$ at the same aggregation level as $q$. Therefore, we can immediately compare the cells of the query to the cells of the belief-set. Algorithm \ref{algo:beliefSameLeveledNovelty} performs the computation of novelty.

\begin{algorithm}[ht]
\DontPrintSemicolon 
\KwIn{A query $q$ and its result; a set of beliefs $B$ over a set of cells $C^B$ at the same aggregation level as $q$; a threshold $\Pi$ for deciding if a cell is eligible for being novel }
\KwOut{The subset of the cells of $q.result$, say $q^{cov}$ that are also part of the space the beliefs cover, as well as its complement $q^{nov}$}
\Begin{

$q^{cov} \gets \emptyset $\;
$q^{nov} \gets q.cells$ \;
$C^\star$ $\gets$ the subset of $C^B$ for which there exists a known belief, i.e., $\{c~|~c \in C^B, \exists~p(M \in m|c) \in B, p(M \in m|c) \geq \Pi \}$\;
\ForAll {$c \in q.cells$}{
    \If {$c^{+}~\in~C^{\star^{+}}$} {
        remove $c$ from $q^{nov}$ and add it to $q^{cov}$\;
    }
}
\Return{$q^{cov}$, $q^{nov}$ }
}
\caption{\sf{Partial Extensional Same-Level Belief-Based Enumeration Of Covered Cells}}
\label{algo:beliefSameLeveledNovelty}
\end{algorithm}

The algorithm, starts by assuming that all cells are novel and none is covered. The first action of the algorithm is to isolate the $\Pi-known$ cells, based on the input set of beliefs, into a set $C^{\star}$. Then, for each cell of the query, it checks whether its signature fits with the signature of any of the cells that belong to $C^{\star}$, and if it does, then it considers the cell to known, adds it to the set of covered cells and removes it from the set of novel cells. The reason for using signatures here, is that the beliefs are expressed with respect to signatures and probabilities for the value range of the measure. Thus, the cell's measure should not be used for assessing the presence of the cell in $C^\star$ (i.e., checking for identity/equality of the measure is not within the spirit of using the beliefs in the first place).

Then, we can compute the \textit{Partial Extensional Same-Level Belief-Based Novelty} (PESLBBN) of the query $q$ as usual:
\[PESLBBN = 
\frac{ |q^{nov}|} {|q^{nov}|~\bigcup~|q^{cov}|}
\]

The check is (a) partial (practically a normalized score), (b) extensional (via cells), and, (c) same-level, i.e., with respect to the actual cells of the involved query. The \textit{Syntactic} version of the algorithm (as contrasted to the \textit{Extensional} one) is quite similar, albeit with the difference that no cells in the query result are needed and all sets and comparisons are performed with respect to the signatures of the queries. The complexity is linear to the result size (assuming the set $C^B$ is fixed) and linear to the size of the set $C^B$ assuming the query result size is fixed. In the case that only the query expression is given as input to the algorithm, and the query result has to be computed, the cost is dominated by the computation of $q.cells$, which is linear to the data cube size.  

\subsubsection{Detailed Belief-Based Novelty}\label{sec:DBBN}
Another (rather extreme) case, assumes that the set of beliefs $B$ is expressed over a set of cells $C^B$ at the most detailed aggregation level. Then, we can compare the cells of $q$ with the cells of $C^B$ by converting them to their detailed equivalents. Algorithm \ref{algo:beliefDetailedNovelty} performs the computation of novelty. 

\begin{algorithm}[ht]
\DontPrintSemicolon 
\KwIn{A query $q$; a set of beliefs $B$ over a set of cells $C^B$ at the most detailed level; a threshold $\Pi$ for deciding if a cell is eligible for being novel }
\KwOut{The subset of the cells of $q^{0}$, say $q^{cov^{0}}$ that are also part of the space the beliefs cover, as well as its complement $q^{nov^{0}}$}
\Begin{
produce $q^{0}.cells$ \;
$q^{cov^{0}} \gets \emptyset $\;
$q^{nov^{0}} \gets q^{0}.cells$ \;
$C^\star$ $\gets$ the subset of $C^B$ for which there exists a known belief, i.e., $\{c~|~c \in C^B, \exists~p(M \in m|c) \in B, p(M \in m|c) \geq \Pi \}$\;
\ForAll {$c^{0} \in q^{0}.cells$}{
    \If {$c^{0^{+}}~\in~C^{\star^{+}}$} {
        remove $c^{0}$ from $q^{nov^{0}}$ and add it to $q^{cov^{0}}$\;
    }
}
\Return{$q^{cov^{0}}$, $q^{nov^{0}}$ }
}
\caption{\sf{Partial Extensional Detailed Belief-Based Enumeration Of Covered Cells}}
\label{algo:beliefDetailedNovelty}
\end{algorithm}

Then, we can compute the 
\textit{Partial Detailed Extensional Belief-Based Novelty} (PDEBBN)
of the query $q$ as usual:
\[PDEBBN =
\frac{ |q^{nov^0}|} {|q^{nov^0}|~\bigcup~|q^{cov^0}|}
\]

The check is (a) partial (practically a normalized score), (b) extensional (via cells), and, (c) detailed, i.e., with respect to the most detailed cells of the data space. The complexity analysis, as well as the discussion of the \textit{Syntactic} variant are homologous to the ones of subsection~\ref{sec:SLBBN}.

\subsubsection{Arbitrary-Level Belief-Based Novelty}\label{sec:ABBN}
\sideNew{Sec. \ref{sec:ABBN}: New}
In this case, the cells of $C^B$ are defined at arbitrary levels of aggregation. Thus, it is not straightforward to compute novelty. For the cells of $C^B$ that are defined at higher levels of aggregation compared to the ones of $q$, even for a single dimension (i.e., even if there is a single dimension $D$ for which the cell of $C^B$ is at a higher level than the aggregation level of $q$), it is clear that we cannot use them for assessing novelty, as they express a coarser computation than the one of the query. Assume that we disqualify these cells and stick to the ones that have their levels at a lower or equal level with respect to the levels of $q$. Again, comparison is not straightforward; converting all to the detailed equivalents is not usable, as the knowledge of an aggregate value does not imply the knowledge of its detailed equivalents. Thus, we need to resort to even stricter measures. 

\begin{algorithm}[bthp]
\DontPrintSemicolon 
\KwIn{A query $q$ and its result; a set of beliefs $B$ over a set of cells $C^B$ at arbitrary aggregation levels; a threshold $\Pi$ for deciding if a cell is eligible for being novel }
\KwOut{The subset of the cells of $q.result$, say $q^{cov}$ that are also part of the space the beliefs cover, as well as its complement $q^{nov}$}
\Begin{

$q^{cov} \gets \emptyset $\;
$q^{nov} \gets q.cells$ \;
$C^\star$ $\gets$ the subset of $C^B$ for which there exists a known belief and all their levels are lower or equal to the respective ones of $q$, i.e., $\{c~|~c \in C^B, \exists~p(M \in m|c) \in B, p(M \in m|c) \geq \Pi \land \forall~dimension~D,~c.D.L^c \preceq q.D.L^q  \}$ \;
\ForAll {$c \in q.cells$}{
    \If {$c~is~fully~covered~by~cells~of~C^{\star}$} {
        remove $c$ from $q^{nov}$ and add it to $q^{cov}$\;
    }
}
\Return{$q^{cov}$, $q^{nov}$ }
}
\caption{\sf{Partial Extensional Arbitrary Belief-Based Enumeration Of Covered Cells}}
\label{algo:beliefArbitraryNovelty}
\end{algorithm}

Of course, Algorithm~\ref{algo:beliefArbitraryNovelty} requires the definition of \textit{full coverage} of a higher-level cell by a set of more detailed cells. 

A cell $c$, defined at a set of levels $L^c$, is \textit{fully covered} by a set of cells $C$, all of which are defined at lower or equal levels that the ones of $L^c$ if:\\

$c^{0^+}$ $\subseteq$ $\bigcup_{i} c_i^{\star^{0^+}}$, for all $c^{\star} \in~C$\\

\noindent i.e., the cells of its detailed area are a subset of the detailed cells that correspond to the members of $C$. For all practical purposes, this means that one can compute $c$ from the more detailed levels of $C$ -- thus, ``knowing" it. The check requires a full scan of $C$ for each cell of $c$ and the determination of coverage. \\

\hrulefill
\begin{remark}
Note that non-probabilistic statements can be treated as having a single cell-expected value statement per cell with probability 1. Also, other variants (e.g., weighted) are eligible.
\end{remark}
\hrulefill

\subsection{Reference Example Revisited}\label{sec:refExNov}
Coming back to the reference example of Section~\ref{sec:refEx}, we can comment on the novelty of query $q$ that is assessed over the presence of a query history $Q$ = $\{q_1, \ldots, q_4\}$. \rem{No example for Beliefs!}
\begin{itemize}
    \item Concerning the \textit{Full Syntactic Same-Level Assessment of Novelty} (FSLSN), it takes the value of 1, as we can observe that no query in the query history has an identical definition with $q$. 
    \item The same would apply for the \textit{Partial Syntactic Same-Level Assessment of Novelty} (PSLSN). For the test to apply, we would require the existence of queries with compatible selection conditions to the ones of $q$ in order for the resulting coordinates to be comparable. However, in contrast to the queries of $Q$, $q$ has no selection filters, therefore, its syntactic novelty is also 1. Similarly for the extensional variant of the same metric.
    \item Concerning the \textit{Full Syntactic Detailed Novelty} (FSDN), to the extent that there is no query that encompasses the entire $q$, the novelty is 1.
    \item Concerning the \textit{Partial Detailed Extensional Novelty} (PDEN), if we run the algorithm, we need to (a) take the union of the detailed areas of the queries of $Q$, say $Q^0$, and (b) intersect it with the detailed area of the $q$, $q^0$. In practice, we detect that 70\% of the cells of $q^0$ do not belong to $Q^0$, thus PDEN = 0.7. 
\end{itemize}

%% file: 06_relevance.tex
 \section{Relevance}\label{sec:relevance}
Relevance is a dimension that pertains to retaining focus towards a specific information goal (or a set of them). The dimension of relevance ensures that the data exploration does not wander around areas of the multidimensional space that are not of interest to the current information acquisition goal.

\begin{figure*}[tbp]
    \centering
    \includegraphics[width=\linewidth]{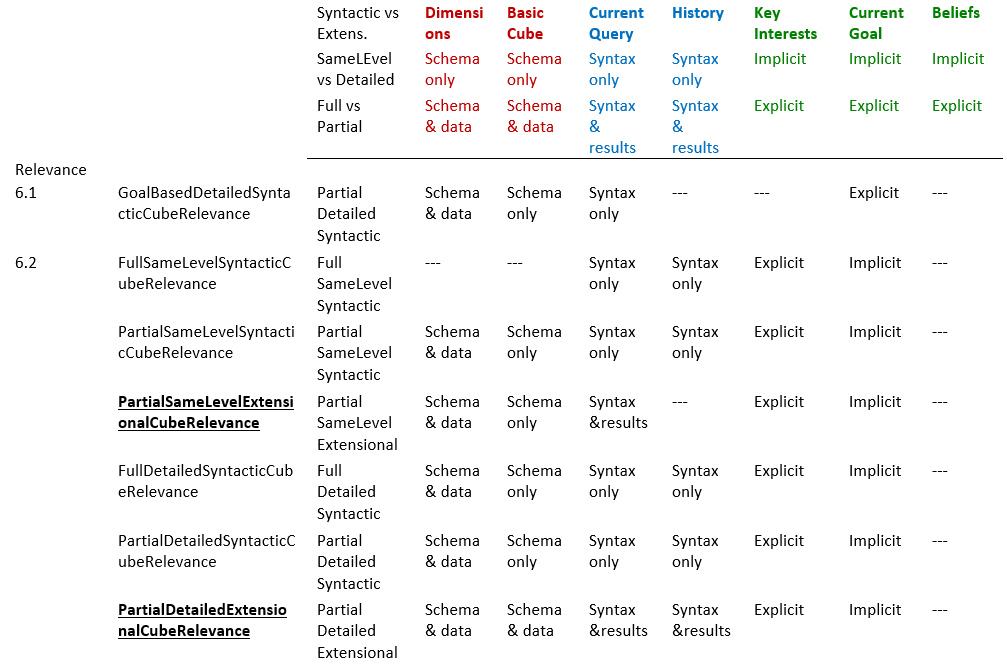}
    \caption{List of Relevance algorithms, characterized with respect to the reference taxonomy (underline: implemented, bold: experimented)}
    \label{fig:ListoAlgoRel}
\end{figure*}

This is particularly the case with business intelligence scenarios, where the need to satisfy an informational gap (either on an ad-hoc or a recurring basis) is the main driver for accessing the database for data. This does not mean that the queries are pre-fixed, however: the quest for an information goal is very often ``open" and an exploration of a certain sub-space of the data, possibly viewed from different angles and at different levels of granularity. In \cite{DBLP:conf/adbis/MarcelPV19} we have named this exploration a ``walk" in the multidimensional space.

As the above discussion demonstrates, a foundation for the assessment of the relevance of a query to an exploratory session or a recommendation to the use is the existence of an informational goal. The goal can be an ad-hoc goal for information, or a recurring one, based on a profile of data that have to be collected to answer recurring questions of the analyst. Specifically, we can discriminate between several cases: (a) the case where the goal is explicitly stated, or, (b) the case where the goal has to be inferred from collateral profile information. In the former case, we will assume that the analyst specifies an area of the information space via a selection predicate (again, the way this is extracted is orthogonal: it can be explicitly requested, it can be inferred from a natural-language expression, it can be part of a query or a KPI, etc). In the latter case, the user has not provided any such information, and the system has to infer the intended goal from other means -- examples include the history of past actions of the analyst, or possibly a profile, or a set of registered KPIs for the analyst.

 \subsection{Relevance assessment in the presence of a declared user goal via a selection predicate}\label{sec:relevanceGoalBased}
 \sideNew{Sec. \ref{sec:relevanceGoalBased}: New}
Assuming, then, that the goal is precisely or approximately specified, the essence of relevance estimation answers the question \emph{how relevant is the query to a user's goal?} The main idea here is we formalize the user's declaration (via an explicit statement) that a specific area of the multidimensional space is of interest to him via a simple selection condition $\phi_G$ that characterizes the user interest.

\begin{algorithm}[ht]
\DontPrintSemicolon 
\KwIn{A query $q$ and, $\phi_G$, a selection condition characterizing an area of the multidimensional space }
\KwOut{The subset of the coordinates of $q^{0}$, say $q^{rel^{0^+}}$ that are also part of the space the detailed proxy of $\phi_G$ covers, as well as its complement $q^{irr^{0^+}}$ of irrelevant cells}
\Begin{
produce $q^{0^+}$ and $\phi^{0^+}_G$ \;
$q^{rel^{0^+}} \gets \emptyset $\;
$q^{irr^{0^+}} \gets q^{0^+}$ \;

\ForAll {$c^{0^+} \in q^{0^+}$}{
    \If {$c^{0^+}~\in~\phi^{0^+}_G$} {
        remove $c^{0^+}$ from $q^{irr^{0^+}}$ and add it to $q^{rel^{0^+}}$\;
    }
}

\Return{$q^{rel^{0^+}}$, $q^{irr^{0^+}}$ }
}
\caption{\sf{Goal-Based Syntactic Enumeration Of Covered Detailed Cells}}
\label{algo:relevanceGoalBased}
\end{algorithm}

Algorithm ~\ref{algo:relevanceGoalBased} computes the subset of the multidimensional space at the most detailed level, i.e., the detailed \textit{signature}, that pertains to the user goal $\phi_G$. Then, it also computes the detailed signature of the query $q$. The algorithm splits the coordinates of the detailed signature of the query in two subsets (a) the ones \textit{relevant to} (or \textit{covered by}) the detailed signature of the user goal, and (b) the \textit{irrelevant}, non-covered ones, represented by the sets $q^{rel^{0^+}}$ and $q^{irr^{0^+}}$, respectively. 

Then, the \textit{Goal-Based Detailed Syntactic Relevance} (GBDSR) of a query is the fraction of its detailed space that overlaps with the user's goal. 

\[GBDSR = 
\frac{ |q^{rel^{0^+}}|} {|q^{irr^{0^+}}|~\bigcup~|q^{rel^{0^+}}|}
\]

The check is  (a) partial (practically a normalized score), (b) syntactical (without using the cells of the query results), and, (c) detailed, i.e., with respect to the detailed levels of the involved cubes. Assuming a fixed goal, and thus a fixed set of signatures for the goal, the complexity of the algorithm is linear with respect to the query result size. Also, to the extent that the test is syntactic, the data size is irrelevant.\\

\hrulefill

\begin{remark}
An extension to a set of multiple goal statements $\Phi$ = $\{\phi_1, \dots, \phi_k\}$ is also possible. The union of the detailed signatures of the goals can provide the equivalent of $\phi^{0^+}_G$ for such an extension. Again, weighted variants can be part of the score evaluation.
\end{remark}
\hrulefill

\subsection{Relevance assessment in the absence of a declared user 
goal}\label{sec:relHisto}
Assume that an explicit goal to study a certain subset of the multidimensional 
space is not available, but instead, 
the system has access to a set of KPIs, expressed as a set of annotated queries 
$Q$ = $\{q_1, \ldots, q_n\}$, which we call \textit{beacon queries}, that \textit{approximate} the user interest. KPIs are explicit expressions of time-invariant interests (rather than a current user goal), so, even if they do not explicate exactly what the user wants to achieve \textit{now}, they act as reference points of relevance for the user's interest. 


As a side-note, observe that, in extremis, one could even resort to the user's history for indications of relevance. We emphasize that past queries are last-resort, coarse manifestations  of relevance, as they are only in the past and not necessarily linked to what the user explores now, or, they could be erroneous, or playful, or 
eventually irrelevant, etc. However, despite all these valid reservations, it could be the case that this is the only thing that the system knows about the user's idea of what is relevant. \\

\textbf{Intuition}. What we want to assess is how much a new query $q$ overlaps with the set $Q$ of beacon queries. Observe that all the methods that we define assess the overlap 
of levels and coordinates between $q$ and the queries of $Q$; \textit{measures and aggregate functions are not involved in the assessment of relevance, as the idea is to ``highlight" the subset of the multidimensional space that seems relevant to the user}.

In the rest of this subsection, we simplify the discussion by avoiding aging factors and possible weights of the different queries and 
considering a single input for the interestingness assessment algorithm:
a set of beacon queries which we (approximately) deem to be relevant. We will also use the notion of \textit{coverage}, 
already discussed for novelty, 
aiming towards finding the overlap of the area covered by the beacon set and the area pertaining the current query.\\

\textbf{The special case where all queries are defined at the same level}. 
Assuming all cubes of $Q$ and $q$ are  at the same level, we can assess relevance via  (a) a full syntactic check returning true/false, as \textit{Full Syntactic Same-Level  Relevance} (FSSLR):
%
\begin{multline*}
FSSLR = \begin{cases}
        1 & \text{if a } q_i \equiv q \text{ exists } \\
        0 & \text{otherwise}
     \end{cases} 
\end{multline*}

\noindent and (b) a partial check returning a \textit{Partial Syntactic Same-Level Relevance} (PSSLR) score

%
%
\[PSLSR =
\frac{ |q^{cov^{+}}|} {|q^{nov^{+}}|~\bigcup~|q^{cov^{+}}|} = \frac{ |q^{cov^{+}}|} {|q^{+}|}
\]

\emph{It is very important to stress that the same-level relevance can only be applied in the case where all the cubes are at the same level of abstraction.} Overall, the idea is that the beacon-set provides a homogeneous space for query evaluation at the same level, and thus, we can compute relevance without having to resort to the detailed space. 

The \emph{Extensional} counterpart of relevance (e.g., $PartialSameLevelExtensionalRelevance$) is defined equivalently, with cells of the query result instead of signatures. 

Now, we are ready to move on to the fundamental definitions of relevance that are based on the detailed level.\\

\textbf{Foundations of history-based relevance assessment}. The most fundamental definition for relevance comes from the space of detailed cells, as \textit{Full Detailed Syntactic Relevance} (FDSR).

\[FDSR =
1 - FullDetailedSyntacticNovelty
\]

The most fundamental assessment method of all is to compare the union of the detailed signatures of the queries of $Q$ with the signature of $q$. The amount of overlap signifies the relevance of the new query.

To characterize the cells of the result of $q$ (in fact: their coordinates) as previously covered vs novel, we can simply refer to Algorithms \ref{algo:EnumDetailedQueryContainmentSignatureBased} and \ref{algo:EnumDetailedQueryContainment} this time passing all the history as argument, i.e., without the requirement of same measures and aggregate functions.
Equivalently, we can use (a) the detailed proxy of $q$, $q^0$ and (b) the detailed equivalents of the queries of $Q$, $q^0_i$, and pass them as input to the algorithm \textsf{ComputePartialImmmediateCubeCoverage} of \cite{PV21}. Observe, that when working at the detailed level, coordinates and cells are practically of the same cost, esp., since measures are not taken into consideration.
Then, the sets $q^{cov^{0^+}}$ and $q^{nov^{0^+}}$ (respectively, $q^{cov^0}$ and $q^{nov^0}$) are produced. Based on these sets, we can compute \textit{Partial Detailed Syntactic Relevance} (PDSR) and \textit{Partial Detailed Extensional Relevance} (PDER), respectively.

\[PDSR = 
\frac{ |q^{cov^{0^+}}|} {|q^{nov^{0^+}}|~\bigcup~|q^{cov^{0^+}}|} = \frac{ |q^{cov^{0^+}}|} {|q^{0^+}|}
\]

\[PDER = 
\frac{ |q^{cov^0}|} {|q^{nov^0}|~\bigcup~|q^{cov^0}|} =  \frac{ |q^{cov^0}|} {|q^{0}|}
\]
The complexity of computing all these formulas is practically the same with the one of Algorithm~\ref{algo:EnumDetailedQueryContainment}, and therefore, linear with respect to query history and fact table size.

\hrulefill

\begin{remark}
Interestingly, when the assessment is history-based, relevance is practically complementary to novelty. For all variants of syntactic vs extensional, partial vs full, same-level vs detailed, when the assessment is history-based, the following formula holds: $relevance$ + $novelty$ = 1. We can only emphasize that this is an approximation applicable only to the history-based metrics that we have introduced here, and by no means do we insinuate that being relevant precludes being novel. Quite the opposite: in an exploratory phase, when a concrete goal starts to shape in the mind of the analyst, the early queries are both relevant and novel. But this, pertains to the case where the analyst has a concrete goal, against which relevance is assessed.

\end{remark}
\hrulefill

\subsection{Reference Example Revisited}\label{sec:refExRel}
Coming back to the reference example of Section~\ref{sec:refEx}, we can comment on the relevance of the query $q$ that is assessed over the presence of a query history $Q$ = $\{q_1, \ldots, q_4\}$. Basically, the explanations given for the case of novelty, in Section~\ref{sec:refExNov}, also cover the discussion for relevance. 

Given that the syntactic checks give a novelty of 1, as one would expect, syntactic relevance takes a value of zero. So, 
\textit{Full Syntactic Same-Level Assessment of Relevance} (FSSLR), \textit{Partial Syntactic Same-Level Assessment of Relevance} (PSSLR), and \textit{Full Syntactic Detailed Relevance} (FSDR) are all zero.
Concerning the \textit{Partial Detailed Extensional Novelty} (PDER), however, it takes the value of 0.3, to the extent that it is the complement of its Novelty counterpart that took the value of 0.7.\\

%
%

%% file: 07_peculiarity.tex
\section{Peculiarity of a query}\label{sec:peculiarity}

How peculiar is a query? To understand peculiarity we must understand that its essence lies in discriminating a particular object (in our case: a query) from its peers (in our case: a session, history, or just collection of other queries, to be used as the context for the assessment of peculiarity). Beliefs, Key Interests and Goals are not explicitly treated here. However, peculiarity can be evaluated on the grounds of whichever entities can be implicitly represented by queries; to the extent that at least Key Interests can be expressed as queries, peculiarity can be implicitly related to them.\rem{This needs deeper consideration}

\begin{figure*}[tbp]
    \centering
    \includegraphics[width=\linewidth]{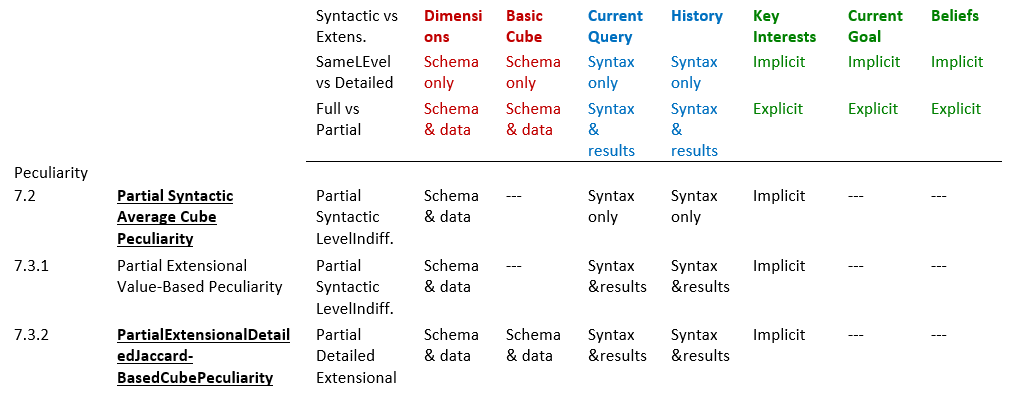}
    \caption{List of Peculiarity algorithms, characterized with respect to the reference taxonomy (underline: implemented, bold: experimented)}
    \label{fig:ListoAlgoPec}
\end{figure*}

Therefore, in the rest of our deliberations, we assume that every query $q$ is going to be assessed against a collection of queries $Q$ = $\{q_{1}, \ldots, q_{n}\}$. This generic setup can cover two alternative situations: (a) a set of KPIs, each expressed via a query, collectively describing a set of static key interests of the user, and, (b) a set of queries in the history (be it the current session, or the history of previous sessions). 

We introduce a variety of methods to assess peculiarity. However, first, we start with a short generic discussion of peculiarity in data mining.

\subsection{Outlierness in Data Mining}

Assume a set of objects $X$ = $\{x_{1}, \ldots, x_{n}\}$ of any kind. When is an item $x$ peculiar? The typical answer to the question, which is pretty much the definition of outlierness, is that $x$ is peculiar whenever it differs a lot from $X$ - $\{x\}$.

\textbf{Definition}. \cite{DBLP:books/sp/Aggarwal15} provides the following definition for outliers: ``\textit{An outlier is an observation which deviates so much from the other observations as to arouse suspicions that it was generated by a different mechanism.}"

\textbf{Methods}. To assess the outlierness, or peculiarity of a data value, \cite{DBLP:books/sp/Aggarwal15} suggests a few nice ideas, including a convex hull algorithm and a K-th nearest neighbor (kNN) distance algorithm. Quoting from \cite{DBLP:books/sp/Aggarwal15}: ``Because outliers are defined as data points that are far away from the ``crowded regions" (or clusters) in the data, a natural and instance-specific way of defining an outlier is as follows:
\textit{The distance-based outlier score of an object O is its \underline{distance to its kth nearest neighbor}. }"


\subsection{Syntactic Peculiarity}\label{sec:syntOutlier}
Assume the query $q$ and a collection of queries $Q$ = $\{q_{1}, \ldots , q_{n}\}$. How different is $q$ from the collection $Q$?

Fundamentally, the question boils down to answering the assessment of the distance of two queries. To support our discussion in the sequel we assume two queries \textit{over the same data set} in a multidimensional space of $n$ dimensions.

\[q^a = \mathbf{DS}^{0},\ \phi^a,\ [L_1^a,\ldots,L_n^a, M_1^a,\ldots,M_{m^a}^a],\ [agg^a_1(M^{a^0}_1),\ldots,agg^a_m(M^{a^0}_{m^a})]\ \]
and
\[q^b = \mathbf{DS}^{0},\ \phi^b,\ [L_1^b,\ldots,L_n^b, M_1^b,\ldots,M_{m^b}^b],\ [agg^b_1(M^{b^0}_1),\ldots,agg^b_m(M^{b^0}_{m^b})]\ \]

To solve the problem of computing the distance of two queries, we use the syntactic formula from \cite{PV21}, which, in turn, is based on results from (see \cite{DBLP:conf/icde/BaikousiRV11}, \cite{DBLP:journals/dss/GolfarelliT14}, \cite{DBLP:journals/kais/AligonGMRT14}).

The syntactic distance of the two queries is expressed by the weighted sum of structural distances between their selection conditions, their grouping levels, and the measures they employ, as:

\[ \delta(q^a, q^b) = w^\phi\delta^\phi(q^a, q^b) + w^L\delta^L(q^a, q^b) + w^M\delta^M(q^a, q^b) ,\]  

\noindent such that the sum of the weights $w^i$ adds up to 1. We follow \cite{DBLP:journals/kais/AligonGMRT14} and recommend the following weights: $w^\phi$: 0.5, $w^L$: 0.35, $w^M$: 0.15.\footnote{For the particularities of the different components of the formula,    we refer the interested reader to \cite{PV21}, Sec.~``Query Distance".}


Given, then, the \cite{PV21} method for computing distance of two queries $\delta(q^a, q^b)$, the computation of the distance of a new query $q$ to a pre-existing collection of queries $Q$ can be computed via several possible methods, out of which we highlight a couple of prominent ones:
\begin{enumerate}
    \item A simple statistic over the distances of the query to the set members, $\delta(q,Q)$ = $\gamma(\delta(q,q_i)), q_i \in Q, \gamma \in \{min, max, average, median\}$.
    \item k-nn distance of the query to the set, $\delta(q,Q)$ = $k$-th smallest $\delta(q,q_i), q_i \in Q$. Practically, this entails ranking all the distances of $q$ to the elements of $Q$ in ascending order and take the k-th one.
\end{enumerate}

The check is (a) partial (practically a normalized score), (b) syntactical (without using the cells of the query results), (c) depending upon the statistic or function that determines the final value of the metric, and, (d) indifferent to the schema levels of the involved cubes.  We can define a \textit{Partial Syntactic Cube Peculiarity} based on which method we pick for the determination of the final value, e.g., \textit{Partial Syntactic Average Cube Peculiarity} uses the average query distance to determine the peculiarity of the measured query. To the extent that we refer to syntactic checks, data size is irrelevant for the complexity of the algorithm. However, the algorithm requires a linear pass from all the queries of the history and a pairwise computation of distance at its first phase, as well as the determination of the final peculiarity (again requiring at most a linear past of all distances): therefore, the complexity is linear with respect to the size of the collection $Q$.\\

\subsection{Value-based Peculiarity}\label{sec:valueOutlier}
When we address the issue of value-based peculiarity assessment, we base the result of the assessment on the actual values of the cells of the result of the query. Then, we treat each query as a set of cells (each cell primarily identified by its coordinates).

\textbf{The general setup of value-based peculiarity}. The general setup of the value-based query peculiarity problem is as follows. Assume a set of background queries $Q$ = $\{q_{1}, \ldots, q_{n}\}$ (either due to the history of a session, or, due to the existence of a set of KPI's). Assume also a new query $q$ that is also submitted to the system.

Algorithm ~\ref{algo:genericValuePeculiarity} provides the generic recipe for computing query peculiarity. Depending on the setup of individual design choices, we can have several configurations of the algorithm.


\begin{algorithm}[ht]
\DontPrintSemicolon 
\KwIn{A set of background queries $Q$ = $\{q_{1}, \ldots, q_{n}\}$; 
a new query $q$ to be assessed over $Q$ for its peculiarity;
a distance function $\delta_q$ for computing the distance of two queries,
an aggregate function to compute the query peculiarity $f_p^{agg}$}
\KwOut{The value-based peculiarity of the query $q$}
\Begin{
The bag of distance values of $q$ over members of $Q$, $V$ = $\emptyset$;

\ForAll{$q_i$ $\in$ $Q$}{
    $V$ = $V$ $\bigcup$ $\delta_q(q_i,q)$;
}
$q.peculiarity$ = $f_p^{agg}(V)$;

\Return{$q.peculiarity$ };
}
\caption{\sf{The general setup of value-based query peculiarity assessment}}
\label{algo:genericValuePeculiarity}
\end{algorithm}

The combination of the query distance function $\delta_q$ and the aggregate function $f_p^{agg}$ can determine the peculiarity of the query. As we will present in the sequel, the two prominent methods for assessing the cube query distance $\delta_q$ are the Hausdorff and the Closest Relative methods, whereas the $f_p^{agg}$ aggregate function can be serviced by any aggregate function like min, max, k-NN, etc.

The check is (a) partial (to the extent that $\delta_q$ returns a score ), (b) extensional (using the cells of the query results), (c) depending upon the aggregate function that determines the final value of the metric, and, (d) indifferent to the schema levels of the involved cubes.
We can define a \textit{Partial Extensional Value-Based Peculiarity} based on which method we pick for the determination of the final value, e.g., \textit{Partial Extensional Hausdorff/ClosestRelative Average/k-NN/Minimum Peculiarity} if we use (a) the Hausdorff or the Closest Relative method for the determination of query distance, and, (b) the average/k-NN/minimum query distance to determine the peculiarity of the measured query. 
The algorithm requires a linear pass from all the queries of the input query set, as well as the determination of the final peculiarity (again requiring at most a linear past of all distances):
therefore, the complexity is linear with respect to the size of the collection $Q$. To the extent that we use query results, we can assume that the size of the query results affects the execution time of the algorithm.\\

\subsubsection{The closest relatives of Hausdorff}\label{sec:Hausdorff}\sideNew{Section ~\ref{sec:Hausdorff}: New}
\textbf{Cell-based Query Distance}. How then, do we compute the distance of two queries? Assume we want to assess how distant are the queries $q^a$ and $q^b$ with

\[q^a.cells = \{c^a_1, \dots, c^a_{n^a}\} \text{ vs. } q^b.cells = \{c^b_1, \dots, c^b_{n^b}\}
\]

Earlier works about comparing queries through their sets of cells,
such as \cite{DBLP:conf/dawak/GiacomettiMN09},
have shown that the distance of these two sets of cells is not straightforward to assess. The reasons can be identified as follows:

\begin{itemize}
\item It is not straightforward how to map the cells of the one query to another; this is especially true if the cardinality of the two queries is not the same;
\item It is possible that the two queries are defined at different levels of aggregation, which means that they are not directly 
comparable;
\item Even if the above problems are not present, deciding a mapping from the cells of $q^a$ to the cells of $q^b$ is not a straightforward task.
\end{itemize}

If we want to exploit the query results, i.e., assuming the cells of the query results are available, we can reuse the results of \cite{DBLP:conf/icde/BaikousiRV11} to derive query distances.
The main idea of \cite{DBLP:conf/icde/BaikousiRV11} was to assess alternative ways of computing the distance of two cubes on the basis of their contents. Two formulae eventually came out as the possible winners of the benchmark, specifically the \textit{Closest Relative} and \textit{Hausdorff} formulae. Before discussing these alternatives, however, we will introduce a cell distance formula, which is necessary for performing the respective cube distance calculations.

\textbf{Cell distance}. How distant are two cells? \cite{DBLP:conf/icde/BaikousiRV11} performs a thorough analysis of several alternatives, out of which, the experimental assessment clearly discriminated the \textit{Weighted Sum of Value distances based on the Least Common Ancestor (LCA)} method as the most appropriate one.

Assume two members of a dimension $D$, say $v$ and $v'$, not necessarily at the same levels. Assume also $v_{LCA}$ is their least common ancestor (could be one of them if they are related with an $anc()$ relation). Then, the \textit{distance of two values of the same dimension} is 
\[
dist(v,v') = \frac{path(v,v_{LCA}) + path(v',v_{LCA})} {2 \times path(ALL,L^0)}
\]
\noindent where $path$ is the number of hops (edges) in the hierarchy between the respective values(levels, respectively).

\begin{figure}[ht]
\centering
\includegraphics[width=0.5\textwidth]{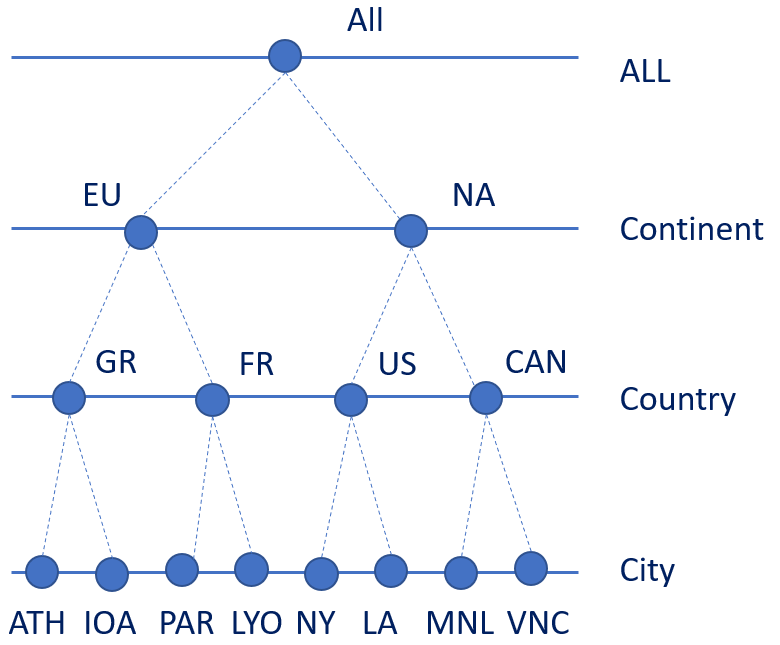}
\caption{A sample geographical dimension}\label{fig:LCA}
\end{figure}

What is the distance of Athens to Canada in figure~\ref{fig:LCA}? The least common ancestor is the $All$ value in level $ALL$ and has distance (number of intervening edges) to Athens equal to 3 and distance to Canada equal to 2. The edges between $L^0$, i.e., $City$, and $ALL$ is 3. Then, $dist(Athens, Canada)$ is $\mathbin{(3+2)/(2 \cdot 3)}$ = 5/6.

To simplify \cite{DBLP:conf/icde/BaikousiRV11}, the \textit{distance of two cells} over the same dimensions is the weighted sum of the distances of their respective values. Given two cells $c: <v_1, \ldots, v_n >$ and $c': <v'_1, \ldots, v'_n >$ their distance is:

\[
dist(c,c') = \frac{1}{n}\sum_{\substack{1 \leq i \leq n }} dist(v_i, v'_i)
\]

\textbf{Closest Relative distance of two cube queries}. The closest relative distance of two cubes \cite{DBLP:conf/icde/BaikousiRV11} is based on mapping the cells of the two cubes in pairs with the minimum distance and taking their average distance. Specifically, the method to compute the closest relative distance of two cube queries $q$ and $q'$ includes the following steps:
\begin{enumerate}
    \item For each cell $c$ of query $q$, find the cell $c'$ in $q'$ with the minimum distance;
    \item Add the respective distance to a bag of values $B_d$;
    \item Once done with all cells of $q$, return the mean value of $B_d$
\end{enumerate}

The intuition of the formula is very simple: we take the average distance between the cells of the two cubes as the distance of the two cubes.

\textbf{Hausdorff distance of two cube queries}. As mentioned in \cite{DBLP:conf/icde/BaikousiRV11}, the Hausdorff distance between two cube queries $q$ and $q'$ can be defined as: 

\begin{multline*}
H(q,q') = max(h(q,q'), h(q',q)),\ \text{ where } \\
h(q,q') = max_{c \in q.cells}(min_{c' \in q'.cells}(\delta(c,c')))\ \text{ and }\\ 
\delta(c,c') \text{ is the distance of any two cells}
\end{multline*}

Function $h(q, q')$ is called the directed Hausdorff distance from $q$ to $q'$, and it is not necessarily symmetric. Practically, to compute $h$ we have to perform the following steps: (a) for every cell $c~\in~q$, we find the cell $c'\ \in\ q'$ with the minimum distance (effectively pairing each cell of cube $q$ to its closest counterpart in $q'$); (b) out of all these distances, we select the maximum one. Then, we do the symmetric computation for the $h(q', q)$ and we take the maximum of the two $h(\cdot)$ values.

\subsubsection{Jaccard-based resolution via cell comparison at the detailed level}
A possible answer to the problem is to address the issue by referring to the detailed cells that pertain to the aggregate cells that constitute the results of the compared queries. Remember that we refer to the set of cells that produce an aggregate cell as the \textit{detailed area} of the cell; the detailed area of a set of aggregate cells is defined respectively. Let $q^0_1.cells$ be the detailed area of $q_{1}$ over $C^{0}$ and $q^0_2.cells$  be the detailed area of $q_{2}$ over $C^{0}$. Then, we can compute the 
Jaccard similarity of the two detailed areas. The distance of the two queries is: 

$distance(q_1, q_2)$ = 1 - $JaccardSimilarity(q^0_1.cells, q^0_2.cells)$. \\

\silence{
\begin{table}[bth]
\centering
\begin{tabular}{|l|}
\hline
\textbf{Algorithm}: \sf{Compute Jaccard-Based Peculiarity} \\
\hline

\textbf{Input}: the query history $Q$, a new query $q$, an integer $k$ for the kNN\\
\textbf{Output:} peculiarity of $q$ to $Q$, $peculiarity(q | Q)$\\

\textbf{Begin} \\
Let $L$ = $\emptyset$ a list of jaccard distances \\

$q^{0}$ = detailed area of interest for the query $q$ \\

For each $q_{i}$ in $Q$ $\{$\\

\quad q$_{i}$$^{0}$ = detailed area of interest for the query $q_{i}$ \\

\quad Jaccard distance $JD_{i}$ = 1 - $\frac{|q^{0}_{i}~\bigcap~q^{0}|}{|q^{0}_{i}~\bigcup~q^{0}|}$ \\   

$\}$ \\

$L_{s}$ = Sort JD$_{i}$ ascending into a sorted list \\

$peculiarity(q | Q)$ = $L_{s}[~k~]$, i.e., the k-th element of the list = q's kNN\\

\textbf{End} \\
\hline
\end{tabular}
\end{table}
}

\begin{algorithm}[tbh]
\DontPrintSemicolon 
\KwIn{A new query $q$, the query history $Q$, and an integer $k$ for picking the k-th neighbor}
\KwOut{the PartialExtensionalDetailedJaccard-BasedCubePeculiarity $valueBasedPeculiarity(q | Q)$}

\Begin{
Let $L$ = $\emptyset$ a list of Jaccard distances \;
Compute $q^{0}$, i.e., the detailed area of interest for the query $q$ \;
\ForAll {$q_{i}$ $\in$ $Q$} {    
    Compute {$q_{i}^{0}$}, i.e., the detailed area of interest for the query {$q_{i}$} \;
    Compute the Jaccard distance $JD_{i}$ = 1 - $\frac{|q^{0}_{i}~\bigcap~q^{0}|}{|q^{0}_{i}~\bigcup~q^{0}|}$ \;
    add {$JD_{i}$} to {$L$} \;  
}
$L_{s}$ = Sort $L$ ascending into a sorted list\;

\Return{$peculiarity(q | Q)$ = $L_{s}[~k~]$}
}
\caption{\sf{Partial Extensional Detailed Jaccard-Based (Value-based) Cube Peculiarity}}
\label{algo:peculiarityJaccardBased}
\end{algorithm}

The intuition of Algorithm~\ref{algo:peculiarityJaccardBased} is based on the idea that the peculiarity of a query is based on how much overlap its detailed cells have with the detailed cells of the queries in the history. 
The check is (a) partial (practically a Jaccard distance), (b) extensional (with the use of the cells of the query results), and, (c) detailed, i.e., with respect to the detailed levels of the involved cubes.Thus, we define the \textit{Partial Extensional Detailed Jaccard-Based Cube Peculiarity} (for short: Value-based Peculiarity) as the result of Algorithm \ref{algo:peculiarityJaccardBased}. 

The execution cost is dominated by the execution of the detailed queries for both the reference queries and the queries of the history $Q$. The complexity of the algorithm is obviously linear with respect to the history size, since there is a single detailed query $q_i^0$ to be executed per member of $Q$. Also, the in-memory check between the results of the queries is also linear with respect to the history size. At the same time, the complexity is also linear with respect to the cube size, assuming that the execution cost for all the queries linearly depends on the cube size (i.e., all the involved queries have their execution time scale linearly with the same scale factor over the cube size).

\hrulefill

\begin{remark}
	A point worth mentioning here, is that the form of peculiarity we have been discussing so far, is signature-based, i.e, defined with respect to the area of the multidimensional space it refers to. As part of future work, research might address peculiarity via a more value-based approach, where the comparison of the queries is more based on values than on signatures. The extent of the issue is vast, since we need to synthesize the combined peculiarity of a query on the basis of its cells, and, to this end, we need dedicated studies on the topic, on how users perceive derived value-based peculiarity. The extent of the problem is such that it places it out of the scope of this paper.
\end{remark}
\hrulefill

\subsection{Reference Example Revisited}

\textit{Partial Syntactic Cube Peculiarity}. The distances of the new query $q$ from the rest of the queries of the history are depicted in Table~\ref{tab:syntPecul}. Then, it is easy to pick either the \textit{Partial Syntactic Average Cube Peculiarity} (as depicted in the Table), or any other aggregate value over the individual distances (e.g., the $k-th$ one).

\begin{table}[tbh]
\begin{tabular}{lrrrr}
$q$ vs  & $\delta^{\phi}$     & $\delta^{L}$         & $\delta^{M}$         & $\delta$                 \\
\hline
$q_1$         & 1.0                  & 0.0                  & 0.0                  & 0.5                      \\
$q_2$         & 1.0                  & 0.0                  & 0.0                  & 0.5                      \\
$q_3$         & 1.0                  & 0.0                  & 0.0                  & 0.5                      \\
$q_4$         & 0.67                 & 0.0                  & 0.0                  & 0.33                     \\
avg           & \multicolumn{1}{l}{} & \multicolumn{1}{l}{} & \multicolumn{1}{l}{} & \multicolumn{1}{l}{0.46}
\end{tabular}
\caption{Syntactic distances of $q$ from the rest of the queries in the reference example. We use the following weights: $w^\phi$: 0.5, $w^L$: 0.35, $w^M$: 0.15.}
\label{tab:syntPecul}
\end{table}

\textit{Partial Extensional Detailed Jaccard-Based Cube Peculiarity}. The basic ingredient for determining the Jaccard based distance is the computation of the quantity $\frac{|q^{0}_{i}~\bigcap~q^{0}|}{|q^{0}_{i}~\bigcup~q^{0}|}$. Assuming we take the $k$ = 2 distance, the  \textit{Partial Extensional Detailed Jaccard-Based Cube Peculiarity} (for short: Value-based Peculiarity) of $q$ is 0.94.

\begin{table}[tbh]
\scriptsize
\begin{tabular}{lrrrr}
q vs  & $|q^{0}_{i} \bigcap q^{0}|$ & $|q^{0}_{i} \bigcup q^{0}|$ & $J$                  & $JD$                     \\
\hline
$q_1$ & 11                          & 190                         & 0.06                 & 0.94                     \\
$q_2$ & 0                           & 137                         & 0.00                 & 1.00                     \\
$q_3$ & 21                          & 123                         & 0.10                 & 0.90                     \\
$q_4$ & 2                           & 117                         & 0.02                 & 0.98                     \\
2-NN  &         &         &  & {0.94}

\end{tabular}
\caption{Jaccard distances of $q$ from the rest of the queries in the reference example. We list the number of cells in the intersection and union of the detailed areas of the involved queries, their fraction $J$, and the Jaccard distance.}
\label{tab:jaccPecul}
\end{table}

%% file: 08_surprise.tex
\section{Surprise}\label{sec:Surprise}
Surprise is an interestingess dimension that depends mainly (if not only) on prior beliefs. The main idea about assessing surprise is to evaluate how far from the beliefs of the analyst do the actual values lie. The two problems that one has to handle are: (a) \textit{what kind of beliefs can we express, and how?}, and, (b) \textit{assuming these beliefs have, somehow, been expressed, how can we compute surprise on their basis?} 

\begin{figure*}[tbp]
    \centering
    \includegraphics[width=\linewidth]{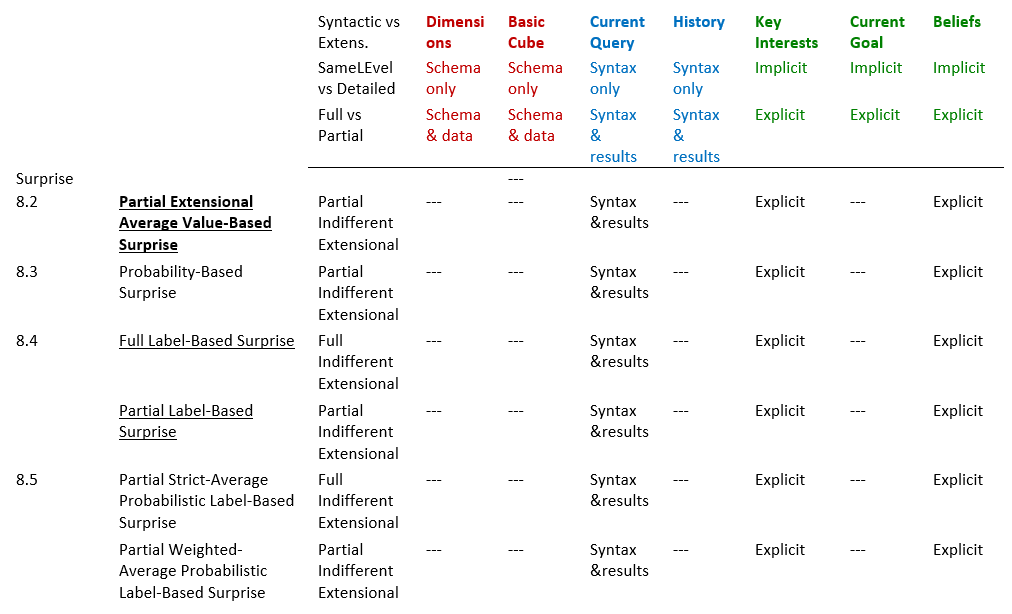}
    \caption{List of Surprise algorithms, characterized with respect to the reference taxonomy (underline: implemented, bold: experimented)}
    \label{fig:ListoAlgoSur}
\end{figure*}

\subsection{Expressing beliefs}
We can express beliefs in a variety of ways: specific values, expected intervals, probabilities; we can even label results and give probabilities for the labels, too as already shown in previous paragraphs. 
Is it necessary, however, for the analysts to express beliefs manually? In the case of KPIs that label performance, this is explicitly done. In the general case, all analysts work with some form of predictions that are automatically derived via methods in the spectrum from a simple regression over past values to elaborate statistical models that economists use.

\subsection{Computing surprise: the overall 
setup}\label{sec:SupriseSetup}
We start with the second problem and assume that for certain cells in the multidimensional space, we can register or compute their expected values for specific measures (several alternatives are discussed in the rest of this section). So, for such a cell, for each of these measures, we have (a) the actual value $m$, and, (b) the expected value $m^{e}$.

Then, the questions that we need to answer are (a) how do we assess the surprise for a specific cell over a specific measure, (b) how do we assess the surprise for a specific cell, with respect to all its measures (assuming multiple such measures exist), and, (c) how do we assess the overall surprise of a query result (which, of course, includes a set of cells)?

Let us start with a single measure for a single cell. Fundamentally, surprise is a function of how far the expected from the actual value lies. Therefore, $surprise(c.M)$ = ($distance(m,m^{e})$) -- for example, $surprise(c.M)$ = $|m-m^{e}|$. 

Assuming now a set of measures per cell, the total surprise of a cell is an aggregate measure computed over the set of surprise values for the various measures of a cell (e.g., the number of measures indicating a non-zero amount of surprise, or maybe the maximum, or the average surprise). Formally:

\begin{multline*}
surprise(c) = f^{agg}_{cell}(surprise(c.M_i)), \\
f^{agg}_{cell} \in \{count, sum, mean, median, max, min,...\}\\
\end{multline*}

Finally, now that we can compute the surprise for each individual cell, we can proceed in computing the surprise for a set of cells, e.g., a query result. The surprise of a set of cells, say $C$ = $\{c_{1}, \dots,c_{n}\}$ is

\begin{multline*}
surprise(C) = f^{agg}(surprise(c_{i})),\\
f^{agg} \in \{count, sum, mean, median, max, min,...\}
\end{multline*}

One possible concern here is what happens if there is no expected value registered for a measure of a cell. Then, there are two ways to handle the situation: (a) this particular measure value does not participate in the rest of the computation, or, (b) a mechanism for computing a derived expected value, against which we will perform the comparison (e.g., the average of the expected values, an interpolation over certain criteria, etc), is introduced. Unless explicitly mentioned otherwise, the former policy of excluding the respective measure value from any computation will be our reaction of choice.

\subsection{Value-based average cell surprise}\label{sec:vbSurpr}
The most simple implementation of the assessment of surprise is to follow the general setup and (a) compute a simple distance of the actual and the expected value per measure, and per cell, (b) aggregate the measures' surprise per cell, and (c) aggregate the different cell surprises to compute the surprise of the set of cells.

Algorithm~\ref{algo:genericValueSurprise} provides the general recipe for computing the surprise according to the general setup. This generic algorithm can be specialized by fixing the involved functions to specific choices. For example, to compute the \textit{Partial Extensional Average Value-Based Surprise}, Algorithm~\ref{algo:avgValueSurprise} works on a single-measured cube, with absolute distance as the distance function to assess how far the actual and the expected measures are, and averaging over all cells with surprise to produce the aggregate cube surprise.   

\begin{algorithm}[t]
\DontPrintSemicolon 
\KwIn{A cube $C$ including a set of cells $\{c_1, \dots, c_n\}$ with a set of measures $\mathbf{M}$; a set of tuples registering the expected values for each cell $E$ = $\{e_1, \ldots, e_n\}$, with each $e_i$ being a tuple of expected measures $e_i = < m^e_1, \ldots, m^e_m >$; a distance function $\delta_M$ for each measure, for computing the distance of the actual from the expected value of a cell's measure; an aggregate function to compute a cell's surprise $f_{cell}^{agg}$; an aggregate function to compute the cube's surprise $f^{agg}$}
\KwOut{The surprise carried by the cube $C$}
\Begin{
The bag of surprise values for $C$, $C.S$ = $\emptyset$;

\ForAll{$c$ $\in$ $C$}{
    The bag of surprise values for this cell $c.S$ = $\emptyset$;
    
    \ForAll{$M~\in~\mathbf{M}$}{
        \If{$\exists$ an expected value $c.m_j^e$ for $c.m_j$, both over measure $M$}{
            $c.S$ = $c.S$ $\bigcup$ $\delta_M(c.m_j, c,m_j^e)$;
            }
    }
    $c.surprise$ = $f_{cell}^{agg}(c.S)$;
    
    $C.S$ = $C.S$ $\bigcup$ $c.surprise$;
}
$C.surprise$ = $f^{agg}(C.S)$;

\Return{$C.surprise$ };
}
\caption{\sf{The general setup of value-based surprise assessment}}
\label{algo:genericValueSurprise}
\end{algorithm}

\begin{algorithm}[t]
\DontPrintSemicolon 
\KwIn{A cube $C$ including a set of cells $\{c_1, \dots, c_n\}$ with a single measure $M$, a set expected values for each cell $E$ = $\{m^e_1, \ldots, m^e_n\}$}
\KwOut{The (average) surprise carried by the cube $C$}
\Begin{
$countOfCellsWithSurprise$ = 0;\\
$C.surprise$ = 0;\\
\ForAll{$c$ $\in$ $C$}{
    $c.surprise$ = null;
    
    \If{$\exists$ an expected value $c.m^e$ for $c.m$}{
        $c.surprise$ = $|c.m - c.m^e|$;\\
        $countOfCellsWithSurprise++$;\\
        $C.surprise$ += $c.surprise$;\\
    }
}
\uIf{$countOfCellsWithSurprise \neq 0$}{
    $C.surprise$ = $C.surprise$/$countOfCellsWithSurprise$;
}
\uElse{
    $C.surprise$ = null;
}
\Return{$C.surprise$ };
}
\caption{\sf{Value-based surprise assessment for a single measured cube by absolute distance for expected values and averaging of cell surprise }}
\label{algo:avgValueSurprise}
\end{algorithm}

The complexity of the algorithm is linear with respect to the result size for the query, assuming a fixed set of expected values $E$.
\silence{
\begin{table}[tb]
\centering
\begin{tabular}{|p{15cm}|}
\hline
\textbf{Algorithm 0}: computeValueSurprise \\
\hline
\textbf{Input}: A cube \textit{C}, with a set of measures \textbf{
M}, the set of expected values for \textbf{M}, say \textbf{V}.\\
\textbf{Output:} surprise of C to V, $surprise(C | V)$ \\

\textbf{Begin} \\

For each cell c $\in$ C, surprise(c) = 0 \\
For each measure M$_{i}$ in \textbf{M} \\
Let c.m$_{i}$ be the actual value for measure $M_{i}$, $v.m_{i}$ the expected one\\
$surprise(c) += |c.m_{i} - v.m_{i}|$ \\
cubeSurprise += surprise(c)\\
Return cubeSurprise / |C| /* i.e., average cell surprise */\\
\textbf{End} \\
\hline
\end{tabular}
\end{table}

} 

\subsection{Expressing expectancy via probabilities of expected values}\label{sec:intervalMeasureProbs}
\sideNew{Section ~\ref{sec:intervalMeasureProbs}: New}
Whereas in the previous section we have assumed that a specific value is available as the expected measure of a cell, in this Section we follow a different approach and register expected values by annotating the expectation for a value to appear via a probability of appearance, and then, measure surprise on the basis of this probability.
We will refer to the surprise metrics that are produced by the alternatives introduced in this Section, as belonging to the category of \textit{Partial Probability-Based Surprise}.

\subsubsection{Probability of values}
Assume a cell $c$ and a certain measure $M$ (for ease of comprehension, we simplify by using just a single measure per cube). Apart from the previously mentioned value-based evaluation, another possibility for assessing surprise is to register probabilities per expected value for the value $m$ = $c.M$. So, we annotate each cell with a set of statements of the form:

\begin{center}
$p(c.M = m)$ = $p^m$, $p^m$ $\in$ $[0..1]$ 
\end{center}

In the above expression, by abuse of notation, we use the term $c$ to refer to the coordinates of the cell $c^{+}$.
In all our deliberations, $p(<expression>)$ expresses the probability of appearance of the parameter of the function $p(\cdot)$.
For example, in a 2 dimensional cube over geography and time, we can 
have:

\begin{eqnarray*}
p(sales=100 ~|~ city=Athens, year=2020) = 20\%\\
p(sales= 80 ~|~ city=Athens, year=2020) = 70\%\\
p(sales= 70 ~|~ city=Athens, year=2020) = 10\%\\
\end{eqnarray*}

Assuming the actual value of the measure is $m$, the strict surprise of 
the cell for a value $m$ is the sum of the probabilities of all the other values $m'$ that are different than $m$.

\begin{center}
$c.StrictSurprise = \sum_{\substack{m' \neq m}} p(c.M = m')$
\end{center}

In the example above, assuming the actual value is 70, the surprise is $20\% + 70\% = 90\%$.

The result of applying Algorithm~\ref{algo:genericValueSurprise} with exact probabilities for the cells' measure will be referred to as \textit{Partial Exact Probability Surprise}.

\subsubsection{Interval-based probability definition}\label{sec:intervalProbability}
A more realistic approach in terms of how we express the probabilities, is that instead of identifying probabilities for individual values, we can assign probabilities to intervals of values. Thus, the statements take the form:

\begin{center}
$p(c.M \in [low \dots high])$ = $p^m$, $p^m$ $\in$ $[0..1]$ 
\end{center}

For example, one could express the statement

\begin{center}
$p(sales \in [100..200] ~|~ city=Athens, year=2020)$ = $20\%$
\end{center}

In the above expression, and in contrast to the setting of the previous subsection, the probability is expressed for a range of measure values, rather than a single one. To facilitate the registration of such expected values, a similar trick can be done, in terms of expression, for the cell coordinates. So, instead of saying

\begin{center}
$p(sales = 100 ~|~ city=Athens, year=2020)$ = $20\%$
\end{center}

one could possibly say

\begin{center}
$p(sales = 100 ~|~ city=Athens, year \in [2018..2020])$ = $20\%$
\end{center}

or even
\begin{center}
$p(sales \in  [100..200] ~|~ city=Athens, year \in [2018..2020])$ = $20\%$
\end{center}

It is important to note, however, that these expressions are no more than syntactic-sugar statements on how we 
express the fundamental statement of assigning probabilities of the form 
$p(M = m ~|~ c)$ = $p$. Therefore, the method for computing surprise does not change, effectively. Specifically, assuming the actual value of the measure is $m$, the strict surprise of the cell for a value $m$ is the sum of the probabilities of all the expressions with ranges $r'$ that do not include $m$.

\begin{center}
$c.StrictSurprise = \sum_{\substack{r'~\not\owns~m}} p(M \in r' | c)$
\end{center}

The result of applying Algorithm~\ref{algo:genericValueSurprise} with probability intervals for the cells' measure will be referred to as \textit{Partial Interval Probability Surprise}.

\subsection{Label-based Surprise Assessment}\label{sec:labelSurprise}
\sideNew{Section ~\ref{sec:labelSurprise}: New}
Inline with interval-based annotation, another possibility is that instead of assessing surprise with respect to the actual measure of a cell, a possibly more convenient and realistic approach is to apply a label to the cell's measure. This allows, not only an easier-to-register mechanism, but also a more robust characterization, as small deviations 
from a measure do not alter the overall assessment. Of course, the price to pay here is that this places the burden of assessing the situation to the labeling mechanism.

Intuitively, labeling turns down the impact of small deviations as captured by measures, and restricts the algorithm to care only on the impact of more significant deviations, as captured by labels. For example, assume that \textbf{we expect} that wine sales in Athens in 2020 will be between 15 and 20 under normal circumstances, and, accordingly we label sales for $city = Athens, product = wine, year = 2020$ as $OK$ if they belong to the interval $[15,20]$, or $Bad$ / $Good$ otherwise, depending on how they turn out to actually be, compared to the expected. Now, we are ready to give a couple of examples, where we contrast this expectancy to the \textbf{actual} values. As a first example, assume that the actual value is 19, resulting also in the label $OK$: this means that there is no surprise really. On the contrary, assume that the actual value is $5$, resulting in a label $Bad$: the difference of expected to actual label signifies a surprising result for this cell.

Formally, for each cell $c$, we require the existence of the function $label(c)$, with $label: C \longrightarrow \Lambda$, $\Lambda$ being a finite set of labels. Unless otherwise specified, we assume the values of $\Lambda$ to be nominal. This also covers the case where they are ordinal, i.e., we can also order the labels and allow the operator $>$. Whenever the domain of labels is of interval type, i.e., we can define the distance of two labels too, this will be explicitly stated. 

We will use the following notation: $c.M.\lambda^e$, or simply $\lambda^e$, is the expected label for the measure $M$ of a cell $c$,  and  $c.M.\lambda$, or simply $\lambda$, is the label for the actual value for the measure $M$ of a cell $c$.

 \subsubsection{Surprise is computed directly over the labels}

We define the following distinction for the computation of the label-based surprise of a cell:
\begin{itemize}
\item \textit{Strict cell surprise}. Assume that, for a certain cell, $c$, there exists a measure $M$, such that $\lambda^e \neq \lambda$. Then, $c.surprise$ = $true$ (equiv., if one insists in a numerical assessment, $c.surprise$ = 100\%)

\item \textit{Loose cell surprise}. We have an interval type of labels, and, thus we can express $c.surprise$ = $distance(\lambda , \lambda^{e})$ via a distance function that accompanies the domain of labels. Assuming a set of measures (as opposed to just one measure), we can combine the different loose surprise evaluations via an aggregate function $f_{cell}^{agg}$ (for example, this functions can be $count(\cdot)$, i.e., we assign as the total surprise of a cell, the number of measures for which a surprise is encountered).

\end{itemize}

In symmetry to the above distinction, we can generalize the computation of surprise for an entire query result, or in general, of a cube $C$ defined as a set of cells, with the same dichotomy:
\begin{itemize}
\item \textit{Strict cube surprise}. If there is even a single surprising cell in the cube's set of cells, the surprise of the cube is $true$; otherwise, it is false. In this case, the result is a \textit{Full Label-Based Surprise} value.

\item \textit{Loose cube surprise}. We can compute an aggregate value of the surprise of the cube's cells, via an aggregate function $f^{agg}$ and return a value (ideally normalized in $[0,1]$). In this case, the result is a \textit{Partial Label-Based Surprise} value.
\end{itemize}


We can think of a generic algorithm (Algorithm~\ref{algo:genericLabelSurprise}) to cover the general case of how to compute the surprise of the entire cube, on the basis of labels. The main idea is as follows. For every cell, and for each of its measures, we try to see whether there is an expected label. If there is such an expected label, we compute the actual label by applying the function $\Lambda_M$ to the measure, and, we contrast it to the expected label via the function $\delta_M$. We retain a composite metadata object for each cell $c$, $c.surprise$ that includes a tuple of comparison results ($c.surprise.Tuple$) and a total surprise score, $c.surprise.Score$, which is computed once all the cell's measures are visited, by applying the function $f_{cell}^{agg}$ to the surprise tuple of the cell. Moreover, we add the cell's surprise metadata object to a global bag $C.S$ that accumulates all such metadata information for all cells. Once all cells have been visited, a global surprise assessment function $f^{agg}$ is applied to this bag, in order to compute the cube's surprise.

\begin{algorithm}[ht]
\DontPrintSemicolon 
\KwIn{A cube $C$ including a set of cells $\{c_1, \dots, c_k\}$ with a set of measures $\mathbf{M}$, 
a set of tuples registering the expected labels for each cell $E$ = $\{e_1, \ldots, e_n\}$, with each $e_i$ being a tuple of expected labels $e_i = < \lambda^e_1, \ldots, \lambda^e_m >$, 
a labeling function $\Lambda_M$ for each measure $M$, 
a distance function $\delta_M$ for computing the distance of a measure's actual from its expected label, 
an aggregate function to compute a cell's surprise $f_{cell}^{agg}$, and, an aggregate function to compute the cube's surprise $f^{agg}$}
\KwOut{The surprise carried by the cube $C$}
\Begin{
The bag of surprise metadata for $C$, $C.S$ = $\emptyset$;

\ForAll{$c$ $\in$ $C$}{
    \ForAll{$M~\in~\mathbf{M}$}{
       \If{$\exists$ an expected value $c.M.\lambda^e$ for $M$}{
            $c.M.\lambda$ = $\Lambda_M$(c.M);\\
            $c.Surprise.Tuple[M]$ = $\delta_M(c.M.\lambda, c.M.\lambda^e)$;
            }
    }
    $c.Surprise.Score$ = $f_{cell}^{agg}(c.Surprise)$;\\
    $C.S$ = $C.S$ $\bigcup$ $c.Surprise$; 
}
$C.surprise$ = $f^{agg}(C.S)$;

\Return{$C.surprise$ };
}
\caption{\sf{The general setup of label-based surprise assessment}}
\label{algo:genericLabelSurprise}
\end{algorithm}

We can come with several algorithms for the assessment of strict and loose surprise, respectively.\\

A potential setup is shown in the following example, that comes with:
\begin{itemize}
    \item A Boolean function $\delta_M$ for each measure, returning $true$ if the expected label is different from the actual, and $false$ otherwise;
    \item A simple function $countTrue()$ for the role of $f_{cell}^{agg}$, counting the number of $true$ values in the tuple of comparison results (assuming we simulate true/false with 1 and 0, a simple sum will suffice)
    \item A normalization function computing the average surprise by dividing the total sum of cell surprises by the amount of cell $\times$ number of measures. (Observe that we intentionally do not normalize each cell's surprise by the number of measures, such that we actually compute the average cell-measure surprise here; however, due to the fact that $c.Surprise$ is a composite object, this is not prohibited)
\end{itemize}

Assuming the \textit{max} aggregate function for $f_{cell}^{agg}$ computing the surprise of a cell, and the \textit{avg} aggregate function for $f^{agg}$ computing the surprise of a cube query, Algorithm~\ref{algo:genericLabelSurprise} computes the \textit{Partial Max-Average Label-Based Surprise} for a cube query.
Algorithm~\ref{algo:labelStrictStrictSurprise} computing a 
\textit{Full Strict-Strict Label-Based Surprise}, provides a double strict version that simplifies the generic algorithm by assuming strict, Boolean semantics for both the cells and the entire cube query. 

\begin{algorithm}[ht]
\DontPrintSemicolon 
\KwIn{A cube $C$ with a set of measures $\mathbf{M}$, including a set of cells $\{c_1, \dots, c_k\}$, 
a labeling function $\Lambda_M$ for each measure $M$, 
an assignment of expected labels $E$ = $\{e_1, \ldots, e_n\}$, with each $e_i$ being a tuple of expected labels $e_i = < \lambda^e_1, \ldots, \lambda^e_m >$
}
\KwOut{The surprise carried by the cube $C$}
\Begin{

\ForAll{$c$ $\in$ $C$}{
    
    \ForAll{$M~\in~\mathbf{M}$}{
        \If{$\exists$ an expected label $c.m.\lambda^e$ for $M$}{
            obtain $c.M.\lambda$ = $\Lambda_M(c.M)$;\\
           \uIf{$c.M.\lambda$ $\neq$ $c.M.\lambda^e$}{\Return true}
            }
    }
}
\Return{ false };
}
\caption{\sf{Label-based surprise assessment with strict semantics for both cells and cubes}}
\label{algo:labelStrictStrictSurprise}
\end{algorithm}

\silence{
\begin{table}[h]
\centering
\begin{tabular}{|p{15cm}|}
\hline
\textbf{Algorithm 0}: computeLabelSurprise \\
\hline
\textbf{Input}: A cube \textit{C}, with a set of actual 
measure-labels \textbf{M}, the set of expected measure-labels for 
\textbf{M}, say \textbf{L}.\textbf{Output:} surprise of C to V, 
surprise(C | V)\textbf{Begin} For each cell c $\in$ C, surprise(c) = 0 
For each measure M$_{i}$ in \textbf{M} Let c.a$_{i}$ be the 
actual label for measure M$_{i}$, c.l$_{i}$ the expected one 
if(c.a$_{i}$ != c.l$_{i}$) surprise(c) += 1 cubeSurprise += 
surprise(c) Return cubeSurprise / |C| /* i.e., average cell surprise */
\textbf{End} \\
\hline
\end{tabular}
\end{table}

} 

\subsection{Expected Labels and Probabilities}\label{sec:labProb}
\sideNew{Sec.~\ref{sec:labProb}: NEW}
\subsubsection{Surprise computed on the basis of probabilities for labels}
If instead of directly using the value of the labeling scheme, we use 
probabilities to express that some values are expected, the 
statements take the form:

\begin{center}
$p(label(c.M)= {\lambda_{i}} ~|~ c) = p$, $p$ $\in$ 
$[0..1]$, $\lambda \in \Lambda$, $\Lambda$: a finite, nominal set of labels
\end{center}

For example: 

\begin{center}
$p(label(sales) = OK ~|~ city=Athens, year=2020)$ = $20\%$
\end{center}

The obvious benefit from the above scheme is that (a) there is a 
significantly more concise set of statements, and (b) any labeling 
scheme can be orthogonally applied to the measures with any degree of 
flexibility and precision fine-tuned by the user.

In this case, the computation of surprise can take any of 
many forms. Again, we can have a strict and a loose form of surprise as 
follows.

\textit{A. Strict surprise}. Assume a set of expressions of the above form exists for a certain cell $c$. Assume also that the actual label $label(c.M)$ is $\lambda$. Then, the strict surprise of the cell $c$ is the sum of the probabilities of all the other labels that are different than $\lambda$.

\begin{center}
$c.StrictSurprise.Score = \sum_{\substack{\lambda' \neq \lambda}} p(label(c.M) = \lambda' ~|~ c)$
\end{center}

Assuming the \textit{avg} aggregate function for $f^{agg}$ computing the surprise of a cube query, Algorithm~\ref{algo:genericLabelSurprise} computes the \textit{Partial Strict-Average Probabilistic Label-Based Surprise} for a cube query.

\textit{B. Loose surprise}. Assume now that we have an interval type of labels, and 
we can express $distance(\lambda, \lambda^{e})$. Then, 
we can use a weighting scheme and assign a weight $w$ for any 
pair of values $(\lambda, \lambda^{e})$, $w$ = $weight(\lambda, \lambda^{e})$. This weighting function can be either the absolute $|\lambda - \lambda^{e}|$
distance of the two values, or any monotone function of $distance(\lambda, \lambda^{e})$, e.g., by normalizing this measure over the max possible 
distance of a cube's cells.

The quantity $w_i \cdot p_i$  expresses two facts: the higher the expected probability of an event, the higher the product $w_i \cdot p_i$ is, and, the 
higher the distance of the actual from the expected label is, the higher 
the product $w_i \cdot p_i$ is. The combination, enforces surprise from both aspects that could generate surprise: missing the actual value, and, missing it with a high probability. 

Then, the surprise is the sum of all probabilities $\lambda_i$ that are different from the actual value, $\lambda$, but this time, these probabilities are weighted by their significance weight $w_i$. 

\begin{center}
$c.LooseSurprise.Score = \sum_{\substack{i}} w_i \cdot p_i$, s.t. $p_i = p(label(c.M) = \lambda_i ~|~ c), \lambda_i \neq \lambda$
\end{center}

Assuming the \textit{avg} aggregate function for $f^{agg}$ computing the surprise of a cube query, Algorithm~\ref{algo:genericLabelSurprise} computes the \textit{Partial Weighted-Average Probabilistic Label-Based Surprise} for a cube query.

Summarizing, taking the possibility to apply labels to the measures of
cells into consideration, we can compute surprise via labels as (a) strict surprise on the basis of the distance from the expected label, (b) loose surprise, also on the basis of the distance from the expected label, (c) strict probability-based surprise, and (d) loose probability-based surprise.

\subsection{Reference Example Revisited}
\textit{Value Based Surprise}. Suppose that we have a set of expected values for the measures of loan amounts regarding the city of Olomouc, as shown in Table~\ref{tab:expVal}. For the computation of the Value Based Surprise of the new query $q$, the absolute distance of measure value of each cell of the results of $q$ that is also found in the expected values is calculated and from all the absolute distances, an average value distance occurs. Finally, in order for the algorithm to always return a result in the scale of 0.0 - 1.0, the average value distance is normalized. 

\begin{table*}[tbh]
\begin{tabular}{llr}
District Name  & Month     & Measure                       \\
\hline
Olomouc	      & 1998-01	             & 22512               \\
Olomouc	      & 1996-09	             & 20048		       \\
Olomouc	      & 1998-09	             & 46666               \\		
Olomouc	      & 1997-05	             & 53212               \\		
Olomouc	      & 1995-07	             & 60005		       \\
Olomouc	      & 1997-10	             & 78696		       \\
Olomouc	      & 1996-12	             & 155616		       \\
Olomouc	      & 1996-05	             & 161496	           \\	
Olomouc	      & 1996-07	             & 187104		       \\
Olomouc	      & 1994-05	             & 193968		       \\
Olomouc	      & 1995-12	             & 263355	 	       \\
Olomouc	      & 1995-09	             & 309552	           \\	
Olomouc 	  & 1997-12	             & 465506	 
\end{tabular}
\caption{Expected values for the measures of loan amounts regarding the city of Olomouc, with the respected month that the loan was granted.}
\label{tab:expVal}
\end{table*}

In Table~\ref{tab:valSurprise}, the cells of the results of $q$ that are also found in the expected values are presented. The table also presents the absolute distance of each cell measure value to the respected one in the expected values. Finally, the table also provides the total and average absolute distance, i.e., average Value Surprise, along with the normalized Value Surprise that is returned from the algorithm.

\begin{table*}[tbh]
\begin{tabular}{llrr}
District Name  & Month               & Measure            & Absolute Distance               \\
\hline
Olomouc	      & 1996-09	             & 29448 		      & 9400                 \\	
Olomouc	      & 1996-12	             & 155616		      & 0                    \\
Olomouc	      & 1996-05	             & 161496	          & 0                     \\	
Olomouc	      & 1996-07	             & 187104		      & 0                      \\
\hline
Sum of Absolute Distances &          &                    & 9400                   \\
Average Absolute Distance &          &                    & 2350                   \\
\hline
Value Based Surprise &                    &                     & 2350 - 0 / 9400 - 0 = 0.25
\end{tabular}
\caption{Results of the new query $q$ that are also found in the expected values as shown in Table 3.}
\label{tab:valSurprise}
\end{table*}

The Valued Based Surprise result as shown in the last row of Table~\ref{tab:valSurprise}, occurs as the normalized Average Absolute Distance of the values of cells of the results that are also found in the expected values. The normalized distance is calculated by deducting the minimum absolute distance of a cell (here, 0) from the average absolute distance (2350) and by dividing it to the maximum absolute distance of a cell (9400) minus the minimum absolute distance (0). The result of this calculation in our example is 0.25, as Table~\ref{tab:valSurprise} shows.

%% file: 09_kaloudis-exps.tex
\section{Experimental Evaluation}\label{sec:exps}
In this Section, we present the experimental result for the assessment 
of several algorithms for assessing different dimensions of 
interestingness. We measure efficiency in terms of time 
performance for the execution of the interestingness assessment 
algorithms, under different conditions of scale.

\subsection{Experimentation methodology}
\begin{figure*}[th]
\centering
\includegraphics[width=\textwidth]{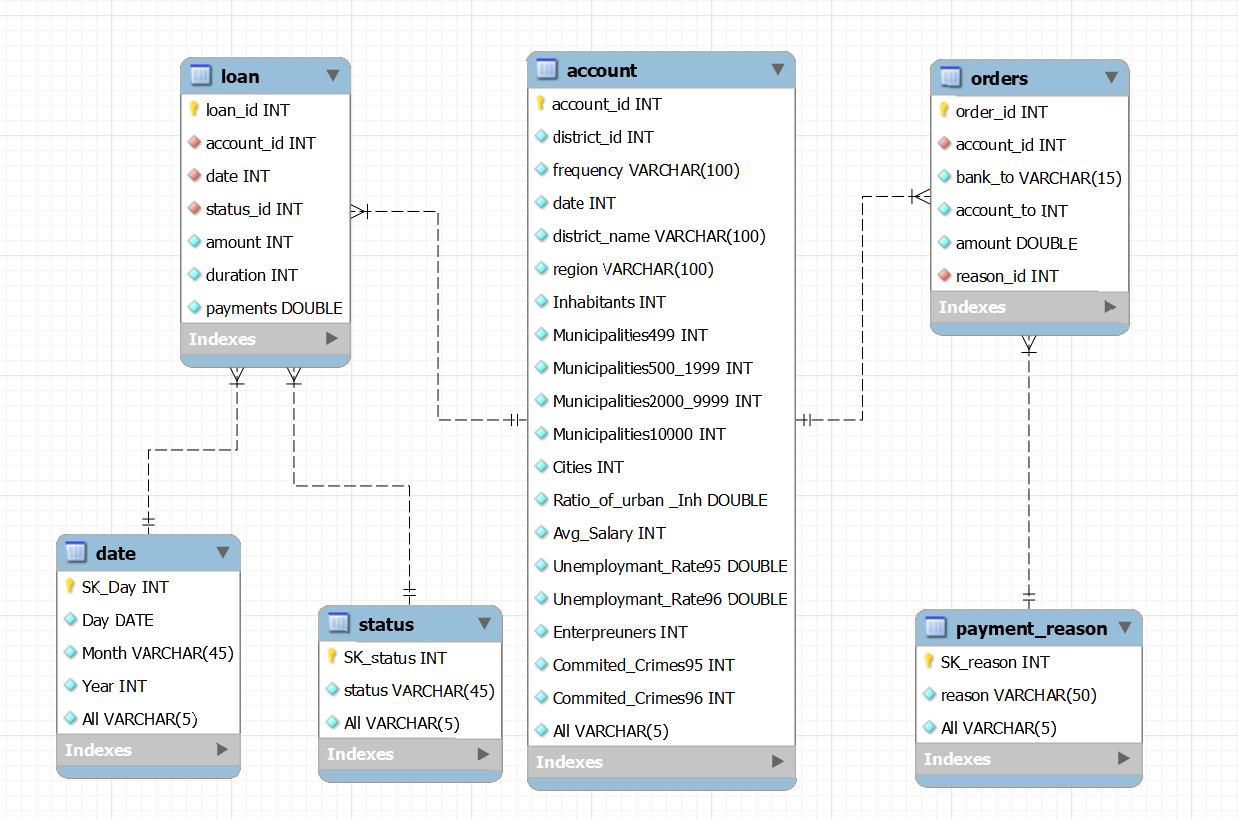}
\caption{Schema of the pkdd99\_star database}\label{fig:pkdd}
\end{figure*}
 
The experiments were performed on the Loan cube of the pkdd99\_star 
database, for which, we artificially generated data of different sizes. 
The contents of the cube were generated with a dedicated random 
generator. The experiments test the scalability of the algorithms along 
two tunable parameters: (a) BASE SIZE, reflecting the number of these in 
the fact table, specifically: 100,000, 1 million, or 10 million records, 
and, (b) HISTORY SIZE, the number of the user's previous queries, i.e., 
the size of the session history, specifically, 1, 5 or 10 past queries. 

The server on which the experiments were performed came with an AMD 
Ryzen 9 5900HS 3.3GHz CPU processor, 16GB of RAM and a 1TB SSD NVMe M2 
hard drive. For all experiments, 8GB of RAM was allocated to the MYSQL 
server, via Workbench 8.0 CE. 
\textit{The experimental goal is to assess the efficiency of the algorithms, via their execution time, by tuning the scale of two parameters of the problem, fact table and history size}.

\subsection{Novelty}
\textbf{Partial Detailed Extensional Novelty}. In this experiment, we study the effect of the 
fact table size and the query history to the execution time of the 
Algorithm for the \textit{Partial Detailed Extensional Novelty}. We have limited 
ourselves to table sizes of 100K, 1M and 10M tuples and query history of 
1, 5 and 10 queries. 

\begin{figure*}[tbh]
\centering
\includegraphics[width=0.8\textwidth]{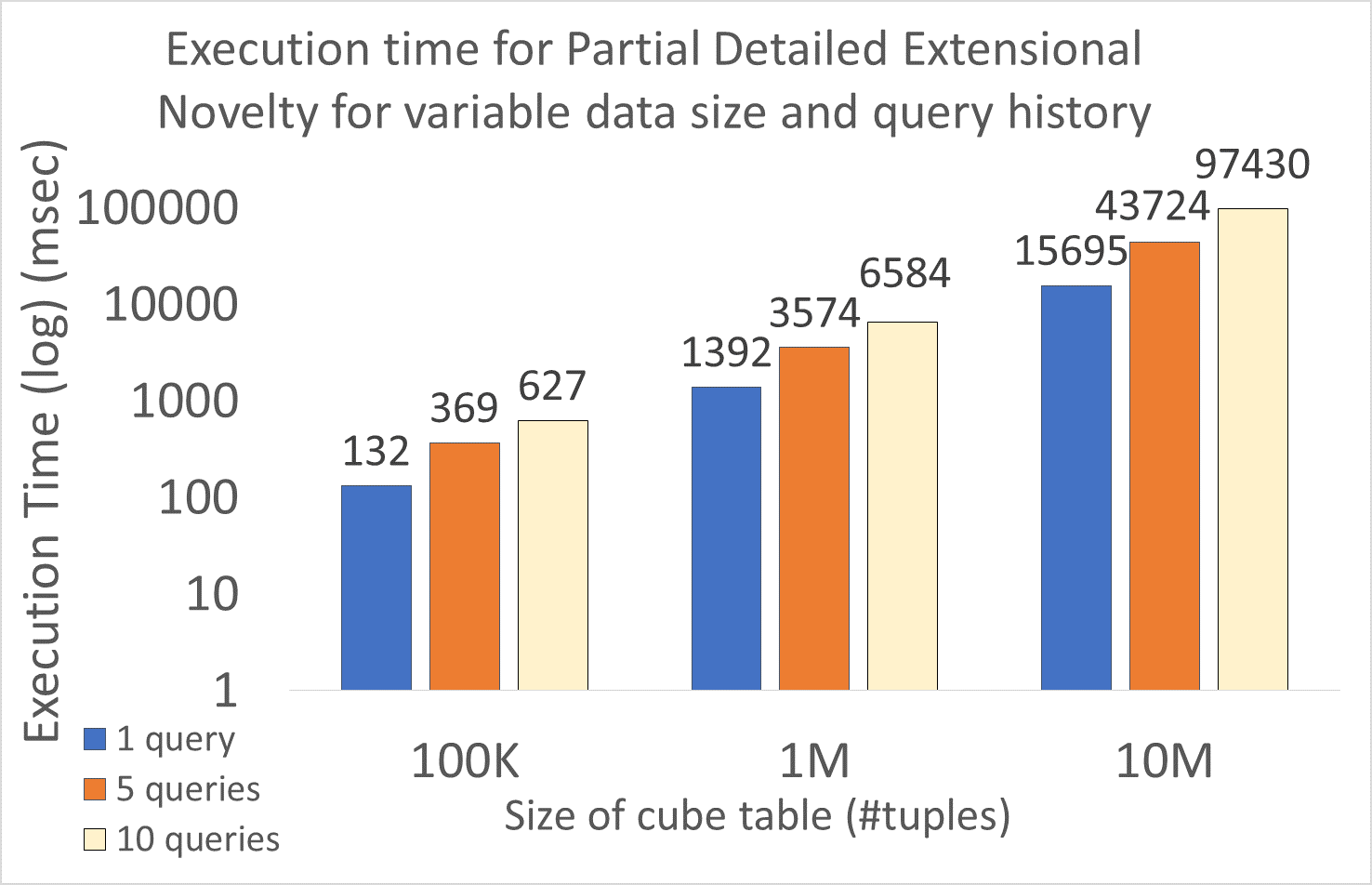}
\caption{Execution time for \textit{Partial Detailed Extensional Novelty} with respect to data size and query history.}\label{fig:exp-2}
	\end{figure*}

\begin{figure*}[tbh]
	\centering
	\includegraphics[width=0.8\textwidth]{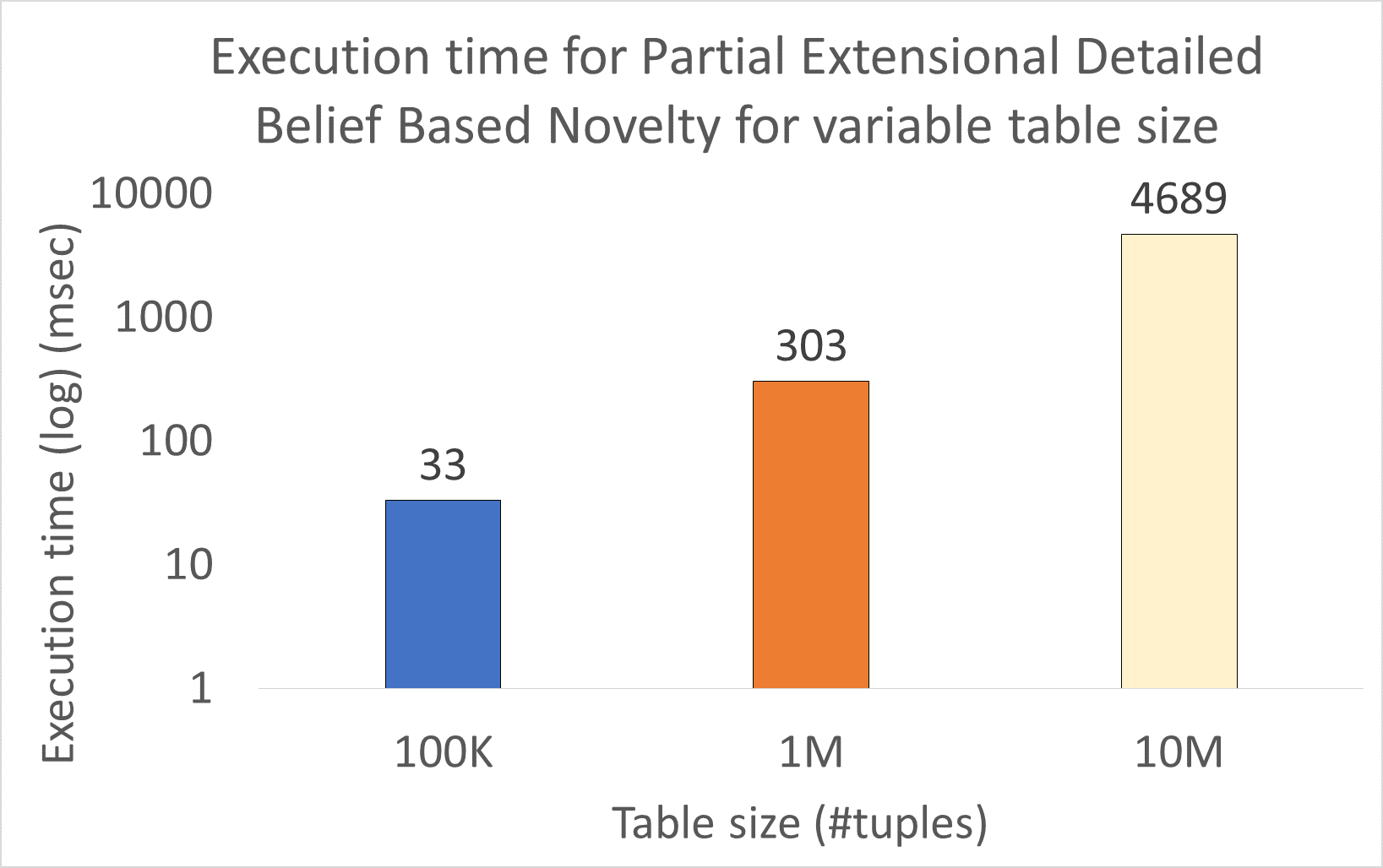}
	\caption{Execution time for \textit{Belief-Based 
			Novelty} with respect to data size and query history.}\label{fig:exp-3}
\end{figure*}

As Figure~\ref{fig:exp-2} shows, both the increase of the table size 
and the size of query history, increase the total execution time of the 
algorithm. The vertical axis is logarithmic. Both the increase of the 
table size and query history size cause a linear increase in the total 
execution time.

\textbf{Belief Based Novelty}. In this experiment, we study the effect of the fact table size to the execution time of the Algorithm for the \textit{Partial Extensional Detailed Belief-Based Novelty}. We have experimented with table sizes of 100K, 1M and 10M tuples. Fig.~\ref{fig:exp-3} demonstrates the results. Bear in mind that the vertical axis is logarithmic and observe that the execution time increases linearly with data size increase, a behavior that agrees with the complexity analysis of the algorithm.  

\textbf{Comparison}. When comparing the two novelty algorithms with each other, it is evident that \textit{Partial Extensional Detailed Belief Based Novelty} is a faster algorithm that \textit{Partial Detailed Extensional Novelty}, due to the fact that the latter is based on the time-consuming procedure of calculating the detailed area of interest of all the queries participating in the query history and comparing them to the one of the given query. This requires additional queries to the database, while on the other hand, \textit{Partial Extensional Detailed Belief-Based Novelty} simply decides if a detailed cell of the result is considered novel based on a set of user’s beliefs.

\subsection{Relevance}
\begin{figure*}[tbh]
\centering
\includegraphics[width=0.8\textwidth]{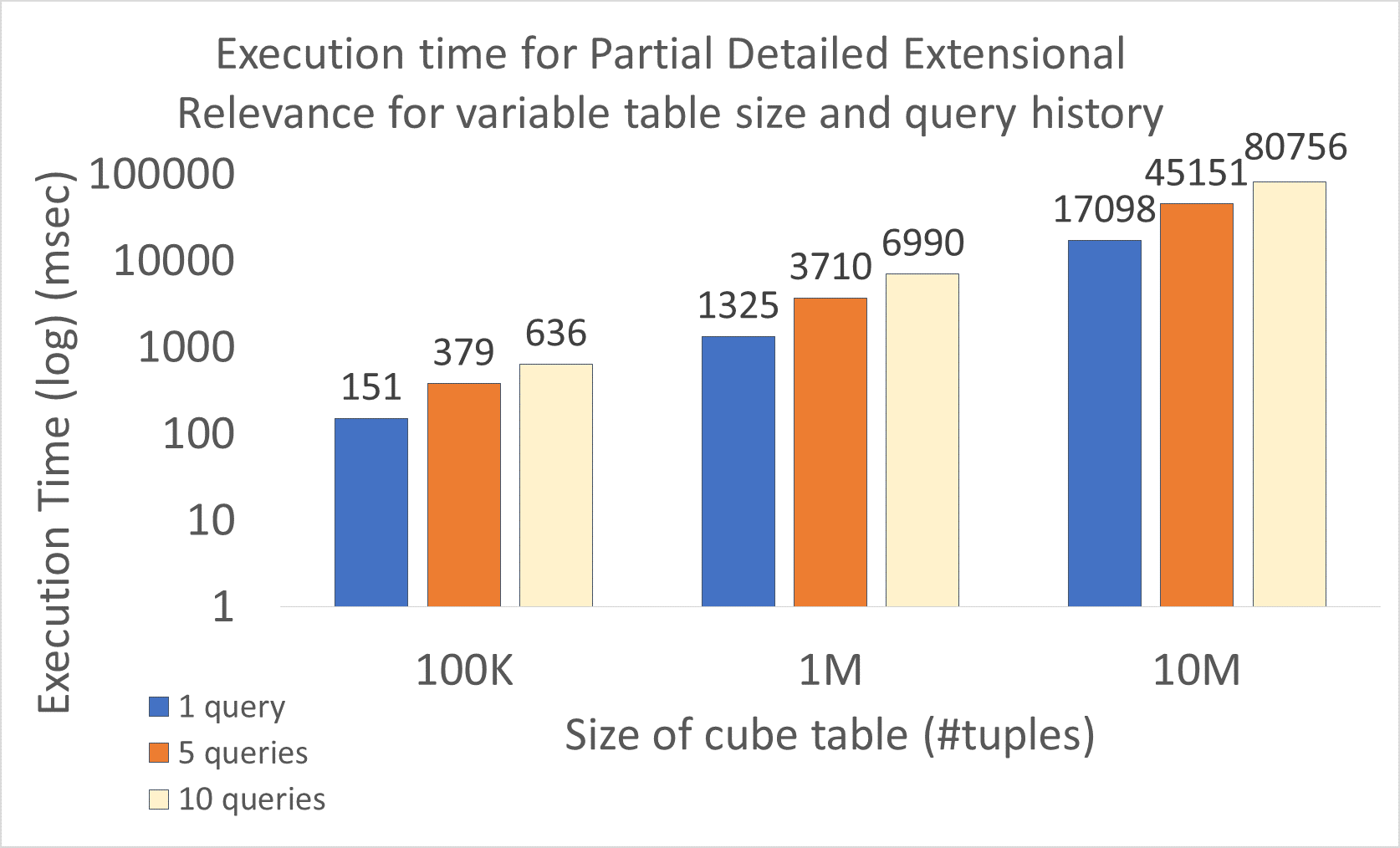}
\caption{Execution time for \textit{Partial Detailed Extensional Cube Relevance} with respect to data size and query history.}\label{fig:exp-4}
\end{figure*}

\textbf{Partial Detailed Extensional Cube Relevance}. In this experiment, we study the effect of the 
increase of the fact table size for a query history of 1, 5 and 10 
queries to the Algorithm for the \textit{Partial Detailed Extensional Cube Relevance}, which is practically assessing relevance with respect to a Detailed Area 
of Interest. We have experimented with 100K, 1M and 10M table sizes 
and query history of 1, 5 and 10 queries. The results are demonstrated 
in the Fig.~\ref{fig:exp-4}. The vertical axis of the figure is logarithmic.

\begin{figure*}[tbh]
	\centering
	\includegraphics[width=0.8\textwidth]{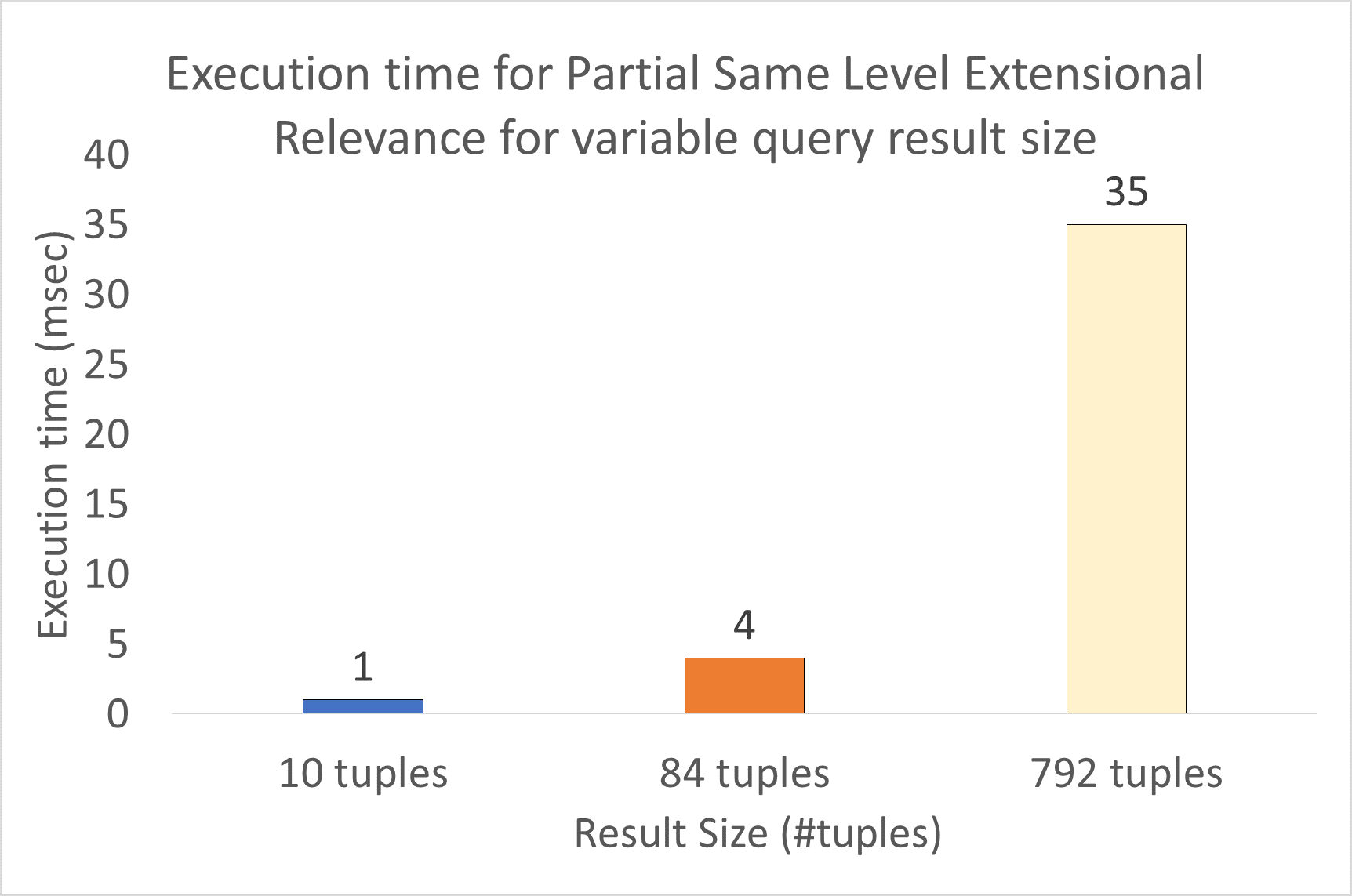}
	\caption{Execution time for \textit{Partial Same Level Extensional Cube Relevance} with respect to result size}\label{fig:exp-5}
\end{figure*}

As Figure~\ref{fig:exp-4} shows, increasing either the table size or 
the query history size results in an increase of the execution time of 
the algorithm. Both the experiment for the table size increase as well 
as the experiment for the query history increase, agree with the 
complexity analysis of the algorithm, which presented that the algorithm 
is depended linearly on the query history size and the table size.

\textbf{Partial Same Level Extensional Cube Relevance}. In this experiment, we study the 
behavior of this goal-based algorithm's execution time when we increase the result 
size of a query, in terms of number of tuples. Specifically, we limit 
ourselves to result sizes of 10, 84 and 792 tuples respectively. Fig. 
~\ref{fig:exp-5} presents the results. Observe that even though that the 
algorithm is relatively fast, the increase of the result size increases 
linearly the execution time of the algorithm.

\textbf{Comparison}. When comparing the results of the two relevance algorithms 
with each other, we find out that \textit{Partial Same Level Extensional Cube Relevance} 
is a much faster algorithm than \textit{Partial Detailed Extensional Cube Relevance}, 
due to the fact that the latter one is calculating the detailed area 
of interest for all the history queries, which includes the execution of 
a set of new queries, while the first algorithms simply calculates the 
coverage of the detailed cells based on the user's goal.

\subsection{Peculiarity}
\textbf{Partial Syntactic Average Cube Peculiarity}. In this experiment, we study the behavior of this 
algorithm's execution time when we increase the number of queries used 
as a query history.

\begin{figure*}[htbp]
\centering
\includegraphics[width=.8\textwidth]{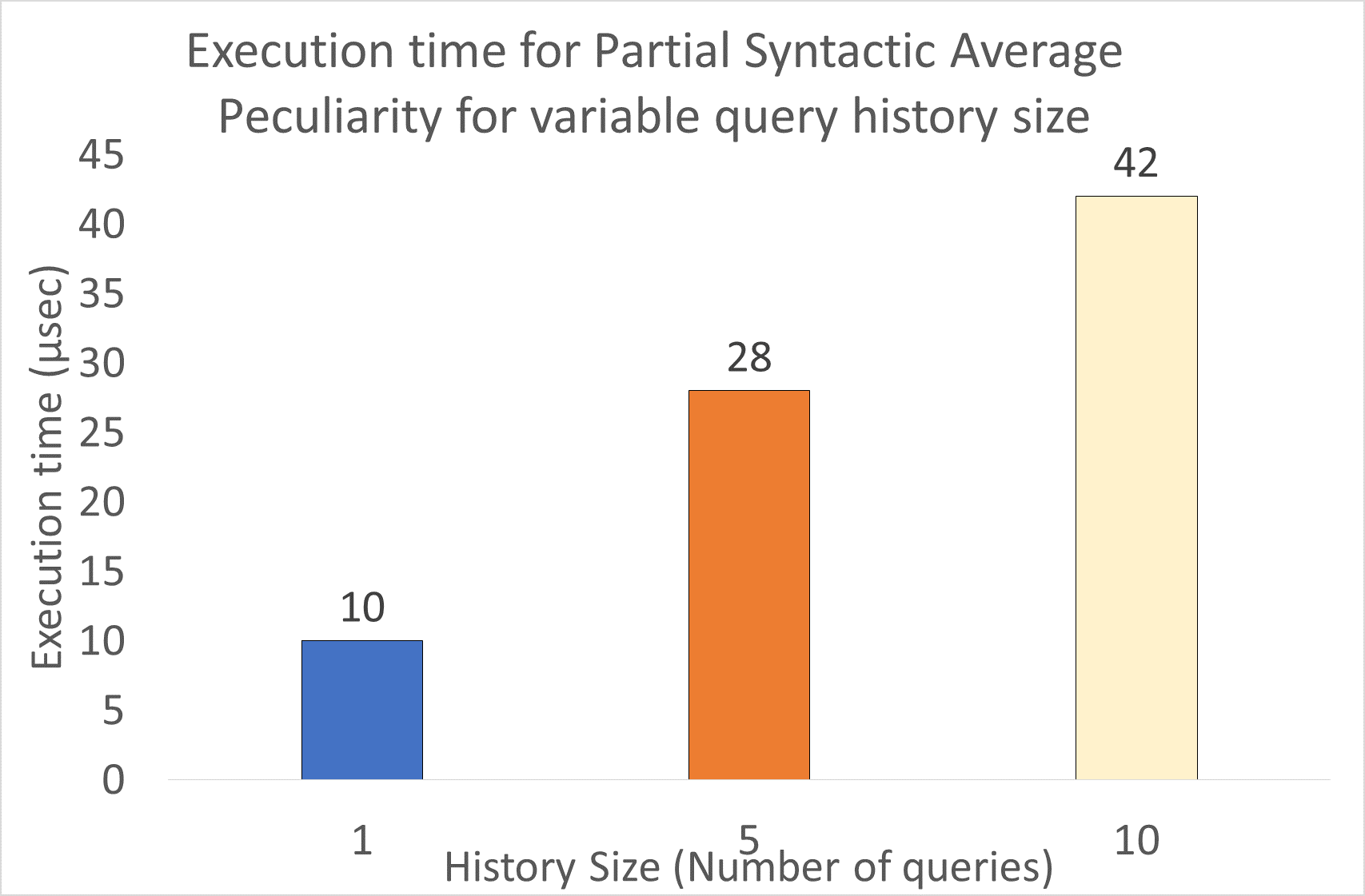}
\caption{Execution time for \textit{Partial Syntactic Average Cube (or, simply: Syntactic) Peculiarity} with respect to query history size}\label{fig:exp-6}
\end{figure*}

\begin{figure*}[htbp]
	\centering
	\includegraphics[width=.8\textwidth]{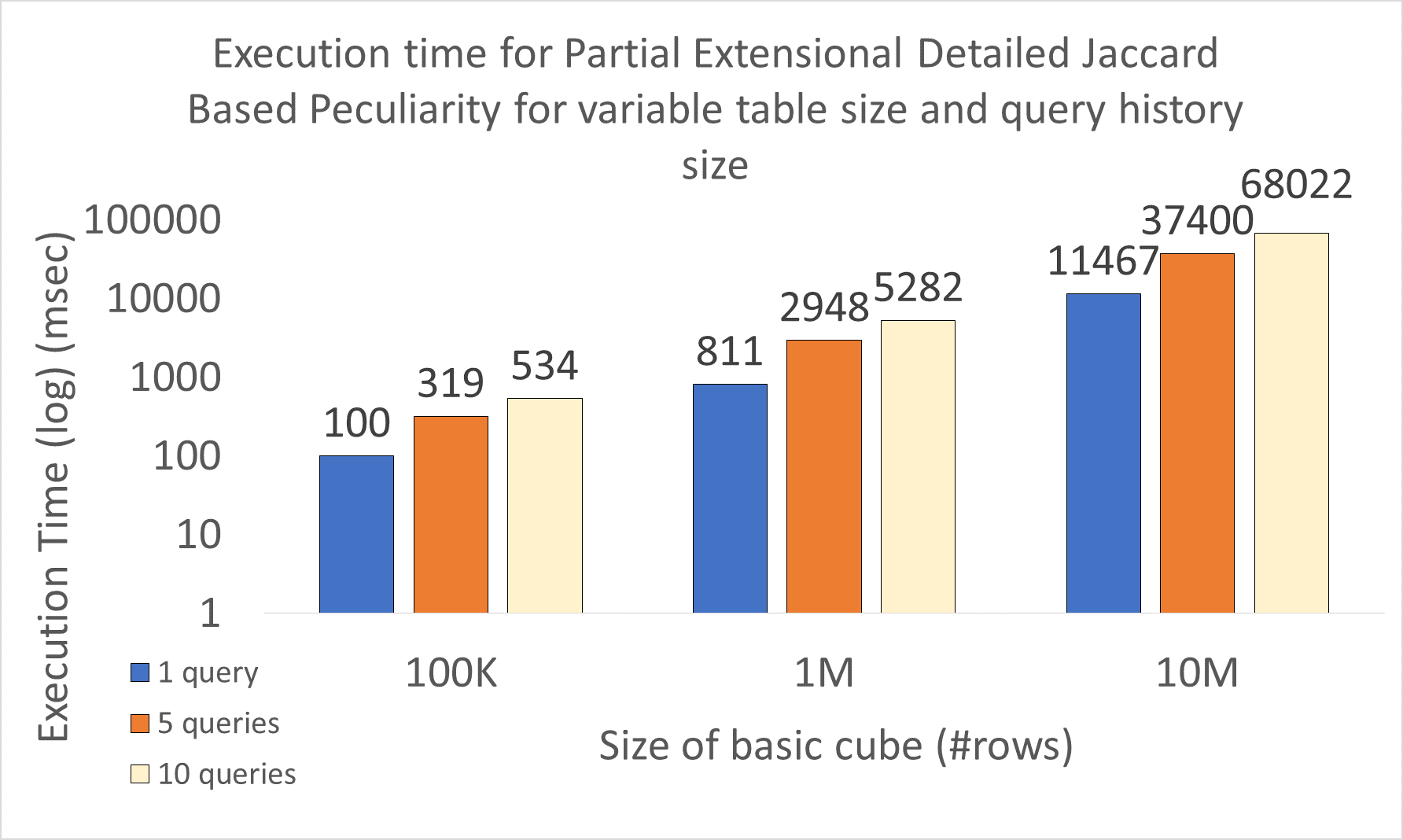}
	\caption{Execution time for \textit{Partial Extensional Detailed Jaccard-Based Peculiarity (or, simply: Value) Peculiarity
		} with respect to table size and query history size}\label{fig:exp-7}
\end{figure*}

The results, as presented in Fig.~\ref{fig:exp-6}, show that the increase of the 
query history size increases linearly increases the total execution time of 
the algorithm, as presented in the complexity analysis of the algorithm 
too.

\textbf{Partial Extensional Detailed Jaccard-Based Peculiarity}. In this experiment, we study the effect of the 
increase of (a) the fact table size, and (b) and the query history to the Partial Extensional Detailed Jaccard-Based Peculiarity algorithm (practically assessing peculiarity on the basis of a Jaccard similarity between the detailed areas of the query and the history of queries).
The assessment is performed for a query history of 1, 5 and 10 queries and fact table sizes of 100K, 1M and 10M tuples. 
Fig.~\ref{fig:exp-7}, with its vertical axis in logarithmic scale, shows the results of the 
experiments. Both table size and query history size increase the 
execution time of the algorithm, but the first one in a much larger 
scale. Even so, both the increases affect linearly the execution time of 
the algorithm

\textbf{Comparison}. When comparing the results of the two Peculiarity 
algorithms, we find out that the \textit{Partial Syntactic Average Cube Peculiarity} is a 
much faster algorithm that the \textit{Partial Extensional Detailed Jaccard-Based Peculiarity}. This is 
caused due to the fact that the first one simply does a syntactic 
analysis of the query and compares it to the already submitted ones, 
while the latter one needs to compute the detailed area of interest of 
all the queries in the history, which hides the execution of a series of 
new queries.

\subsection{Surprise}
\textbf{Partial Extensional Value-Based Surprise}. In this experiment, we study the behavior of this 
algorithm's execution time when we increase the result size of a query, 
in terms of number of tuples. The vertical axis is in logarithmic scale.

\begin{figure*}[htbp]
\centering
\includegraphics[width=.8\textwidth]{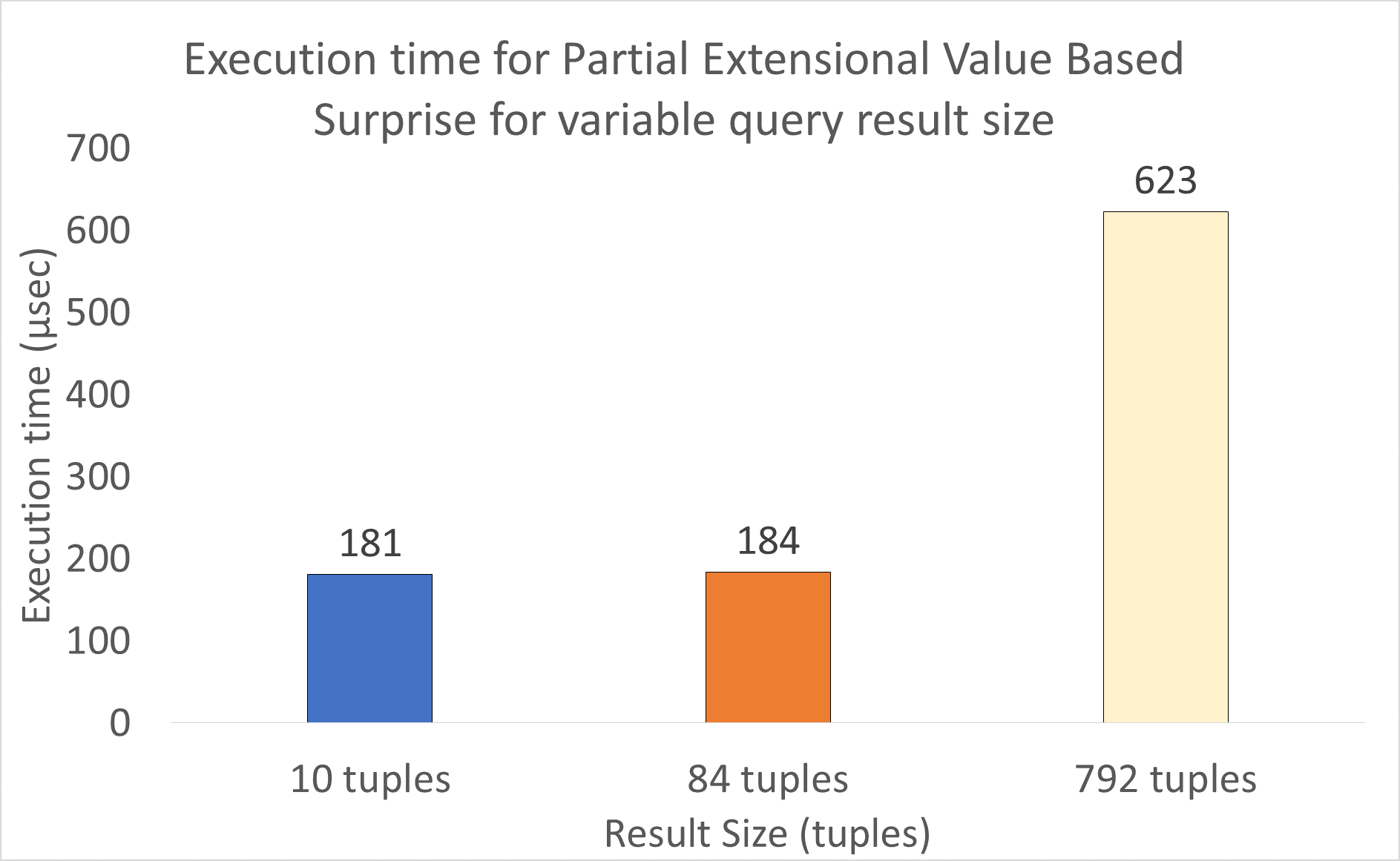}
\caption{The execution time for the \textit{Partial Extensional Value-Based Surprise} with respect to query result size}\label{fig:exp-8}
\end{figure*}

The results, as presented in Fig.~\ref{fig:exp-8}, show that the theoretical lineal increase with respect to the result size is not exactly achieved. The algorithm is quite fast, of course, due to its simple nature that works on top of a query result (remember, surprise cannot work with signatures, and requires the query result and the cells measures to be computed). We attribute the variation of the execution time to the probability of hitting an expected value when the result size of the query is larger, which results in extra CPU time for computing the surprise. 

%% file: 10_userStudy.tex
\section{A User Study on the Interestingness Dimensions}\label{sec:userStudy}
\sideNew{Sec.~\ref{sec:userStudy}: NEW!}
In this Section, we discuss a user study that we conducted in order to 
evaluate how do the introduced interestingness dimensions relate to the 
behavior of people working with cubes and cube queries. All the material of the study, along with our findings, are available via the public repository \url{https://github.com/OLAP3/2023InterestingnessUserStudy}.

\subsection{Goal and Research Questions}
The goal of the study has been to identify whether there are significant 
influences by particular interestingness dimensions, as well as patterns 
of behavior related to these dimensions, when users interact with cubes 
and cube queries.

To solidify this goal, our user study was based on the following 
research questions:

\textit{RQ1.\ \ \ \ Can we rank the Interestingness dimensions in terms of 
significance to the overall interestingness of a cube query? Is there 
any interestingness dimension that dominates the determination of the 
overall interestingness of a cube query?}

\textit{RQ2.\ \ \ \ As a session progresses, does the significance of the 
interestingness dimensions change overtime?}

\textit{RQ3.\ \ \ \ Do participants demonstrate a consistent behavior with 
respect to the ranking of their interestingness dimensions?}

\textit{RQ4.\ \ \ \ Are there patterns of behavior concerning interestingness 
dimensions? Can we form clusters of users based on their preferences?}

To answer these questions, we constructed and executed the experimental 
protocol that is detailed in the sequel.

\subsection{Experimental Protocol}
The user study we conducted was based on asking participants to assess 
how interesting a query result appeared to them, without giving them any 
details on how the query ranked in terms of the four interestingness 
dimensions, namely Relevance, Novelty, Peculiarity and Surprise.

\textbf{Material}. We created a set of cube querying sessions. For 
all the cube querying sessions, we have used the Adult dataset which is 
a census dataset that has 8 dimensions (\textit{Age, Native Country, 
Education, Occupation, Marital Status, Work Class, Gender and Race}) 
and a single measure, \textit{Work Hours Per Week}. 

Each session was constructed as a PowerPoint presentation that was given 
to the participants. The presentation started with a set of slides 
giving a description of the dataset structure and semantics. Then, the 
participants were given the goal of finding out which are the categories 
of working people with the significantly higher and lower average 
working hours per week, depending on a set of data dimensions of the 
data set, like education, occupation, work class, age, and in the 
context of this task, we were giving them pre-computed queries along 
with their results to help them determine the answer to the task. 

The following parts of the presentation given to the participants 
included a warm-up slide and 3 slides of 4 queries. The single warm-up 
slide contained query results that give a broad description of how work\_hours 
are related to the various dimensions that we use in that specific 
querying session. This served as a contextualization of the participants 
in the data of in the data set. To make the participants pay attention to these 
data, we also asked them to write a short memo of what their original 
impression was on who works more.

\begin{figure*}[tbh]
\centering
\includegraphics[width=\linewidth]{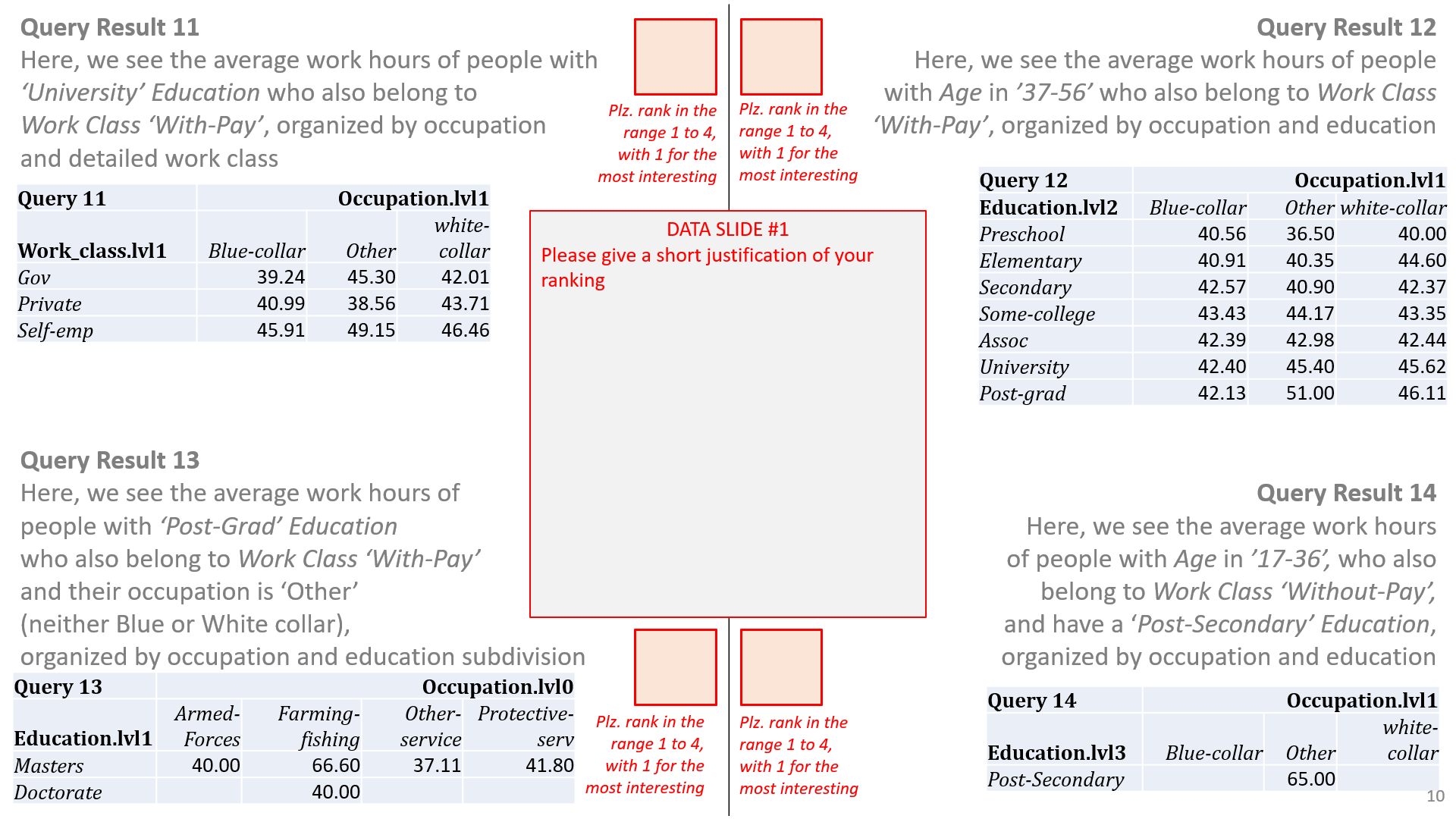}
\caption{A slide as presented to the participants}
\label{fig:slide1}
\end{figure*}

Subsequently, the report contained 3 slides and, in each of these 3 
slides, 4 queries were presented. The queries of each slide of the 
session were expressed in natural language and were presented along with 
their resulting tuples, without any additional information about 
interestingness dimensions or values (Fig~\ref{fig:slide1}).

We asked the participants to rank the queries that they faced in each 
slide on a scale of 1 to 4, with 1 being the most interesting and 4 
being the least interesting. The ranking was based on the users' 
personal criteria, with respect to the specific target that we gave 
them, i.e., to find out which are the characteristics of people that 
work the most and the least in a week time period. To avoid any bias, we 
never referenced any of the interestingness dimensions to the 
participants. Thus, they were fully ignorant of the overall goal of the 
study and the underlying assessment that we were making. To make the 
users pay more attention to the data, we also asked them to write a 
short memo per slide on their rationale.

\begin{figure*}[tbh]
	\centering
	\includegraphics[width=\linewidth]{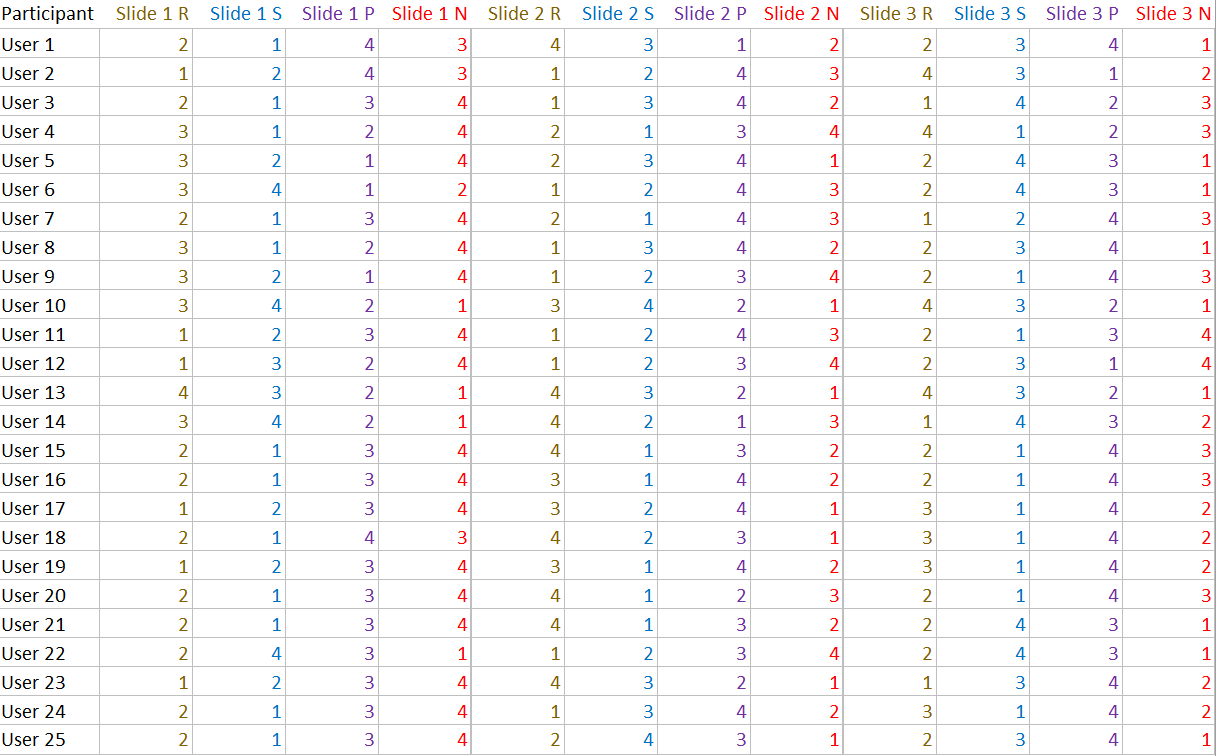}
\caption{Participants and their rankings of interestingness dimensions, per slide}
\label{fig:participants}
\end{figure*}

The trick, unknown to the participants was that each of the 4 queries maximized the value of an interestingness dimension. \textit{Thus, by ranking queries, the participants also ranked interestingness dimensions without knowing}. Practically, in each slide we had 4 queries-representatives of the interestingness dimensions. In simpler words, in each slide we presented to the user a highly Relevant, a highly Novel, a highly Peculiar and a highly Surprising query at a random order. The score of each interestingness dimension for each query was the 
result of an algorithm that we selected to run as the dimension-representative. Specifically, for each query, we selected to run \textit{Partial Detailed Extensional Relevance}, \textit{Partial Detailed Extensional Novelty}, \textit{Partial Detailed Jaccard-Based Extensional Peculiarity} and \textit{Partial Extensional Value Based Surprise}. The rankings given by all participants are listed in Figure~\ref{fig:participants}. 

The final step of the process was that once they had worked with the presentation and made their decisions and comments, the participants had to record them in a Google Form whose link was also given to them, along with the instructions and the presentation. The participants were given the fairly large time interval of an entire day to conduct the experiment. The results were collected from the Google Form's back-stage spreadsheet for further processing.

\textbf{Population}. The participants of the experiment were 25, and specifically, 7 PhD and 11 MSc students, all trained in the concepts of cubes, dimensions and business intelligence, as well as 7 undergraduate students with significantly less exposure to BI concepts. All participants were volunteers from France and Greece. 

Interestingly enough, in the subsequent study of the collected data, we did not observe any particular differentiation between the educational levels. The simplicity of the multidimensional model, as well as the textual description of queries have obviously made the data analysis work smooth. Therefore, we report all the results collectively, independently of the educational level.

\textbf{Anti-Bias and Integrity measures}. To preserve the integrity 
of the study, several measures were taken.
\begin{itemize}
    \item To randomize the experiment, we grouped the participants at random in 
one of the two sessions that we had previously constructed.
    \item To forestall any technical difficulties impeding any student whatsoever, all 
the query expressions were presented in natural language.
    \item To involve participants in the data set, we asked them to construct 
short memos per slide, which we later checked. We found no frivolous 
behavior from the part of the participants.
    \item To avoid any ordering bias, we shuffled the position of the queries in 
the slides.
    \item By asking each user to rank 12 queries overall, we addressed the issue 
of volume, too.
\end{itemize}

\subsection{Research Question: significance of individual 
interestingness dimensions}
After collecting the participants' responses, the analysis phase begun. 
The first task to address was to answer the question on the significance 
of individual interestingness dimensions to the overall interestingness 
of an individual query.

 The input to the analysis was a matrix where for every participant, for 
every slide and for every query in the slide, there was a rank between 1 to 
4. We had instructed participants to avoid ties, and indeed we had a 
clean vote from this respect. To synthesize the results, we resorted in 
a Borda scoring of the ranks. A Borda count \cite{ElHelaly19} is a 
simple process for synthesizing ranking preferences. The idea is that 
you have N candidates, and voters rank them. Then, for every rank, you 
give a score which is produced by the formula \textit{score} = \textit{N}+1 - \textit{rank}. For 
example, with 4 candidates to be voted per slide, the query with rank 1 
gets 4 points, whereas the query with rank 4, gets 1 point. Then, the 
scores are simply summed up per candidate.

Here, the candidates are the interestingness dimensions, hidden behind 
the queries that are voted. Once we added all the scores, the results 
were demonstrating a layering of preferences.

\begin{table}[tbh]
\centering
\begin{tabular}{lr}

Int. Dim & Borda score \\
\hline
Peculiarity & 151 \\
Novelty & 183 \\
Relevance & 203 \\
Surprise & 213 \\
\end{tabular}
\caption{Borda score for the different dimensions of interestingness, after composing individual rankings in our user study}\label{tab:borda}
\end{table}

\textit{The results suggest that no particular interestingness dimension drives the 
overall interest single-handedly. However, there are differences, with Surprise and Relevance being most significant, Novelty coming third at a distance, and Peculiarity being the least significant.}

Surprise came first and Relevance second, with close distance to one 
another. Surprise was the dimension that was ranked (i) first most times 
than any other dimension, and, (ii) last, less than any other dimension. 
Closely following Surprise, Relevance ranked typically first or second, 
and rarely third or fourth. So, this practically instructs us that if 
recommending queries to users a-priori, or assessing them a-posteriori, 
surprise and relevance seem stable choices.

\begin{table}[tbh]
\centering
\begin{tabular}{lrrrr}
Int. Dim & 1 & 2 & 3 & 4 \\
\hline
Peculiarity & 7 & 14 & 27 & 27 \\
Novelty & 20 & 16 & 16 & 23 \\
Relevance & 19 & 28 & 15 & 13 \\
Surprise & 29 & 17 & 17 & 12 \\
\end{tabular}
\caption{Occurrence per rank, for each of the interestingness dimensions (position 1 is the most appreciated, position 4 the less)}
\label{tab:occPerDim}
\end{table}

On the other hand, Novelty is practically equally distributed in all 
ranks (as we will see, not equally over time though). We believe that 
this is a result closely related to the setup of the study: users were 
given a specific task, as well as a contextualization warm-up, meaning 
that there was not a phase of exploring without any particular focus in 
search for interesting pieces of information. But, what we learn on the 
other hand, is that in these occasions, where a clear focus has been set 
early on, novelty is not so important as we originally expected. 
Finally, peculiarity went particularly low in terms of preferences. 
Again, we relate this to the previous discussion on novelty: digressions 
from the central task are not particularly appreciated once the focus 
has been set.

Interestingly, a statistical analysis of correlation between the 
measurements found a couple of interesting anti-correlations. We 
measured the pairwise Pearson correlation for all the four 
interestingness dimensions. Surprise is anticorrelated with Novelty, 
with a score of -0.62 and Relevance is anticorrelated with Peculiarity 
with a score of -0.50. The effect for the rest of the pairs was weaker.

\subsection{Research Question: does interest change over time?}

\begin{figure*}[tbh]
\centering
\includegraphics[width=\linewidth]{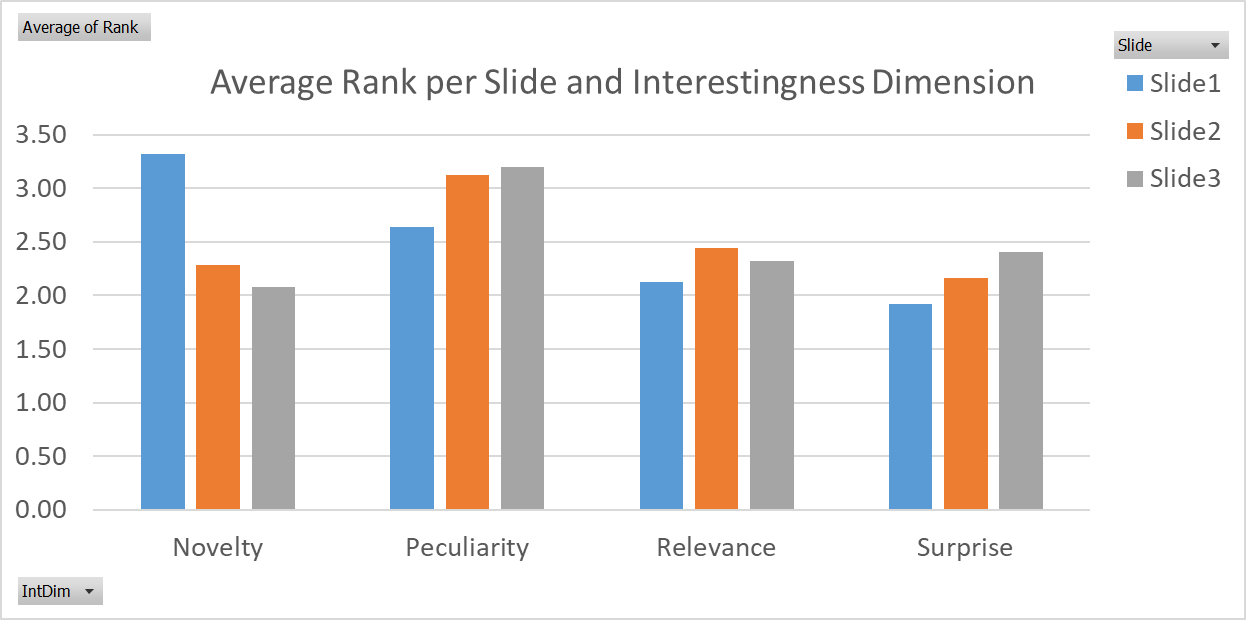}
\caption{Average Rank of Interestingness Dimensions per Slide}
\label{fig:slideInTime}
\end{figure*}

Another question we asked was if participants appreciated the 
interestingness dimensions differently as time passes. To the extent 
that we have a set of slides ordered over time, we assess the effect of 
time via the position of the respective slides. In Figure~\ref{fig:slideInTime}, we depict 
the average rank per slide, for each of the interestingness dimensions. 
Beware these are ranks, not scores: so, in the Figure, the higher the 
bar, the less appreciated a dimension is.

Unsurprisingly, \textit{Surprise and Relevance seem rather unaffected from the 
position of the slide, although as time passes, surprise becomes slightly less of importance. Peculiarity also seems to lose interest as time 
passes, especially between slides 1 and 2. What is most revealing, 
though, is the sharp decline of Novelty in rank over time}. At the 
beginning, Novelty is not that interesting, ranking top (thus, least 
appreciated) among all interestingness dimensions. From slide 2, though, 
Novelty starts being more appreciated by the participants. 
Novelty was probably considered out-of-scope at the beginning, right 
after contextualization had taken place, but later, it picked up in 
stature.

\subsection{Research Question: Do participants demonstrate a consistent behavior with 
respect to the ranking of their interestingness dimensions?}

The next research question concerned the existence of a 
constant behavior of the participants with respect to how they ranked 
the different interestingness dimensions. We will broadly use the term 
\textit{consistency} to refer to the tendency of a participant to 
place the same rank to the same interestingness dimension in different 
slides.

\textbf{Ranking Data and Comparisons}. Before proceeding with the 
definitions of the metrics used to quantify consistency, let us briefly 
summarize the available data. Remember that each participant gives 4 
rankings for each slide, in the range 1 - 4, one per interestingness 
dimension. Since there are 3 such slides, eventually each participant 
comes with a vector of 12 rankings. 

Moreover, there are 3 comparisons to be made: (i) slides 1 and 2, (ii) 
slides 1 and 3, and, (iii) slides 2 and 3. This is important as we have 
a vector of 12 comparisons for the rankings given by the participants: 4 
comparisons (one per dimension) for each of the cases (i) - (iii). We call 
this vector the \textit{comparison vector}.

\textbf{Definitions}. To address this question, we resort to two 
different metrics, point-based and score-based consistency. To be able 
to define them, we define the following metric:

\textit{Average Point-based Consistency} is the total number of 
comparisons where the participant gave the same rank to the same 
interestingness dimension in the two compared slides, normalized by the 
number of comparisons. 

To define score-based consistency, we need a couple of auxiliary 
metrics:

\textit{Score inconsistency} is the absolute difference of two 
rankings of the same interestingness dimension in a comparison - 
practically the absolute value of a cell in the comparison vector.

\textit{Normalized comparison score-based inconsistency} is the 
normalized sum of the 4 cells of the comparison vector that pertain to a 
comparison between two specific slides. We sum the inconsistencies for 
the four different measures and normalize by 8 which is the maximum 
amount of inconsistency for the 4 rankings within a slide. Thus, we have 
3 normalized inconsistency scores, one per case (i) - (iii). \textit{Normalized comparison score-based 
consistency }is defined as its complement: 1 - normalized comparison score-based inconsistency.

Then, \textit{Average score-based consistency} is the average of the 
three normalized comparison score-based consistencies for cases (i) - 
(iii).

\textbf{Intuition}. Practically speaking, the two metrics handle 
consistency from two different points of view.

Point consistency is a "\,Boolean"-based metric: if the participant gave 
the same ranking to the same dimension, it raises a true flag, otherwise 
a false one. Practically, we count how many times there was a 
coincidence of rankings. We normalize the count of coincidence occasions 
by the number of comparisons (here: 12) and we get a score within 0 and 
1 (1 meaning the participant gave always the same rankings).

Score-based consistency goes one step further, as the value of the rank 
is used. Assume you compare novelty in slide 1 with novelty in slide 3. 
If the participant gave a rank of 1 to the former and a rank of 4 to the 
latter, this is more inconsistent compared to the case where the participant gave 2 
and 3, respectively.

\begin{figure*}[tbh]
	\centering
	\includegraphics[width=\linewidth]{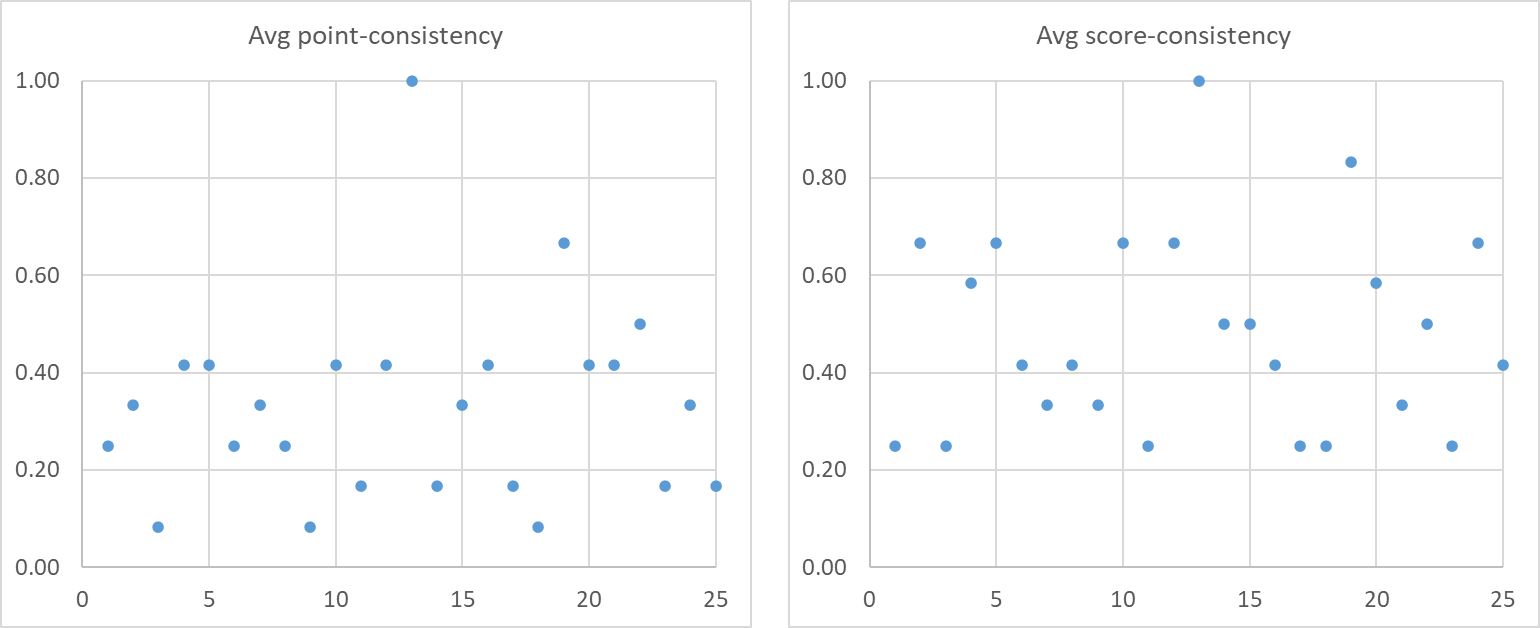}
\caption{Consistency scatterplots: The horizontal axis refers to the participant id (thus, 
	each point is a different participant), and the vertical axis to the 
	average point-based and score-based consistency of the participant}
\label{fig:consistScatt}
\end{figure*}

\begin{figure*}[tbh]
\centering
\includegraphics[width=\linewidth]{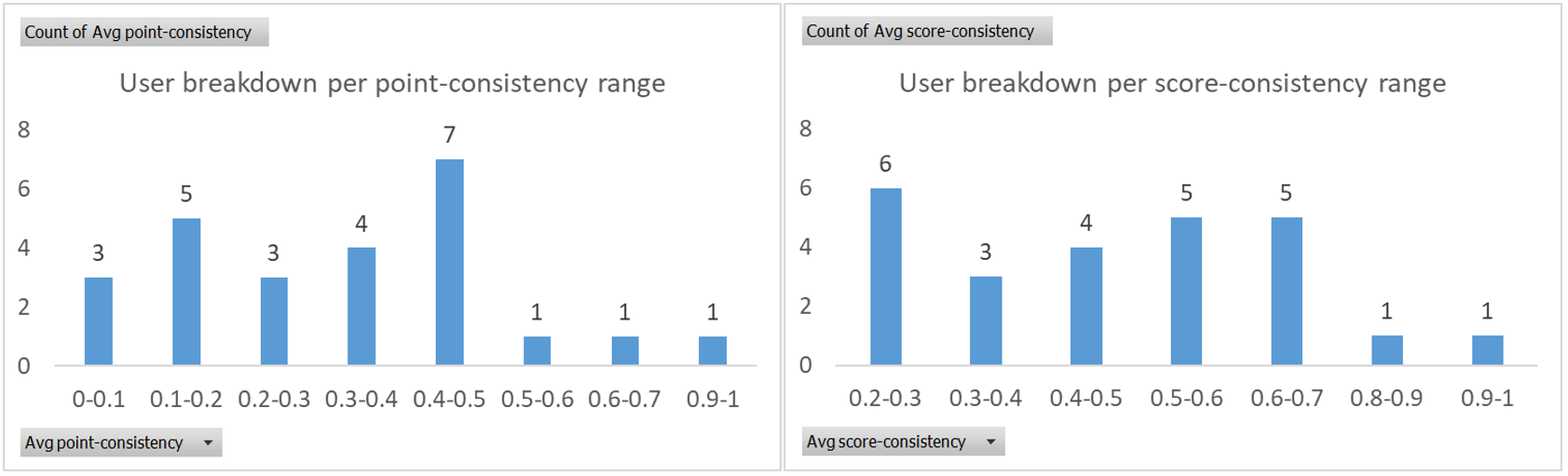}
\caption{Breakdown of participants in 0.1 ranges of consistency}
\label{fig:consistHist}
\end{figure*}

\textbf{Evaluation}. When it comes to evaluating the consistency of 
individual users the situation is depicted in Figures~\ref{fig:consistScatt} and \ref{fig:consistHist}.

Both scatterplots demonstrate a similar behavior of points randomly 
spread in a band of values. The two plots provide a different evaluation 
of the situation however. When we assess consistency in a strict, 
Boolean way, the participants find themselves spread in a band between 
0.1 and 0.4 (with the exception of a single user with a consistency of 
exactly 1). This is an indicator that more often than not, the rankings 
of the same interestingness dimension are different.

At the same time, the score-based consistency tells us that they are not 
entirely different after all: the band of points lies between 0.2 and 0.7, 
with 13 participants below 0.5 and 12 participants above 0.5. In other 
words, although they may not coincide exactly, the rankings used are 
not that far.

\textit{In summary: the participants did not exhibit a strong bias towards a particular ranking of the interestingness dimensions, although the rankings are not completely arbitrary.}

\subsection{Research Question: Are there any clusters of participant behavior?}
Another test we applied was to attempt and cluster participants on the basis of their behavior. We employed two methods of clustering: (a) k-Means and (b) Louvain clustering on the grounds of two versions of the measurements: (i) the original 12 rankings given by each user, and, (ii) the average value of each interestingness dimension per user (thus, with a vector of 4 values per user instead of 12, in an attempt to reduce dimensionality).

\textit{The results are quite indicative on the absence of clusters}. All clustering methods returned low Silhouette coefficients (0.252 for the k-means clustering of the original and 0.323 for the k-means clustering of the averaged data), and their Silhouette plots indicate that clusters are not very cohesive.

\begin{figure*}[tbh]
\centering
\includegraphics[width=0.75\linewidth]{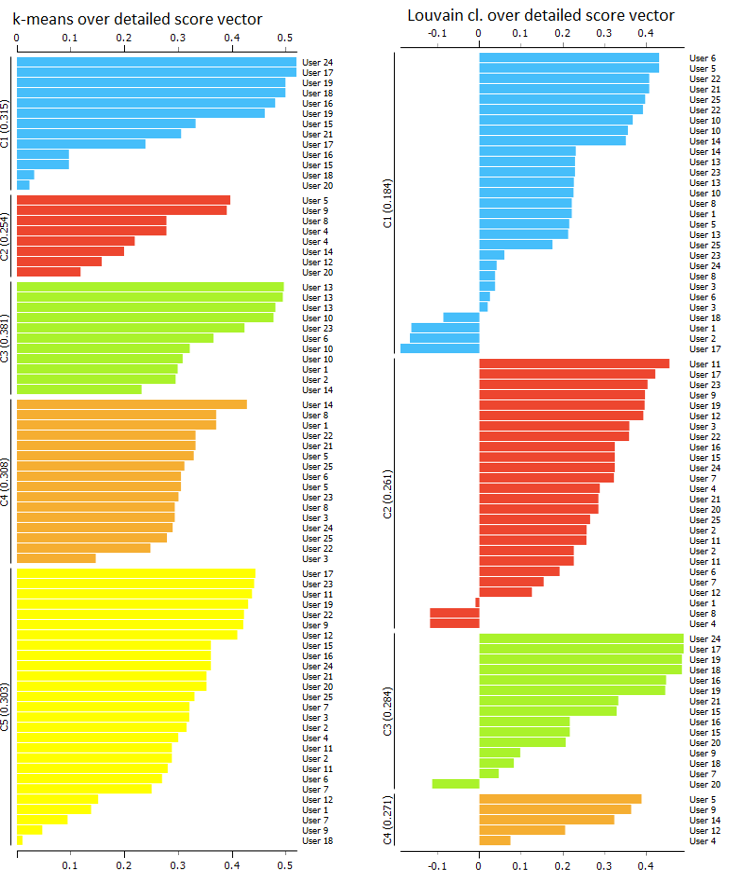}
\includegraphics[width=0.75\linewidth]{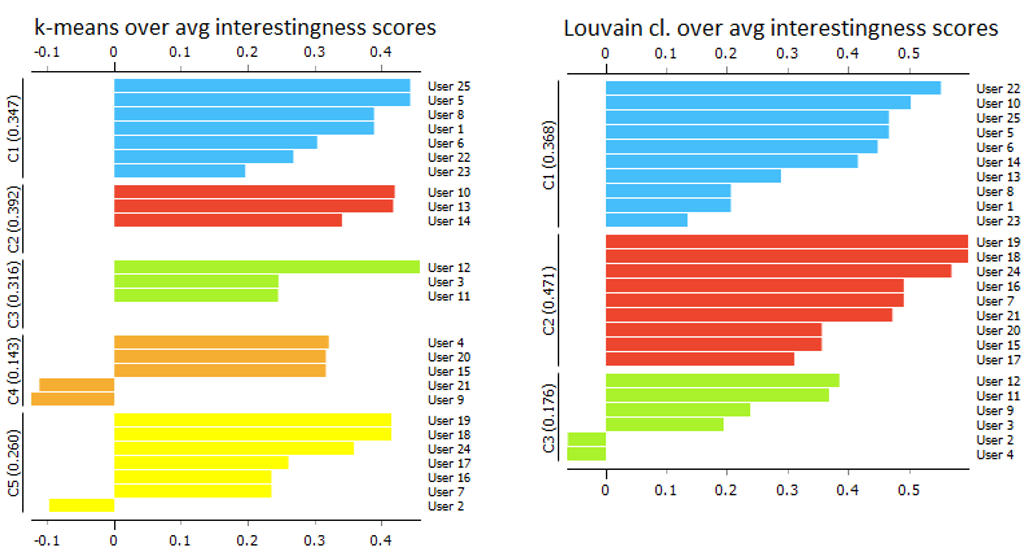}
\caption{Silhouette plots for all our clustering attempts}
\label{fig:clust}
\end{figure*}

\subsection{Threats to validity}
In this section, we discuss threats to the validity of our study.\\

\textbf{Construct Validity}. We have taken care to check that the answers given by the participants were valid. We had no violations of the scoring constraints and the memos returned by the users demonstrated a valid level of engagement to the study. We made sure that the true purpose of the study was not revealed to the participants. Therefore, when they ranked queries, they had no idea about interestingness dimensions that could affect their scoring. Based on the above, we can state that the ranking of queries in terms of overall interestingness was valid. 

At the same time, a potential threat might appear from the hiding of the dimensions behind queries. With the exception of relevance, we made sure that each query that secretly represented a dimension was either on very high values of the respective metric, or with significant difference for this metric against the others. For relevance this was not possible, as we explore a fairly well "fenced" area of the multidimensional space. However, the queries with high relevance were very low in all other metrics, which means that if selected, only relevance could be the reason for selecting them.
Based on the above, we can claim that the rank of a query can be validly mapped to a rank of the respective interestingness dimension.

\textbf{Internal validity}. Internal validity refers to cause-and-effect relationships. We do not measure any interventions, so we do not search for hidden variables that can override the effect of an intervention, as typical internal validity checks should do. 

We attribute the significance assessment of the different dimensions to the scope of the study (see also external validity). Other factors that could influence the behavior of the participants towards specific dimensions might be applicable, although we cannot think of any. The same applies for the behavior of Novelty over time. The lack of extreme consistency and clusters of participants is probably a good rather than a bad sign towards the integrity of the study: the population demonstrated variability in behaviors which means there is no threat of bias of some sort. Apparently, although some difference in significance exists, overall, all dimensions play a role. Interestingly, a potential threat to the internal validity of the study involves other dimensions, that we have not thought of, that might coexist with the ones we study.


\textbf{External validity}. What is the scope of the study? How generalizable are our findings? 
The existence of 25 participants is not overwhelming, but still adequate enough to allow the drawing of conclusions. 

Is the use of students as participants a problem? We believe we satisfy the most important properties of \cite{DBLP:journals/ese/Tichy00} that supports accepting the validity of tests with students: students were trained in the context of the study, and were adequately well equipped to perform the study, and, at the same time, both our positive and negative results are novel and a starting point for further research.

Concerning the scope, we have to be clear that the study did not have an exploratory nature: the participants were given query descriptions and results around a very specific topic. Therefore,  our results are restricted in the case of focused assessment of information around a specific topic and task.

%% file: 11_conclusions.tex

\section{Conclusions} \label{sec:concl}
In this paper, we have addressed the problem of assessing the interestingness of a cube query in the context of a hierarchical multidimensional database with cubes and level hierarchies. We have performed an extensive survey of the related work, both in the area of computer science and in the area of the study of human behavior. We have focused the discussion on 4 interestingness dimensions, specifically, relevance, surprise, novelty, and peculiarity. For these dimensions of interestingness, we have also proposed specific measures and algorithms for assessing them in a quantitative fashion. We take care to discriminate between result-based algorithms, after the query has been executed and syntax-based algorithms, before the query is executed. We have also explored the runtime behavior of such algorithms, over different sizes and session histories. Finally, we have conducted a user study to determine the significance, as well as the evolution over time, of the different interestingness dimensions.

Future work can continue in different roads. First, although the dimensions of interestingness that we discuss in this paper provided a principled and well-founded setup of how interestingness can be handled, one can only expect that a deeper study -- esp., of the fundamentals, in the area of human behavior-- can reveal more perspectives to the essence of interestingness. \textit{Conciseness} in one such dimension, although defined at a meta-level: the shorter the query description and the query result are, the more easy to comprehend them. The effect of conciseness and aggregation level (which can go hand-in-hand with the size of the query result) has not been studied either.
Another notable dimension concerns the \emph{expression} aspect, in which data are contextualized with respect to the medium used to expressed it (and not displayed) -- e.g., a cube can be described by the set of cells, or by a query, or by a visualization, etc.

Moreover, even for the presented dimensions, it is clear that the presented algorithms are only a first attack to the problem. More algorithms and metrics are possible for the aforementioned dimensions. We have been particularly interested in syntactic checks in this paper, as they allow the prediction of the interestingness of a query without actually executing it. More value-based algorithms, however, are certainly possible. The role of time (but also space, and in general, dimensional context) is also worth pursuing: what is interesting now for an analyst, might be indifferent some time later. Aging, decay factors can be introduced in the assessment of interestingness when queries are compared to the history of the user, or, other users as a matter of fact.

Personal profiles, crowd-wisdom and log mining can be employed to best model user beliefs. We refer the interested reader to \cite{DBLP:conf/kdd/Bie11,DBLP:conf/ida/Bie13} for a starting point, but of course, the problem of belief estimation is a large research territory that can fit gracefully with our taxonomical framework.

The scope of our user study has not studied highly interactive user sessions. The extent that interactivity affects the assessment of interestingness is yet another unexplored territory for future research.

%% file: 99_CommentsKept.tex
\isDraft{

\section{Not Addressed, no time to do it}

5.1.1 EQUALITY OR IDENTITY
\textbf{Full Syntactic Same-Level Assessment of Novelty}.
In this case, the question to be answered is: Given $q$ and $Q$ = $\{q_{1}, \ldots, q_{n}\}$, is there any $q_i~\in~Q$ such that $q~=~q_i$? \rem{PM: = or $\equiv$? PV used = for short}

\hrulefill

5.1.2 THE FORMULA

\textbf{Partial Extensional Detailed Assessment of Novelty}. In this case, the question to be answered is practically the same, albeit with a different means to compute the answer, specifically, cells instead of signatures: Given $q$ and $Q$ = $\{q_{1}, \ldots, q_{n}\}$, can we identify which part of the results of the detailed area of $q^0$ are already covered by the detailed areas of the queries of $Q$?
\sticky{PM: see src for formula!!!}

\textbf{NEW}: The equivalent expression to the algorithm is given by the formula $\frac{|q^{nov^0}|}{|q^0.result|}$,

\noindent where $q^{nov^0}=q^0.result \setminus \bigcup_{i} q_i^{0}.result$, and $q^{cov^0}$ being $q^0.result \setminus q^{nov^0}$


\hrulefill

WE DO \textbf{NOT} HAVE THIS IMPLEMENTATION
\begin{remark}
It is easy to introduce a \textbf{weighted variation} of Algorithm~\ref{algo:EnumDetailedQueryContainment}. \rem{see comment below} Observe that the Algorithm~\ref{algo:EnumDetailedQueryContainment} computes the union of the detailed areas of the queries with set semantics.
\end{remark}

\textbf{COMMENT}: apparently Eirini's algo has no weights, as it has no occurrence counters. Spyros' is the same algo.

\hrulefill

WE DO \textbf{NOT} HAVE THIS IMPLEMENTATION, WE DO NOT HAVE GOALS AS A CONCEPT IN DELIAN

6.1 Then, the \textit{Goal-Based Detailed Syntactic Relevance} (GBDSR) of a query is the fraction of its detailed space that overlaps with the user's goal. \rem{see comment}

\textbf{COMMENT}: used to assume this is Spyros algo 3.1.2.1.a -- goal-based; not really. We do not have anything close to a goal as a concept in Delian. A Goal is a $\phi$ "stand-alone".

\hrulefill

6.2 \textbf{Foundations of history-based relevance assessment}. 

$\ldots$

Then, the sets $q^{cov^{0^+}}$ and $q^{nov^{0^+}}$ (respectively, $q^{cov^0}$ and $q^{nov^0}$) are produced.\rem{ see below } Based on these sets, we can compute \textit{Partial Detailed Syntactic Relevance} (PDSR) and Partial Detailed Extensional Relevance (PDER), respectively.

\textbf{COMMENTS}: Eirini's algo is PDER. 
Spyros' relevance algo's at the code are two. Both are partial and cell-based (extensional) and differ: one is sameLevel and another is detailed.
The detailed one is, therefore, PDER also (as Eirini's).
The same level is \textit{Partial Same-Level Extensional Relevance}.

\hrulefill

8.3 \textbf{Value-based average cell surprise}

\rem{see below} Algorithm~\ref{algo:genericValueSurprise} provides the general recipe for computing the surprise according to the general setup. This generic algorithm can be specialized by fixing the involved functions to specific choices. For example, to compute the \textit{Partial Extensional Average Value-Based Surprise}, Algorithm~\ref{algo:avgValueSurprise} works on a single-measured cube, with absolute distance as the distance function to assess how far the actual and the expected measures are, and averaging over all cells with surprise to produce the aggregate cube surprise.

\textbf{COMMENT} Algorithm~\ref{algo:genericValueSurprise} is almost the same for Spiros and Eirini. Spyros' 3.1.2.3 hashes expected values first + checks the cell existence in the expected values by searching it in the h/m; Eirini searches it in an A/L and has an extra flag if the cell is found. Spiros returns the result normalized in [0.0 - 1.0], Eirini doesn't.

We use Spiros' version after all.

\hrulefill

8.5.1\textbf{ Surprise computed over the labels}

We can think of a generic algorithm (Algorithm~\ref{algo:genericLabelSurprise}) to cover the general case of how to compute the surprise of the entire cube, on the basis of labels.\\

$\ldots$ and much later $\ldots$\\

Assuming the \textit{max} aggregate function for $f_{cell}^{agg}$ computing the surprise of a cell, and the \textit{avg} aggregate function for $f^{agg}$ computing the surprise of a cube query, Algorithm~\ref{algo:genericLabelSurprise} computes the \textit{Partial Max-Average Label-Based Surprise} for a cube query.
Algorithm~\ref{algo:labelStrictStrictSurprise} computing a 
\textit{Full Strict-Strict Label-Based Surprise}, \rem{Eirini's algo strict label surprise} provides a double strict version that simplifies the generic algorithm by assuming strict, Boolean semantics for both the cells and the entire cube query. \\

COMMENTS
\textbf{ MUST FIX BOTH LABEL ALGOS TO WORK WITH LABELS, NOT WITH MEASURES!!!} \\

IMPLEM. ERRORS

We need to fix both algorithms to work with labels and not with the measure of each cell, like they currently do.

\hrulefill

A GENERIC COMMENT

We need to address the concepts of (a) labels, (b) user's goals and (c) user's beliefs. For labels we need to make the algorithms work with labels properly (see the previous comment). As for goals and beliefs, we need to properly implement them as concepts in Delian.

\hrulefill

} 

%% file: 00_ToDo.tex
\section{Schedule of work \& Deadlines}

\begin{CheckList}{Goal}
    \Goal[deadline=mid July]{open}{Jesus Christ help us}
	 \begin{CheckList}{Task}
		 \Task{done}{Address all other reviewer comments} 
          \Task{done}{Write a letter}
          \Task{open}{Produce an arxiv long v.}
          \Task{open}{Shrink to a reasonable size}
          \Task{open}{Proofread!}
		 \Task{open}{Submit!}
	\end{CheckList} 

    \Goal[deadline=mid June]{achieved}{User study} 
        \begin{CheckList}{Task}
            \Task{done}{Create cubes in slides}
            \Task{done}{Package the whole thing into 1-2 pptx CR}
            \Task{done}{Prepare the xls to inject data}
            \Task{done}{Execute the user study}
            \Task{done}{Conduct the analysis of the xls}
            \Task{done}{WriteUp in the text}
         \end{CheckList}
	  
	\Goal[deadline=end June]{achieved}{Writings}
	  \begin{CheckList}{Task}
            \Task{done}{Format cleanly as Elsevier 2 column}
            \Task{done}{Involve the apriori/posteriori more in the paper}           
            \Task{done}{Definition of interestingness (either intro or sec 3) (R1.1)}
            \Task{done}{Merge the two different R.W. (R1.2)}
            \Task{done}{In the R.W. -- if space-- discuss how people combine different inter. dimensions; otherwise, covered by user study (R1.4)}
            \Task{done}{In the R.W. do sth for Sintos' paper (R2.2)}
            \Task{done}{In the R.W. explain why prior work does not apply to cubes (R3.W2)}
            \Task{done}{Involve a reference example +   an example per algo (R1.6, R1.7, R2.1)} 
            \Task{done}{Add a section on the parameters of the problem}
	  \end{CheckList}

    \Goal[deadline=mid July]{open}{[Optionally?] More beef wrt use and performance}
    \begin{CheckList}{Task}
        \Task{open}{Handle better user beliefs: (a) metamodel? (b) declaration via language? (c) elicitation from past sessions? (R3.W4)}
        \Task{open}{Extend the problem framework with more dimensions (??)}
	\Task{open}{optimizations to speed up the detailed areas ?}
	\Task{open}{Mat. views, indexes ?}
    \end{CheckList}

\end{CheckList} 

%% file: __article_CubeInterestingness.bbl
\newcommand{\etalchar}[1]{$^{#1}$}
\begin{thebibliography}{AGG{\etalchar{+}}15b}

\bibitem[Agg15a]{DBLP:books/sp/Aggarwal15}
Charu~C. Aggarwal.
\newblock {\em Data Mining - The Textbook}.
\newblock Springer, 2015.

\bibitem[AGG{\etalchar{+}}15b]{DBLP:journals/dss/AligonGGMR15}
Julien Aligon, Enrico Gallinucci, Matteo Golfarelli, Patrick Marcel, and
  Stefano Rizzi.
\newblock A collaborative filtering approach for recommending {OLAP} sessions.
\newblock {\em Decision Support Systems}, 69:20--30, 2015.

\bibitem[AGM{\etalchar{+}}14]{DBLP:journals/kais/AligonGMRT14}
Julien Aligon, Matteo Golfarelli, Patrick Marcel, Stefano Rizzi, and Elisa
  Turricchia.
\newblock Similarity measures for {OLAP} sessions.
\newblock {\em Knowl. And Inf. Syst.}, 39(2):463--489, 2014.

\bibitem[AKS{\etalchar{+}}21]{DBLP:journals/vldb/AbuzaidKSGXSASM21}
Firas Abuzaid, Peter Kraft, Sahaana Suri, Edward Gan, Eric Xu, Atul Shenoy,
  Asvin Ananthanarayan, John Sheu, Erik Meijer, Xi~Wu, Jeffrey~F. Naughton,
  Peter Bailis, and Matei Zaharia.
\newblock {DIFF:} a relational interface for large-scale data explanation.
\newblock {\em {VLDB} J.}, 30(1):45--70, 2021.

\bibitem[Bie11]{DBLP:conf/kdd/Bie11}
Tijl~De Bie.
\newblock An information theoretic framework for data mining.
\newblock In {\em Proceedings of SIGKDD}, pages 564--572, 2011.

\bibitem[Bie13]{DBLP:conf/ida/Bie13}
Tijl~De Bie.
\newblock Subjective interestingness in exploratory data mining.
\newblock In {\em Proceedings of IDA}, pages 19--31, 2013.

\bibitem[BRV11]{DBLP:conf/icde/BaikousiRV11}
Eftychia Baikousi, Georgios Rogkakos, and Panos Vassiliadis.
\newblock Similarity measures for multidimensional data.
\newblock In Serge Abiteboul, Klemens B{\"{o}}hm, Christoph Koch, and
  Kian{-}Lee Tan, editors, {\em Proceedings of the 27th International
  Conference on Data Engineering, {ICDE} 2011, April 11-16, 2011, Hannover,
  Germany}, pages 171--182. {IEEE} Computer Society, 2011.

\bibitem[CCD{\etalchar{+}}19]{DBLP:conf/dolap/ChansonCDLM19}
Alexandre Chanson, Ben Crulis, Krista Drushku, Nicolas Labroche, and Patrick
  Marcel.
\newblock Profiling user belief in {BI} exploration for measuring subjective
  interestingness.
\newblock In {\em {DOLAP}}, volume 2324 of {\em {CEUR} Workshop Proceedings}.
  CEUR-WS.org, 2019.

\bibitem[CLM{\etalchar{+}}22]{DBLP:conf/edbt/ChansonLMRT22}
Alexandre Chanson, Nicolas Labroche, Patrick Marcel, Stefano Rizzi, and Vincent
  T'kindt.
\newblock Automatic generation of comparison notebooks for interactive data
  exploration.
\newblock In {\em {EDBT}}, pages 2:274--2:284. OpenProceedings.org, 2022.

\bibitem[DDL{\etalchar{+}}19]{DjedainiDLMPV19}
Mahfoud Djedaini, Krista Drushku, Nicolas Labroche, Patrick Marcel,
  Ver{\'{o}}nika Peralta, and Willeme Verdeau.
\newblock Automatic assessment of interactive {OLAP} explorations.
\newblock {\em Inf. Syst.}, 82:148--163, 2019.

\bibitem[{D.E}54]{berlyne54}
{D.E. Berlyne}.
\newblock A theory of human curiosity.
\newblock {\em British Journal of Psychology}, 45(3):180 -- 191, 1954.

\bibitem[DHX{\etalchar{+}}19]{DBLP:conf/sigmod/DingHXZZ19}
Rui Ding, Shi Han, Yong Xu, Haidong Zhang, and Dongmei Zhang.
\newblock {QuickInsights}: Quick and automatic discovery of insights from
  multi-dimensional data.
\newblock In {\em Proceedings of SIGMOD}, pages 317--332, Amsterdam, The
  Netherlands, 2019.

\bibitem[DLMP17]{DBLP:conf/adbis/DjedainiLMP17}
Mahfoud Djedaini, Nicolas Labroche, Patrick Marcel, and Ver{\'{o}}nika Peralta.
\newblock Detecting user focus in {OLAP} analyses.
\newblock In {\em {ADBIS}}, pages 105--119, 2017.

\bibitem[EAPS14]{DBLP:journals/tkde/EirinakiAPS14}
Magdalini Eirinaki, Suju Abraham, Neoklis Polyzotis, and Naushin Shaikh.
\newblock {QueRIE}: Collaborative database exploration.
\newblock {\em {IEEE} Trans. Knowl. Data Eng.}, 26(7):1778--1790, 2014.

\bibitem[EH19]{ElHelaly19}
Sherif El-Helaly.
\newblock {\em The Mathematics of Voting and Apportionment: An Introduction}.
\newblock Springer International Publishing, 2019.

\bibitem[EMS20]{DBLP:conf/sigmod/ElMS20}
Ori~Bar El, Tova Milo, and Amit Somech.
\newblock Automatically generating data exploration sessions using deep
  reinforcement learning.
\newblock In {\em Proceedings of SIGMOD}, pages 1527--1537, Portland, OR, USA,
  2020.

\bibitem[FF01]{DBLP:conf/sbbd/FabrisF01}
Carem~C. Fabris and Alex~Alves Freitas.
\newblock Incorporating deviation-detection functionality into the {OLAP}
  paradigm.
\newblock In {\em SBBD}, pages 274--285, 2001.

\bibitem[FMG10]{FoMG10}
Jens F\"{o}rster, Janina Marguc, and Marleen Gillebaart.
\newblock Novelty categorization theory.
\newblock {\em {Social and Personality Psychology Compass}}, 4(9):736 -- 755,
  2010.

\bibitem[FMPR22]{DBLP:journals/isf/FranciaMPR22}
Matteo Francia, Patrick Marcel, Ver{\'{o}}nika Peralta, and Stefano Rizzi.
\newblock Enhancing cubes with models to describe multidimensional data.
\newblock {\em Inf. Syst. Frontiers}, 24(1):31--48, 2022.

\bibitem[GH06]{DBLP:journals/csur/GengH06}
Liqiang Geng and Howard~J. Hamilton.
\newblock Interestingness measures for data mining: A survey.
\newblock {\em {ACM} Comput. Surv.}, 38(3):9, 2006.

\bibitem[GKM{\etalchar{+}}23]{DBLP:conf/dolap/GkitsakisKMPMV23}
Dimos Gkitsakis, Spyridon Kaloudis, Eirini Mouselli, Ver{\'{o}}nika Peralta,
  Patrick Marcel, and Panos Vassiliadis.
\newblock Assessment methods for the interestingness of cube queries.
\newblock In {\em Proceedings of the 25th International Workshop on Design,
  Optimization, Languages and Analytical Processing of Big Data {(DOLAP)}
  co-located with the 26th International Conference on Extending Database
  Technology and the 26th International Conference on Database Theory
  {(EDBT/ICDT} 2023), Ioannina, Greece, March 28, 2023}, pages 13--22, 2023.

\bibitem[GMN09]{DBLP:conf/dawak/GiacomettiMN09}
Arnaud Giacometti, Patrick Marcel, and Elsa Negre.
\newblock Recommending multidimensional queries.
\newblock In {\em DaWaK}, volume 5691 of {\em Lecture Notes in Computer
  Science}, pages 453--466. Springer, 2009.

\bibitem[GS09]{DBLP:journals/jmlr/GunawardanaS09}
Asela Gunawardana and Guy Shani.
\newblock A survey of accuracy evaluation metrics of recommendation tasks.
\newblock {\em Journal of Machine Learning Research}, 10:2935--2962, 2009.

\bibitem[GT14]{DBLP:journals/dss/GolfarelliT14}
Matteo Golfarelli and Elisa Turricchia.
\newblock A characterization of hierarchical computable distance functions for
  data warehouse systems.
\newblock {\em Decis. Support Syst.}, 62:144--157, 2014.

\bibitem[GVM15]{DBLP:journals/is/GkesoulisVM15}
Dimitrios Gkesoulis, Panos Vassiliadis, and Petros Manousis.
\newblock Cinecubes: Aiding data workers gain insights from {OLAP} queries.
\newblock {\em Inf. Syst.}, 53:60--86, 2015.

\bibitem[HKTR04]{DBLP:journals/tois/HerlockerKTR04}
Jonathan~L. Herlocker, Joseph~A. Konstan, Loren~G. Terveen, and John Riedl.
\newblock Evaluating collaborative filtering recommender systems.
\newblock {\em {ACM} Trans. Inf. Syst.}, 22(1):5--53, 2004.

\bibitem[KB17]{DBLP:journals/tiis/KaminskasB17}
Marius Kaminskas and Derek Bridge.
\newblock Diversity, serendipity, novelty, and coverage: {A} survey and
  empirical analysis of beyond-accuracy objectives in recommender systems.
\newblock {\em TiiS}, 7(1):2:1--2:42, 2017.

\bibitem[KGB{\etalchar{+}}08]{KUMAR2008}
Navin Kumar, Aryya Gangopadhyay, Sanjay Bapna, George Karabatis, and Zhiyuan
  Chen.
\newblock Measuring interestingness of discovered skewed patterns in data
  cubes.
\newblock {\em Decision Support Systems}, 46(1):429 -- 439, 2008.

\bibitem[KH15]{Kidd15}
Celeste Kidd and Benjamin~Y. Hayden.
\newblock The psychology and neuroscience of curiosity.
\newblock {\em Neuron}, 88:449--460, 2015.

\bibitem[KMT99]{KLEMETTINEN1999}
M.~Klemettinen, H.~Mannila, and H.~Toivonen.
\newblock Interactive exploration of interesting findings in the
  telecommunication network alarm sequence analyzer (tasa).
\newblock {\em Information and Software Technology}, 41(9):557 -- 567, 1999.

\bibitem[Lit05]{Litm05}
Jordan Litman.
\newblock Curiosity and the pleasures of learning: Wanting and liking new
  information.
\newblock {\em {Cognition and Emotion}}, 19(6):793--814, 2005.

\bibitem[Loe94]{Loewenstein1994}
George Loewenstein.
\newblock The psychology of curiosity: a review and reinterpretation.
\newblock {\em Psychological Bulletin}, 116(1):75--98, 1994.

\bibitem[MDHZ21]{DBLP:conf/sigmod/00040HZ21}
Pingchuan Ma, Rui Ding, Shi Han, and Dongmei Zhang.
\newblock {MetaInsight}: Automatic discovery of structured knowledge for
  exploratory data analysis.
\newblock In {\em Proceedings of SIGMOD}, pages 1262--1274, 2021.

\bibitem[MPV19]{DBLP:conf/adbis/MarcelPV19}
Patrick Marcel, Ver{\'{o}}nika Peralta, and Panos Vassiliadis.
\newblock A framework for learning cell interestingness from cube explorations.
\newblock In {\em 23rd European Conference on the Advances in Databases and
  Information Systems ({ADBIS} 2019), Bled, Slovenia, September 8-11, 2019},
  volume 11695 of {\em Lecture Notes in Computer Science}, pages 425--440.
  Springer, 2019.

\bibitem[MS20]{DBLP:conf/sigmod/MiloS20}
Tova Milo and Amit Somech.
\newblock Automating exploratory data analysis via machine learning: An
  overview.
\newblock In {\em {SIGMOD}}, 2020.

\bibitem[MTM17]{DBLP:journals/dke/MateTM17}
Alejandro Mat{\'{e}}, Juan Trujillo, and John Mylopoulos.
\newblock Specification and derivation of key performance indicators for
  business analytics: {A} semantic approach.
\newblock {\em Data Knowl. Eng.}, 108:30--49, 2017.

\bibitem[PAB{\etalchar{+}}21]{DBLP:conf/cikm/PersonnazABFS21}
Aur{\'{e}}lien Personnaz, Sihem Amer{-}Yahia, Laure Berti{-}{\'{E}}quille,
  Maximilian Fabricius, and Srividya Subramanian.
\newblock {DORA} {THE} {EXPLORER:} exploring very large data with interactive
  deep reinforcement learning.
\newblock In {\em {CIKM}}, 2021.

\bibitem[RMN12]{ReMN12}
R.~Reisenzein, W.-U. Meyer, and M.~Niepel.
\newblock Surprise.
\newblock In V.~S.~Ramachandran(chief editor), editor, {\em Encyclopedia of
  Human Behavior}. Elsevier, 2nd ed. edition, 2012.

\bibitem[RS14]{2014RouSu}
James Rounds and Rong Su.
\newblock The nature and power of interests.
\newblock {\em Current Directions in Psychological Science}, 23(2):98--103,
  2014.

\bibitem[SAM98]{DBLP:conf/edbt/SarawagiAM98}
Sunita Sarawagi, Rakesh Agrawal, and Nimrod Megiddo.
\newblock Discovery-driven exploration of {OLAP} data cubes.
\newblock In {\em EDBT}, pages 168--182, 1998.

\bibitem[Sar99]{DBLP:conf/vldb/Sarawagi99}
Sunita Sarawagi.
\newblock Explaining differences in multidimensional aggregates.
\newblock In {\em Proceedings of VLDB}, pages 42--53, 1999.

\bibitem[Sar00]{DBLP:conf/vldb/Sarawagi00}
Sunita Sarawagi.
\newblock User-adaptive exploration of multidimensional data.
\newblock In {\em Proceedings of VLDB}, pages 307--316, 2000.

\bibitem[SAY19]{DBLP:journals/pvldb/SintosAY19}
Stavros Sintos, Pankaj~K. Agarwal, and Jun Yang.
\newblock Selecting data to clean for fact checking: Minimizing uncertainty vs.
  maximizing surprise.
\newblock {\em Proc. {VLDB} Endow.}, 12(13):2408--2421, 2019.

\bibitem[SGS18]{DBLP:conf/sigmod/SalimiGS18}
Babak Salimi, Johannes Gehrke, and Dan Suciu.
\newblock Bias in {OLAP} queries: Detection, explanation, and removal.
\newblock In {\em {SIGMOD}}, pages 1021--1035, 2018.

\bibitem[Sil08]{2008Silvia}
Paul~J. Silvia.
\newblock Interest: The curious emotion.
\newblock {\em Current Directions in Psychological Science}, 17(1):57--60,
  2008.

\bibitem[SS01]{DBLP:conf/vldb/SatheS01}
Gayatri Sathe and Sunita Sarawagi.
\newblock Intelligent rollups in multidimensional {OLAP} data.
\newblock In {\em Proceedings of VLDB}, pages 531--540, 2001.

\bibitem[SSR19]{Su201911}
Rong Su, Gundula Stoll, and James Rounds.
\newblock The nature of interests: Toward a unifying theory of trait-state
  interest dynamics.
\newblock In Christopher Nye and James Rounds, editors, {\em Vocational
  Interests in the Workplace: Rethinking Behavior at Work}, page 11 – 38.
  Taylor and Francis, 2019.

\bibitem[THY{\etalchar{+}}17]{DBLP:conf/sigmod/TangHYDZ17}
Bo~Tang, Shi Han, Man~Lung Yiu, Rui Ding, and Dongmei Zhang.
\newblock Extracting top-k insights from multi-dimensional data.
\newblock In {\em {SIGMOD} Conference}, pages 1509--1524. {ACM}, 2017.

\bibitem[Tic00]{DBLP:journals/ese/Tichy00}
Walter~F. Tichy.
\newblock Hints for reviewing empirical work in software engineering.
\newblock {\em Empir. Softw. Eng.}, 5(4):309--312, 2000.

\bibitem[Vas22]{PV21}
Panos Vassiliadis.
\newblock {A Cube Algebra with Comparative Operations: Containment, Overlap,
  Distance and Usability}.
\newblock {\em CoRR}, abs/2203.09390, 2022.

\bibitem[VM18]{DBLP:conf/dolap/VassiliadisM18}
Panos Vassiliadis and Patrick Marcel.
\newblock The road to highlights is paved with good intentions: Envisioning a
  paradigm shift in {OLAP} modeling.
\newblock In {\em Proceedings of DOLAP}, 2018.

\bibitem[VMR19]{DBLP:journals/is/VassiliadisMR19}
Panos Vassiliadis, Patrick Marcel, and Stefano Rizzi.
\newblock Beyond roll-up's and drill-down's: An intentional analytics model to
  reinvent {OLAP}.
\newblock {\em Information Systems}, 85:68--91, 2019.

\bibitem[WSZ{\etalchar{+}}20]{DBLP:journals/tvcg/WangSZCXMZ20}
Yun Wang, Zhida Sun, Haidong Zhang, Weiwei Cui, Ke~Xu, Xiaojuan Ma, and Dongmei
  Zhang.
\newblock Datashot: Automatic generation of fact sheets from tabular data.
\newblock {\em {IEEE} Trans. Vis. Comput. Graph.}, 26(1):895--905, 2020.

\bibitem[YCY06]{DBLP:series/sci/YaoCY06}
Yiyu Yao, Yaohua Chen, and Xue~Dong Yang.
\newblock A measurement-theoretic foundation of rule interestingness
  evaluation.
\newblock In Tsau Young~Lin, Setsuo Ohsuga, Churn-Jung Liau, and Xiaohua Hu,
  editors, {\em Foundations and Novel Approaches in Data Mining}, pages 41--59.
  Springer Berlin Heidelberg, 2006.

\bibitem[ZSZ{\etalchar{+}}17]{DBLP:conf/sigmod/ZhaoSZBUK17}
Zheguang Zhao, Lorenzo~De Stefani, Emanuel Zgraggen, Carsten Binnig, Eli Upfal,
  and Tim Kraska.
\newblock Controlling false discoveries during interactive data exploration.
\newblock In {\em SIGMOD}, pages 527--540, 2017.

\bibitem[ZZZK18]{DBLP:conf/chi/ZgraggenZZK18}
Emanuel Zgraggen, Zheguang Zhao, Robert~C. Zeleznik, and Tim Kraska.
\newblock Investigating the effect of the multiple comparisons problem in
  visual analysis.
\newblock In {\em Proceedings of CHI}, page 479, Montreal, QC, Canada, 2018.

\end{thebibliography}
